\documentclass[11pt]{article}
\pdfoutput=1 % if your are submitting a pdflatex (i.e. if you have images in pdf, png or jpg format)
\usepackage{jheppub}

\usepackage{graphicx}
\usepackage{amsfonts,amsmath,amssymb}
\usepackage{tikz}
\usepackage{framed}
\usepackage{verbatim}
\usepackage{array,setspace,mathrsfs,yfonts,dsfont,bbm,colonequals,amscd}
\usepackage{relsize,suffix,mathtools,cancel,bbm}

\newcommand{\be}{\begin{eqnarray}}
\newcommand{\ee}{\end{eqnarray}}
\newcommand{\bea}{\begin{eqnarray}}
\newcommand{\eea}{\end{eqnarray}}

\newcommand{\bn}{\begin{enumerate}}
\newcommand{\en}{\end{enumerate}}

\title{Compactifications of ADE  conformal matter on a torus}

\preprint{}

\author[a]{Hee-Cheol Kim,\!}
\author[b]{Shlomo S. Razamat,\!}
\author[c]{Cumrun Vafa,\!}
\author[d]{Gabi Zafrir}

\affiliation[a]{Department of Physics, POSTECH, Pohang 790-784, Korea}
\affiliation[b]{Department of Physics, Technion, Haifa, 32000, Israel}
\affiliation[c]{Jefferson Physical Laboratory, Harvard University, Cambridge, MA 02138, USA}
\affiliation[d]{IPMU,  University of Tokyo,  Kashiwa, Chiba 277-8583, Japan}

\emailAdd{heecheol1@gmail.com}
\emailAdd{razamat@physics.technion.ac.il}
\emailAdd{vafa@physics.harvard.edu}
\emailAdd{gabi.zafrir@ipmu.jp}

\abstract
{In this paper we study compactifications of ADE type conformal matter, $N$ M5 branes probing $ADE$ singularity, on torus with flux for global symmetry.  We systematically construct the four dimensional theories by first going to five dimensions and studying interfaces. We claim that certain interfaces can be associated with turning on flux in six dimensions. The interface models when compactified on a circle comprise building blocks for constructing four dimensional models associated to flux compactifications of six dimensional theories on a torus. The theories in four dimensions turn out to be quiver gauge theories and the construction implies many interesting cases of IR symmetry enhancements and dualities of such theories.
}

\begin{document} 

\maketitle
\flushbottom

%%%%%%%%%%%%%%%%%%%%%%%%%%%%%%%%

\section{Introduction}

Often one can construct  conformal field theories as 
fixed point models of several different RG flows. RG flows might explicitly exhibit some of the properties of the fixed point CFT while other properties might only emerge
in the deep IR. These explicitly exhibited properties can be very different depending on the flow.
  
A very rich plethora of examples of flows, terminating in interesting conformal field theories in four dimensions with some supersymmetry, 
 is given by compactifications of $(1,0)$ theories on Riemann surfaces. 
The compactification depends first on the chosen $(1,0)$ model of which we have a wide but controlled variety of examples \cite{HMRV,ZHTV,Bhardwaj:2015xxa}.
 The CFTs inherit symmetry properties of the six dimensional model preserved by the details of the compactification. The details which can have an effect on the symmetry are the background gauge fields one can turn on. These involve holonomies and fluxes, with the latter giving a discrete set of different models while the former often parametrizing the conformal manifolds of the fixed point.
In some cases the same CFT can be obtained as an IR description of a UV complete four dimensional asymptotically free theory. This description might exhibit the same symmetry
properties as the flow starting with six dimensional model,  or they can appear only in the IR. In this paper we discuss a huge variety of examples of such relations between six dimensional and four dimensional flows. 

In particular we consider compactifications of $(G,\widetilde G)$ conformal matter on a torus with flux for the global symmetry for the cases when $G$ is the same as $\widetilde G$. These models can be engineered as the low energy description of $M5$ branes probing transverse $G$ type singularity of the corresponding ALE space.  Such compactifications were considered before for various special instances of $G$. For example, $A_0$ 
\cite{Gaiotto:2009we,Benini:2009mz,Bah:2012dg}, $A$ \cite{Gaiotto:2015usa,Razamat:2016dpl,Bah:2017gph}, $D$ \cite{Kim:2017toz,Kim:2018bpg}, and $ADE$ on a torus with no flux \cite{OSTYKafffguy,OSTYKdf,DelZotto:2015rca}. Here we will perform a uniform analysis for all ADE cases with flux in the $G\times G$ symmetry by  realizing that there is a natural way to get the models in four dimensions by first going through five dimensions. In five dimensions  the theories are given by $G$ type affine quiver theories when the six dimensional models are put on a circle with proper choices of holonomies. We will argue that the flux for the global symmetry can be obtained in five dimensions as a sequence of duality interfaces relating affine quiver models  with different mass parameters. The non obvious part of the statement is to find the description of the four dimensional theories living on the interfaces. In the cases relevant for us we will identify these as constructed from weakly coupled fields.  Upon reduction to four dimensions we then will obtain theories having Lagrangians.  These involve pairs of  quiver theories in the shape of affine Dynkin diagrams with ${\cal N}=1$ matter content and where the links of the quiver are chiral bifundamentals. We will discover that there are certain choices which define the interface theory, which in turn determine the details of the chiral matter content of the theory. 
%There are $rank(G)$ independent choices for the chirality of the fields, as there are $rank(G)$ links, where $r$ is the rank of the corresponding ADE. 
% In addition there are bifundamental field between the nodes of the two pair of affine quivers whose chirality gives us additional $r$ choices. 
 Altogether there are  $2Rank(G)$ independent choices and they correspond to fluxes which we believe will cover arbitrary flux in the $G\times G$ global symmetry, as long as the flux is integral\footnote{By integral flux, we mean fluxes obeying the flux quantization condition. It is possible to also have fluxes that do not obey the quantization condition, which we shall refer to as fractional fluxes, if they are accompanied by additional elements compensating for it, see \cite{Kim:2017toz} for examples and details.}.  
 We show that this is indeed the case in many examples.
 It would be interesting to clarify whether we get all possible fluxes in this way which we leave for future work.  In the (A,A) case there is an additional $U(1) $ and we do not know how to construct interfaces corresponding to it.  In fact the flux in the $U(1)$ symmetry of class ${\cal S}$, that is $A_0$ compactification,  do not have known weakly coupled Lagrangian, see for example \cite{Nardoni:2016ffl,Fazzi:2016eec, Bah:2017gph}, so we expect naively this should be rather non-trivial in general.

\begin{figure}[htb]
\begin{center}
\includegraphics[scale=0.46]{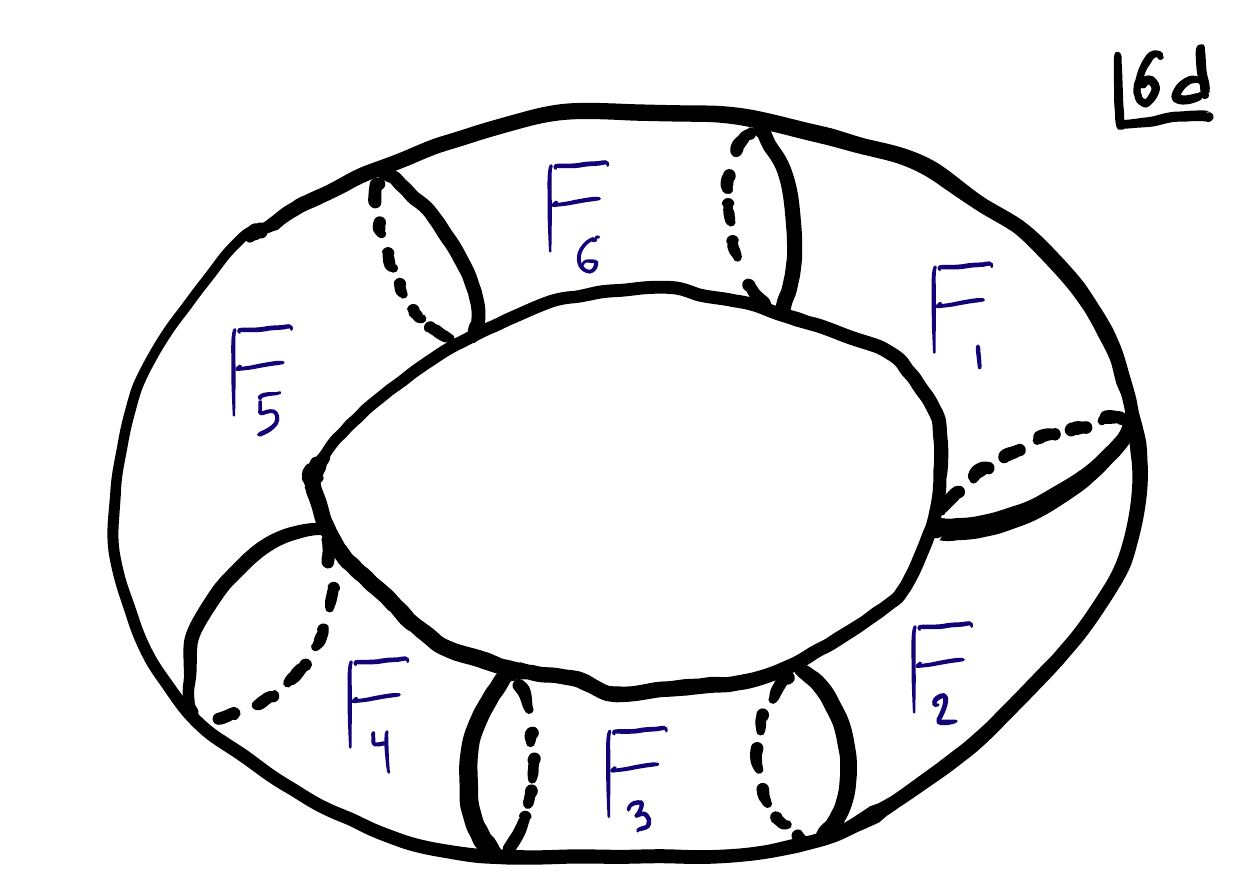}\;\;\,\,
\includegraphics[scale=0.48]{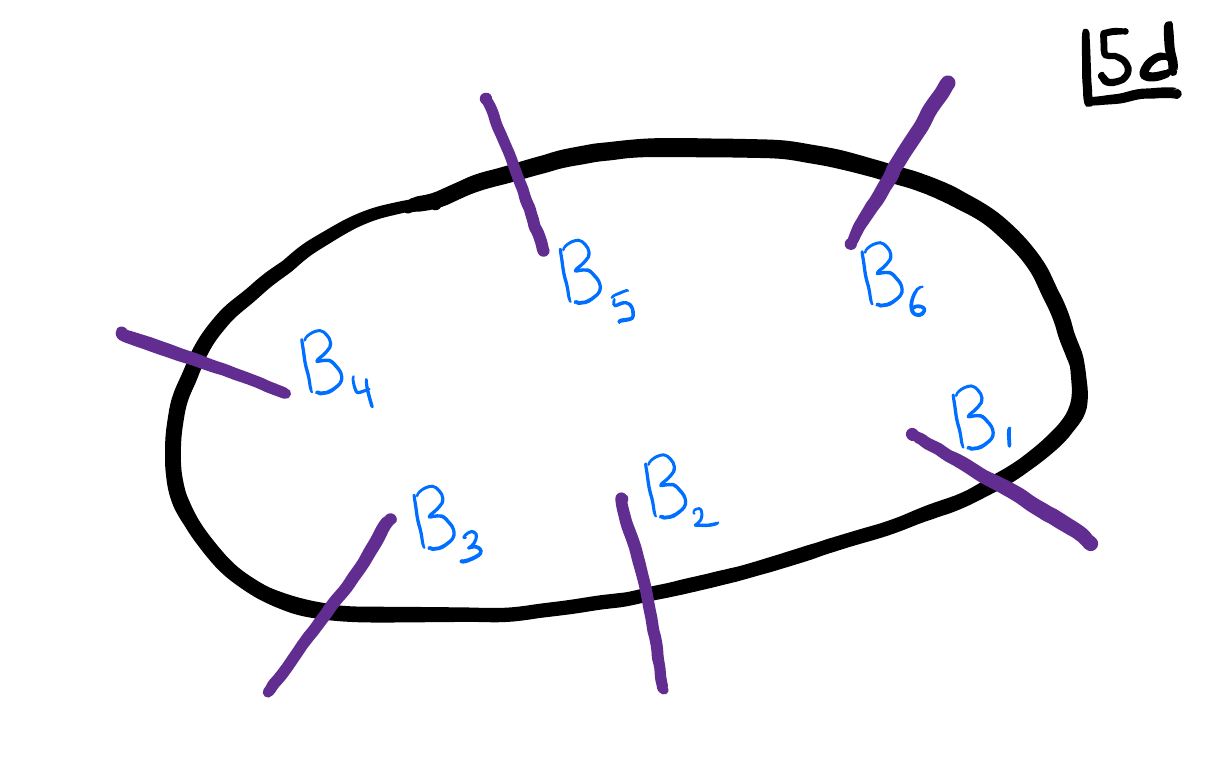}
\caption{Compactification on a torus in six dimensions with flux $F$ for the global symmetry is constructed as a combination of blocks. Each block is associated with a tube and flux $F_j$ such that $\sum F_j =F$. The blocks are obtained by first going to five dimensions, considering then interfaces $B_j$, and then compactifying on an additional circle. This provides a systematic way to construct compactifications. 
}\label{fhur}
\end{center}
\end{figure}

We will engineer theories corresponding to compactification on torus with flux $F$ by combining together block theories to which we associate flux $F_i$ such that $\sum_j F_j = F$.  See Figure \ref{fhur}. The block theories will exhibit only abelian symmetries corresponding to the Cartan of the six dimensional model. For general values of flux this is also the expected symmetry of the theory compactified on the torus. However, for special values of flux the symmetry will contain non abelian factors.  The typical situation for us is that we have a conformal manifold for the corresponding conformal theories each having (or arising from the IR limit of) weak coupling gauge theories, involving distinct quiver like theories. In some cases the dynamics of the  gauge theories turns out to be rather interesting. For example, a way to view the quiver theories will be as a sequence of flows starting from weak coupled UV free theory flowing to IR which is strongly coupled, and then gauging additional global symmetries.  The enhancement to the non abelian symmetry will emerge in this way of obtaining the models only at certain strongly coupled points.
We will thus define a  dictionary between four dimensional quiver theories and six dimensional compactifications. The check of this dictionary will involve anomaly computations and observation of the expected symmetry. Moreover, for the consistency of the considerations certain dualities should hold true. In some cases these are well known IR equivalences, while in other we will obtain novel types of dualities.   

Let us here mention an important puzzle we do not resolve in this paper. Although our procedure passes all the tests for closed Riemann surfaces and tubes with integer flux, our basic minimal   blocks, naively associated to tubes with fractional flux, 
do not pass the check of anomaly matching with six dimensional computation. There are two possible resolutions of this puzzle. One is that the minimal blocks do not correspond to tubes and only combining several of them such that the flux is integer corresponds to a tube. Second would be that there are subtleties with anomaly computation that we miss. We will define precisely our conjectures and leave this interesting puzzle for future work.

This paper is a third in a sequence following \cite{Kim:2017toz} and \cite{Kim:2018bpg}. In the former we analyzed the case of $D_4$ minimal conformal matter, rank one E-string, on a torus and on general surfaces. The latter discussed minimal $D$ conformal matter on a torus but using a different five dimensional description than we do here. The different five dimensional descriptions lead to the interesting novel dualities we have mentioned.

The paper is organized as follows. In section two we discuss the six dimensional models and general issues of their reduction to five dimensions. In section three we discuss the six dimensional models on a circle. 
We will discuss the interface models and formulate the general conjecture of the relation of these to compactifications down on an additional circle. In section four we perform checks of the conjecture in four dimensions.

\section{Six dimensions}\label{sec:sixdim}

We consider the $6d$ SCFT living on $N$ M$5$-branes probing a transverse ${\mathbb C}^2/\Gamma$ singularity. Here $\Gamma$ is a discrete subgroup of $SU(2)$, which is known to have an ADE classification. We shall use the notation $G$ for the ADE Lie group associated with $\Gamma$. 

We next summarize some of the properties of these SCFTs that will be useful later. The most important property of the SCFTs that we need is their global symmetries. The Lie algebra of the global symmetry of these SCFTs is known to be $G \times G$, with the $A$ case having an extra $U(1)$\footnote{The symmetry is also enhanced in some special cases, as will become apparent from the low-energy gauge theory descriptions of these SCFTs that shall be discussed momentarily.}. To get a better understanding of both the global structure, and the $4d$ expectations from the compactification, we should also consider some elements of the operator spectrum of these theories.  

For this it is useful to consider a different representation of these SCFTs. Besides the string theory construction, these theories can also be realized as UV completions of gauge or semi-gauge quiver theories, which can be employed to uncover some of their properties. In this description a special role is played by the $N=1$ cases, the so called minimal $(G,G)$ conformal matter \cite{ZHTV}. The reason for that is that the generic $N$ cases can be built by taking $N$ minimal $(G,G)$ conformal matter theories and connecting them by gauging the symmetry $G$.

For example, take the $A_{k-1}$ case. Here the $N=1$ case is just a theory of $k^2$ free hypermultiplets, that can be grouped to form an $SU(k)\times SU(k)$ bifundamental. The $N=2$ case is then given by taking two such bifundamentals and connecting them by identifying and gauging an $SU(k)$ group. This leads to the $6d$ gauge theory $SU(k)$ with $2k$ fundamental hypermultiplets. For generic $N$ we have $N$ bifundamentals connected via $SU(k)$ gauging, leading to the $6d$ quiver gauge theory containing $N-1$ $SU(k)$ gauge groups connected by bifundamental hypermultiplets, with $k$ fundamental hypermultiplets for each of the groups at the ends of the quiver.

In the $D_k$ case, the minimal conformal matter theory is a $USp(2k-8)$ gauge theory with $2k$ hypermultiplets in the fundamental representation. Therefore, the general $N$ case is now an alternating $SO-USp$ $6d$ quiver gauge theory. The low-energy description for the $E$ theories can also be constructed in this way, though the minimal conformal matter theories get progressively more involved. We refer the reader to \cite{ZHTV} for a complete description of the low-energy theories for every $G$.

From the low-energy descriptions it is possible to read some of the operator spectrum of the SCFTs, where we shall concentrate on the feature shared for every $G$. First there are the moment map operators, which contain a scalar in the adjoint of $G \times G$ and in the $\bold{3}$ of $SU(2)_R$, the R-symmetry of the theory. Additionally, all the SCFTs contain a bifundamental scalar operator in the $(\bold{F}_G, \bold{F}_G)$ of $G \times G$ where $\bold{F}_G$ is the fundamental representation of $G$. This operator transform in the $\bold{N p_G+1}$ dimensional representation of $SU(2)_R$, where $p_G$ is a group dependent constant whose values for the various groups is given in table \ref{GrpData}.   

Besides these, there are various other operators which are group specific. For instance, in the $A$ case we naively have baryon operators\footnote{For a study of the Higgs branch chiral ring operators in the $A$ type case, which are an interesting subset of the operators of the SCFT, see \cite{HZ2018}.}. In the $D$ case, it is known that the minimal conformal matter theory possesses a non-perturbative state in the spinor of the $SO$ group\cite{HM2018,Kim:2018bpg}, and it is thus expected to lead to bispinor states in the non-minimal case. While it may be interesting to gain a better understanding of the operator spectrum of these SCFTs, we shall not follow this further here.

One interesting observation that follows from our studies so far is that the global symmetry group of these SCFTs appear to be $\frac{G \times G}{Z_G}$. Here $Z_G$ stands for the center of $G$, and the modded group is the diagonal center of the two groups. For the readers convenience we have summarized these discrete groups for the relevant choices of $G$ in table \ref{GrpData}.

\begin{table}[h!]
\begin{center}
\begin{tabular}{|c||c||c||c|c|c|}
  \hline 
    & $SU(k)$ & $SO(2k)$ & $E_6$ & $E_7$ & $E_8$ \\
\hline\hline
  $p_G$ & $1$ & $2$  & $4$ & $6$ & $12$ \\ 
\hline
  $Z_G$ &
   ${\mathbb Z}_k$ & \begin{tabular}{@{}c@{}}  ${\mathbb Z}_2 \times {\mathbb Z}_2$, $k$ even \\ ${\mathbb Z}_4$, $k$ odd \end{tabular} & ${\mathbb Z}_3$ & ${\mathbb Z}_2$ & $1$ \\ 
\hline
  $|\Gamma|$ & $k$ & $4k-8$  & $24$ & $48$ & $120$ \\
 \hline
  $r_G$ & $k-1$ & $k$  &
   $6$ & $7$ & $8$ \\
 \hline
   $d_G$ & 
   $k^2 - 1$ & $k(2k-1)$  & $78$ & $133$ & $248$ \\
 \hline
   $d_F$ & 
   $\frac{1}{2}$ & $1$  & $3$ & $6$ & $30$ \\
 \hline
   $t_{G}$ & 
   $2k$ &
    $2k-8$  & $0$ & $0$ & $0$ \\
 \hline
   $u_{G}$ & $2$ & $4$  & $6$ & $8$ & $12$ \\
 \hline
   $h^{\vee}$ & $k$ & $2k-2$  & $12$ & $18$ & $30$ \\
 \hline
\end{tabular}
 \end{center}
\caption{Various data used in this paper. Here $Z_G$, $r_G$, and $d_G$ are the center, rank and dimension of the group $G$ respectively. $|\Gamma|$ is the order of the finite group $\Gamma$. $d_F$ and $h^{\vee}$ are the Dynkin index of the fundamental representation and the dual Coxeter number of the group $G$. $p_G, t_{G}$ and $u_{G}$ are various group dependent constants.  }
\label{GrpData}
\end{table}

\

\subsection*{Anomalies from $6d$}

 We can estimate the anomalies of the $4d$ theories resulting from the compactification of the $6d$ theory, using the anomaly polynomial of the $6d$ SCFT. For that we first need the expression for it, which was evaluated in \cite{OSTY}. The result can be written down for any group $G$ where it reads:

\bea
&I  =&  \frac{1}{24} (|\Gamma|^2 N^2 - 2 N (|\Gamma| r_G + |\Gamma| - 1 ) + d_G - 1 ) C^2_2 (R)\nonumber\\ &-  & \frac{1}{48} (N (|\Gamma| r_G + |\Gamma| - 2 ) - d_G + 1 ) p_1 (T) C_2 (R) \\ & - & \frac{(|\Gamma| N - h^{\vee})}{4 d_F} C_2 (R) \left( C_2 (G_1)_{\bold{F}} + C_2 (G_2)_{\bold{F}} \right) + \frac{h^{\vee}}{48 d_F} p_1 (T) \left( C_2 (G_1)_{\bold{F}} + C_2 (G_2)_{\bold{F}} \right) \nonumber \\  & + & \frac{(36 N u_G + d^2_F N t_G - 3)}{24 N d^2_F} \left( C^2_2 (G_1)_{\bold{F}} + C^2_2 (G_2)_{\bold{F}} \right) - \frac{1}{4 N d^2_F} C_2 (G_1)_{\bold{F}} C_2 (G_2)_{\bold{F}} \nonumber \\ & - & \frac{t_G}{12}\left( C_4 (G_1)_{\bold{F}} + C_4 (G_2)_{\bold{F}} \right) + \frac{(30N+7d_G-23) p^2_1 (T) - 4(30N+d_G-29) p_2 (T)}{5760} \nonumber
\eea

Here $C_2 (R)$ stands for the second Chern class in the fundamental representation of $SU(2)_R$, and $p_1 (T), p_2 (T)$ stand for the first and second Pontryagin classes respectively. We also employ the notation $C_n(G)_{\bold{R}}$ for the n-th Chern class of the global symmetry $G$, evaluated in the representation $\bold{R}$ (here $F$ stands for fundamental). The rest of the symbols are various group theoretic constants whose values are given in table \ref{GrpData}.

Here we only write the anomalies for symmetries that appear generically. As previously mentioned, in the $A$ case there is an extra $U(1)$ and the expression can be extended to include it. This case was studied extensively in \cite{Bah:2017gph}, and we refer the reader there for more information. 

We next consider compactifying the theory on a torus and turning on non-trivial flux under various $U(1)$ subgroups of the global symmetry $G \times G$. By integrating the anomaly polynomial $8$-form of the $6d$ theory on the Riemann surface we get the anomaly polynomial $6$-form of the resulting $4d$ theory\cite{Benini:2009mz}. 

To do this we first need to decompose the various characteristic classes to those of the symmetries preserved in the presence of flux. First, the flux breaks half of the supersymmetry so that out of the original $8$ supercharges only $4$ remain. This corresponds to $\mathcal{N} = 1$ in $4d$. This also leads to the $SU(2)_R$ symmetry of the $6d$ theory being broken down to its $U(1)$ Cartan, which becomes an R-symmetry in $4d$. At the level of characteristic classes, these two are related by $C_2 (R) = - C^2_1 (R)$.

We also need to decompose the flavor symmetry characteristic classes to those of the symmetry preserved by the flux. In general, a symmetry $G$ is broken to $G\rightarrow (\prod U(1)_i) \times (\prod G'_a)$, where $G'_a$ are assumed to be non-abelian. In that case we can decompose:

\be
C_2 (G)_{\bold{F}} = -2 \sum_{i,j} \xi_{i j} C_1 (U(1)_i) C_1 (U(1)_j) + \sum_a \mathfrak{i}_a C_2 (G'_a)_{\bold{F}},
\ee

\bea
C_4 (G)_{\bold{F}} & = & -2 \sum_{i,j,k,l} \lambda_{i j k l} C_1 (U(1)_i) C_1 (U(1)_j)C_1 (U(1)_k) C_1 (U(1)_l)\\ &+& \sum_{i,j}\sum_{a} \tau^{a}_{i j} C_1 (U(1)_i) C_1 (U(1)_j) C_2 (G'_a)_{\bold{F}} +\sum_{i}\sum_{a} \rho^{a}_{i} C_1 (U(1)_i) C_3 (G'_a)_{\bold{F}} + ..., \nonumber
\eea 
with the additional terms integrating to zero. 

We next need to take the flux into account. This is done by setting $C_1 (U(1)_i) = - z_i t + \epsilon_i C_1 (R) + C_1 (U(1)^{4d}_i)$, where $t$ is a unit $2$-form on the torus. The first term then takes the flux into account as $\int_{T^2} C_1 (U(1)^{4d}_i)=-z_i$. The other terms then account for the $4d$ curvature of the $U(1)$, particularly the third term. The second term can be introduced to take account of the possible mixing of the $U(1)$ with the R-symmetry. With this terms $C_1 (R)$ measures the curvature of $U(1)^{6d}_R + \sum_i \epsilon_i U(1)^{4d}_i $. If one desires, the anomalies for the superconformal R-symmetry can be evaluated this way, with $\epsilon_i$ determined via a-maximization.

All that remains is to evaluate the various constants appearing in the decomposition and perform the integration.  We will not detail these computations as they are quite straightforward. In what follows we will only quote the result in various specific instances of various reductions from six dimensions. Reader interested in more details on the integration of anomaly polynomials from six to four dimensions can consult for example \cite{Benini:2009mz} and \cite{Razamat:2016dpl,Kim:2017toz}.

%where here 

\section{Five dimensions}

Let us consider 6d $(G,G)$ conformal matter theories compactified on a long cylinder. When the circle radius is small and with certain choices of holonomies for the global symmetries, the conformal matter theories reduce to affine ADE quiver gauge theories in 5d \cite{ZHTV}. We can also consider flavor flux along the cylinder in 6d. As studied in \cite{Chan:2000qc, Kim:2017toz, Kim:2018bpg}, the 6d flux introduces interfaces, which we call {\it flux domain walls}, in the 5d gauge theories. In this section, we propose Lagrangian constructions of these flux domain walls in the 5d quiver gauge theories. The five dimensional models then will be compactified to four dimensions leading to Lagrangians for torus or tube compactifications of the conformal matters.

\subsection{A-type domain walls}\label{sec:a-type-walls}
We begin with flux domain walls in affine $A_{k-1}$ quiver gauge theories. For $N$ M5-branes, the 5d theory is a circular quiver gauge theory consisting of $k$ $SU(N)$ gauge groups connected via bifundamental hypermultiplets of $SU(N)_i\times SU(N)_{i+1}$ symmetry (with $SU(N)_{k+1}=SU(N)_1$). Classically, this theory has $U(1)^{k}$ flavor symmetries of $k$ bifundamental hypermultiplets and $U(1)^k$ topological instanton symmetries for the $k$ gauge nodes. We however expect that these classical abelian symmetries, when combined together, enhance in the UV to the $SU(k)_\beta\times SU(k)_\gamma\times U(1)_t$ symmetry of the 6d $(SU(k)_\beta,SU(k)_\gamma)$ conformal matter theory by quantum instanton states. Here one $U(1)$ global symmetry is identified with the Kaluza-Klein (KK) symmetry along the 6d circle which will be ignored in what follows. In our notation, the $i$-th bifundamental hypermultiplet carries charges $(Q_{\beta_i},Q_{\gamma_i},t) = (1,-1,1)$ under the $U(1)_{\beta_i}\times U(1)_{\gamma_i}\times U(1)_t \subset SU(k)_\beta\times SU(k)_\gamma\times U(1)_t$ flavor symmetry.

\subsection*{Interfaces}

Domain walls in 5d theories can be constructed by joining two 5d theories by a certain 4d interface which is defined with boundary conditions of 5d fields and their couplings to extra degrees of freedom living at the interface. Since the 6d fluxes we are interested in preserve  one half of the supersymmetries, the corresponding flux domain walls in 5d must be 1/2 BPS domain walls.
We first suggest a type of 4d interfaces which can consistently couple to 5d 1/2 BPS boundary conditions and then identify this domain wall configuration with the flux domain wall of the 6d theory.
The domain wall construction discussed in this subsection works also for other domain walls in the D- and E-type cases with minor changes.

The first step is to impose 1/2 BPS boundary conditions at the interface ($x^4=0$) for 5d theories of the  two chambers $x^4<0$ and $x^4>0$ respectively. We will choose Neumann boundary condition for the vector multiplets which sets the gauge fields at $x^4=0$ as
\begin{equation}\label{eq:Neumann}
	\partial_{4}A_\mu = 0 \ (\mu=0,1,2,3) \ , \quad A_4 = 0 \ .
\end{equation}
The 5d vector multiplets with this boundary condition reduce to 4d $\mathcal{N}=1$ vector multiplets at $x^4=0$. Therefore, we have $G\times G'$ gauge symmetries at the interface coming from the 5d gauge fields in the left chamber (for $G=SU(N)^k$) and in the right chamber (for $G'=SU(N)'{}^k$) respectively. For non-minimal D and E cases which we will discuss later, the gauge symmetry at the interface is a pair of two affine $D$- and $E$-type quiver gauge symmetries, respectively. When $N=1$, on the other hand, the $SU(1)$ gauge nodes in the affine quiver diagrams are replaced by two fundamental hypermultiplets for the adjacent gauge nodes.

For each bifundamental hypermultiplet with scalar fields $\Phi=(X,Y)$, we have two choices of boundary conditions:
\begin{eqnarray}\label{eq:bchyper}
	1) \ \partial_4 X = Y=0 \quad {\rm or} \quad 2) \ \partial_4 Y=X =0 \ .
\end{eqnarray}
Under this 1/2 BPS boundary condition, a 5d hypermultiplet reduces to a 4d $\mathcal{N}=1$ chiral multiplet at $x^4=0$ involving the scalar field, $X$ or $Y$, with Neumann boundary condition.
We will denote the first boundary condition by $+$ sign and the second boundary condition by $-$ sign. So the boundary condition of $k$ bifundamental matters is labeled by a vector $\mathcal{B}=\{s_1,s_2,\cdots,s_k\}$ with $k$ signs $s_i=\pm$. Since there are two 5d theories ending on the interface from both sides, we need a set of boundary conditions $(\mathcal{B},\mathcal{B}')$ for the 5d hypermultiplets in the first and the second chambers of the 5d theory. For our domain walls, we shall impose the same boundary conditions $\mathcal{B}= \mathcal{B}'$.

We now couple 4d degrees of freedom at the interface to the 5d boundary conditions. First, we introduce at the interface 4d chiral multiplets $q_i$ in $(\bar{\bf N},{\bf N})$ representation of $SU(N)_{i}\times SU(N)_{i}'$ symmetry for $i=1,2,\cdots,k$.  
In addition, we add 4d bifundamental chirals $\tilde{q}_i$ of $SU(N)_{i+1}\times SU(N)_{i}'$ or $SU(N)_{i}\times SU(N)_{i+1}'$ coupled to the other fields by the cubic superpotential of the forms
\begin{equation}\label{eq:domainwall-W}
	\mathcal{W}_{x^4=0}=\sum_{i=+} \left( \tilde{q}_iq_i X_i + q_{i+1}\tilde{q}_iX_i'\right) +\sum_{i=-} \left( Y_i \tilde{q}_i q_{i+1}+Y_i'  q_i \tilde{q}_i\right) \ ,
\end{equation}
where $X_i,Y_j$ and $X_i',Y_j'$ stand for the 4d chiral multiplets involving 5d bifundamental scalars with Neumann boundary condition in the first and the second chambers, respectively. This superpotential equates the boundary conditions on two sides, i.e. $\mathcal{B}=\mathcal{B}'$, as expected. Lastly, we add flip chiral fields coupled to the baryonic operators of the 4d chirals $q_i$.

We can also consider similar domain walls by replacing the representations of 4d chiaral fields $q_i$ by $({\bf N},\bar{\bf N})$  and by coupling 4d fields $q_i$ and $\tilde{q}_i$ to the 5d boundary conditions through the superpotential of the form (\ref{eq:domainwall-W}) accordingly. We remark here that these two choices of 4d fields $q_i$ in either $(\bar{\bf N},{\bf N})$ or $({\bf N},\bar{\bf N})$ lead to two different types of domain walls: the former gives domain walls for the flux on $SU(k)_\beta$, and the latter leads to domain walls for the flux on $SU(k)_\gamma$. We will distinguish these two types of domain walls by the subscript $\mathcal{T}=\beta$ or $\gamma$.
We will first discuss the domain walls for $\mathcal{T}=\beta$ with $q_i$ in $(\bar{\bf N},{\bf N})$ and then discuss the domain walls for $\mathcal{T}=\gamma$ with $q_i$ in $({\bf N},\bar{\bf N})$ later.

\begin{figure}[htbp]
\begin{center}
\includegraphics[scale=0.35]{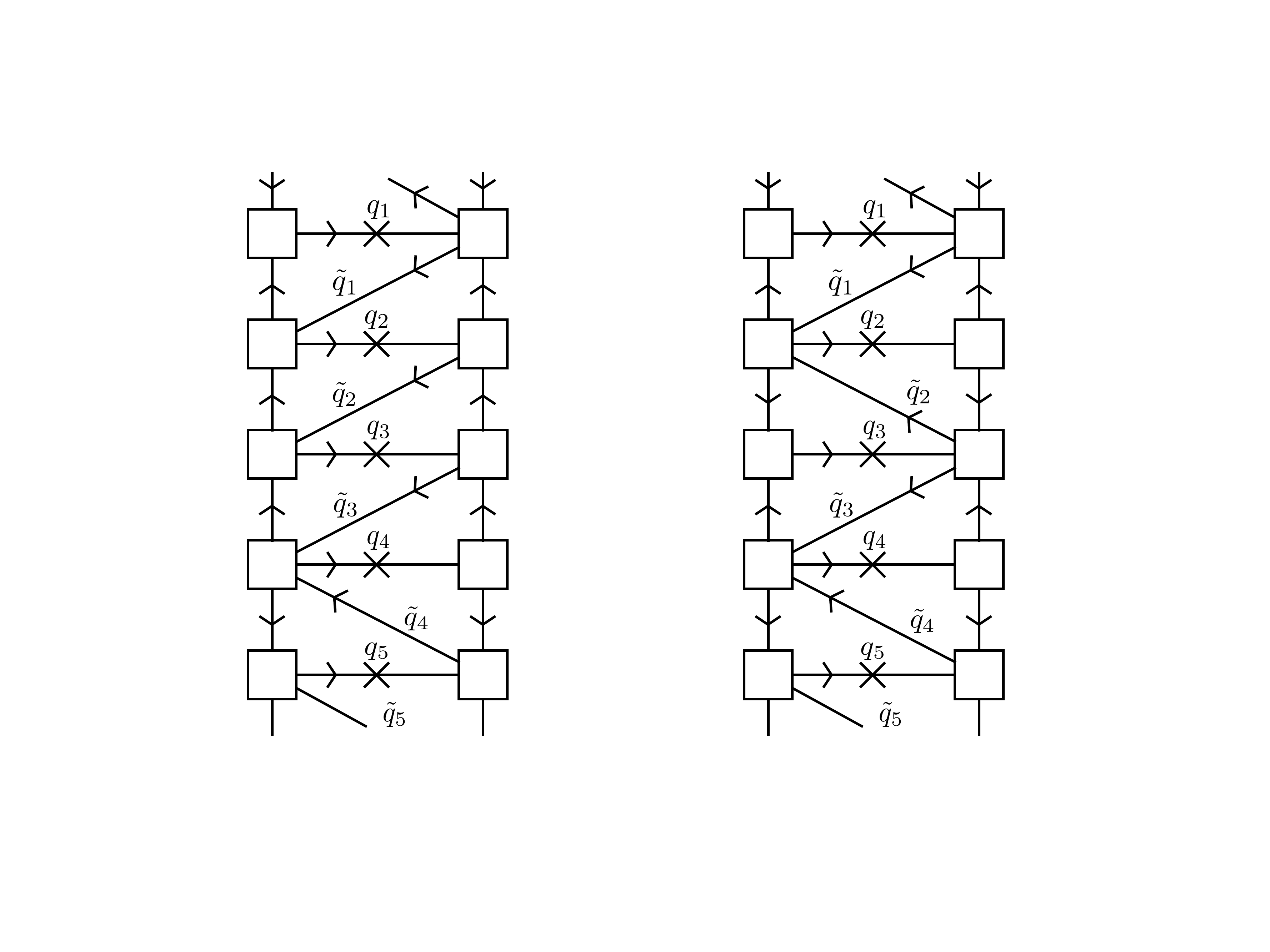}
\caption{Quiver diagrams for the domain walls in the $A_4$ quiver gauge theory. The domain wall on the left is for the boundary condition $\mathcal{B}=\{+,+,+,-,-\}$ and the domain wall on the right is for $\mathcal{B}=\{+,-,+,-,-\}$. The square boxes denote the $SU(N)^5\times SU(N)'{}^5$ gauge symmetries of 5d gauge theories on both sides of the walls. The symbol $\times$ denotes flip fields coupled to the baryonic operator made from $q_i$.}\label{fig:an-walls}
\end{center}
\end{figure}

Figure \ref{fig:an-walls} depicts two domain wall examples with boundary conditions $\mathcal{B}=\{+,+,+,-,-\}$ and $\mathcal{B}=\{+,-,+,-,-\}$ in the $A_4$ quiver gauge theory. There are cubic superpotentials of the form (\ref{eq:domainwall-W}) for the triangles in the quiver diagrams. The boxes in the quiver diagrams represent the $G\times G'$ symmetries at the interface and these symmetries will be gauged by the 5d vector multiplets with Neumann boundary condition in two chambers\footnote{We shall generically use boxes for 4d global symmetries and circles for gauge symmetries. When discussing interfaces in 5d we use boxes for symmetries gauged by 5d vector multiplets as, later when we discuss the reduction to 4d on intervals, these become 4d global symmetries.}.

The boundary conditions and the 4d couplings at the interface define a domain wall in the 5d gauge theory. Let us now check if this domain wall is consistent with the 5d gauge theory. The boundary conditions of the 5d bulk fields induce non-trivial 4d gauge anomalies at the interface. For being a consistent domain wall, these gauge anomalies must be canceled by the extra 4d fields living on the boundary.

Let us first discuss cubic anomalies of the $SU(N)^k$ gauge symmetries. The $i$-th hypermultiplet with boundary condition $s_i=\pm$ leaves a bifundamental chiral multiplet of $SU(N)_{i}\times SU(N)_{i+1}$ gauge symmetry at the boundary. This chiral multiplet leads to cubic gauge anomalies of the $SU(N)_{i}$ and $SU(N)_{i+1}$ symmetries given by
\begin{equation}
	Tr(SU(N)_{i}^3)=\frac{s_iN}{2}\ , \quad Tr (SU(N)_{i+1}^3)=-\frac{s_iN}{2}\ .
\end{equation}
Remember that we always need to multiply by the factor $\frac{1}{2}$ to all anomaly contributions from the 5d hypermultiplets at boundaries \cite{Gaiotto:2015una,Horava:1996ma,Horava:1995qa}. This comes from the fact that the anomaly contributions of a chiral multiplet coming from the 5d boundary condition equals one half of those from  a 4d chiral multiplet with the same charges.

We also need to take into account the anomalies from the 4d bifundamental chiral multiplets $q_i$ and $\tilde{q}_j$.
The 4d chiral field $q_i$ contributes to the $SU(N)_{i}$ anomaly as $Tr(SU(N)_{i}^3) = -N$. Another chiral field $\tilde{q}_{i}$ has cubic gauge anomaly $Tr(SU(N)_{i+1})=N$ for $s_i=+$ and $Tr(SU(N)_{i})=N $ for $s_i=-$.
 One can easily see that the total cubic gauge anomalies in the domain wall vanish when we sum over all anomaly contributions from the boundary conditions and the 4d chiral multiplets. The cubic gauge anomalies of $SU(N)_{i}'$ in the other chamber are canceled in the same way.

We then move on to the gauge-global mixed anomalies at the interface. Firstly, there are anomaly inflow contributions from the 5d bulk gauge theory. The boundary condition of the $i$-th bifundamental hypermultiplet with $s_i$ in the first chamber induces the following anomalies at the boundary
\begin{eqnarray}
	&&Tr(U(1)_tSU(N)_{i}^2)=Tr(U(1)_tSU(N)_{i+1}^2)=\frac{s_iN}{4} \ , \nonumber \\ 
	&& Tr(U(1)_{\beta_i}SU(N)_{i}^2)=Tr(U(1)_{\beta_i}SU(N)_{i+1}^2) =\frac{s_iN}{4} \ ,  \nonumber  \\
	&& Tr(U(1)_{\gamma_i}SU(N)_{i}^2)=Tr(U(1)_{\gamma_i}SU(N)_{i+1}^2) = -\frac{s_iN}{4} \ ,
\end{eqnarray}
with $\sum_iU(1)_{\beta_i}=\sum_i U(1)_{\gamma_i}=0$.
Also, the 5d $SU(N)_{i}$ vector multiplet with Neumann boundary condition leads to the anomaly inflow contributions toward the 4d boundary as
\begin{equation}
	Tr(U(1)_R SU(N)_{i}^2) = \frac{N}{2} \ ,
\end{equation}
where $U(1)_R \subset SU(2)_R$.

In addition, there are anomaly inflows from the gauge kinetic terms $\frac{4\pi^2}{g_i^2}Tr(F^2_i)$. These terms can be considered as the 5d $\mathcal{N}=1$ mixed Chern-Simons terms between the $U(1)_{I_i}$ instanton symmetry and the $SU(N)_i$ gauge symmetry with background scalar field $\frac{4\pi^2}{g_i^2}$ in the $U(1)_{I_i}$ vector multiplet. In the presence of the 4d boundary, these CS-terms generate anomaly inflows toward the boundary. 

It should be noted that the contribution of this term is novel in this construction, and did not appear in previous discussions of 5d domain walls in relation to the compactifications of 6d SCFTs to 4d, like in \cite{Kim:2017toz,Kim:2018bpg}. The distinguishing feature in the cases discussed here is that the 5d gauge theories contain more then one gauge group. Generically the topological symmetries of the 5d gauge theory, together with the flavor symmetry, appear to form an affine version of the global symmetry of the SCFT, where the affine extension being associated with the Kaluza-Klein tower of the 5d conserved current, which is expected to build the 6d one. Therefore, these contain one additional $U(1)$ which does not survive the 4d reduction. In cases with a single gauge group in 5d, the topological $U(1)$ is usually related to this symmetry, and so the contribution of the gauge kinetic term is unneeded as we are only concerned with anomalies of 4d symmetries after the 4d reduction. However, in the cases we consider here, the 5d gauge theory has many gauge groups, and their topological symmetry should be related to symmetries appearing in 4d, with the exception of one combination. Therefore, the 5d gauge kinetic terms should contribute to the anomalies of the 4d theories and must be taken into account.
In fact, the 4d chiral fields $q_i$ and $\tilde{q}_i$ also carry the charges of this Kaluza-Klein symmetry and these charges are uniquely fixed by the gauge-global mixed anomaly cancellation and cubic superpotential terms. We will however ignore these charges as we are interested only in 4d symmetries.

 The instanton number $I_i$ and the baryon symmetry $B_i$ for the $i$-th gauge node are related to the Cartan generators $H_{i,\pm}$ of the enhanced $SU(k)\times SU(k)$ symmetry as \cite{Tachikawa:2015mha,Yonekura:2015ksa}
\begin{equation}\label{eq:Cartans-symmetry}
	H_{i,\pm} = \frac{1}{4}\sum_{j}A_{ij}I_j \pm \frac{B_i}{2N} \ ,
\end{equation}
where $A_{ij}$ is the Cartan matrix of $A_{k-1}$ symmetry.
The mass parameters $m_{i,\pm}$ for the Cartans $H_{i,\pm}$ are associated to the gauge couplings $g_i$ and the mass parameters $m_{B,i}$ for $B_i$ as
\begin{equation}\label{eq:coupling-baryon}
	\frac{8\pi^2}{g_i^2} = \frac{1}{2}\sum_i A_{ij}(m_{j,+} + m_{j,-}) \ , \quad m_{B,i} = \frac{m_{i,+}-m_{i,-}}{N_i} \ ,
\end{equation}
where $N_i$ is $h^\vee_i$ for $i$-th gauge node.
This implies that the kinetic term for the $SU(N)_{i}$ symmetry induces the 4d anomaly inflows as
\begin{eqnarray}\label{eq:instanton-anomaly}
	&&Tr(U(1)_{\beta_i} SU(N)_{i}^2) = Tr(U(1)_{\gamma_i} SU(N)_{i}^2) = \frac{N}{4} \ , \nonumber \\
	&&Tr(U(1)_{\beta_i} SU(N)_{i+1}^2) = Tr(U(1)_{\gamma_i} SU(N)_{i+1}^2) = -\frac{N}{4} \ .
\end{eqnarray}
We have similar anomaly inflow contributions for the $SU(N)_{i}'$ gauge symmetries from the 5d boundary conditions in the other chamber.

\begin{figure}[htbp]
\begin{center}
\includegraphics[scale=0.4]{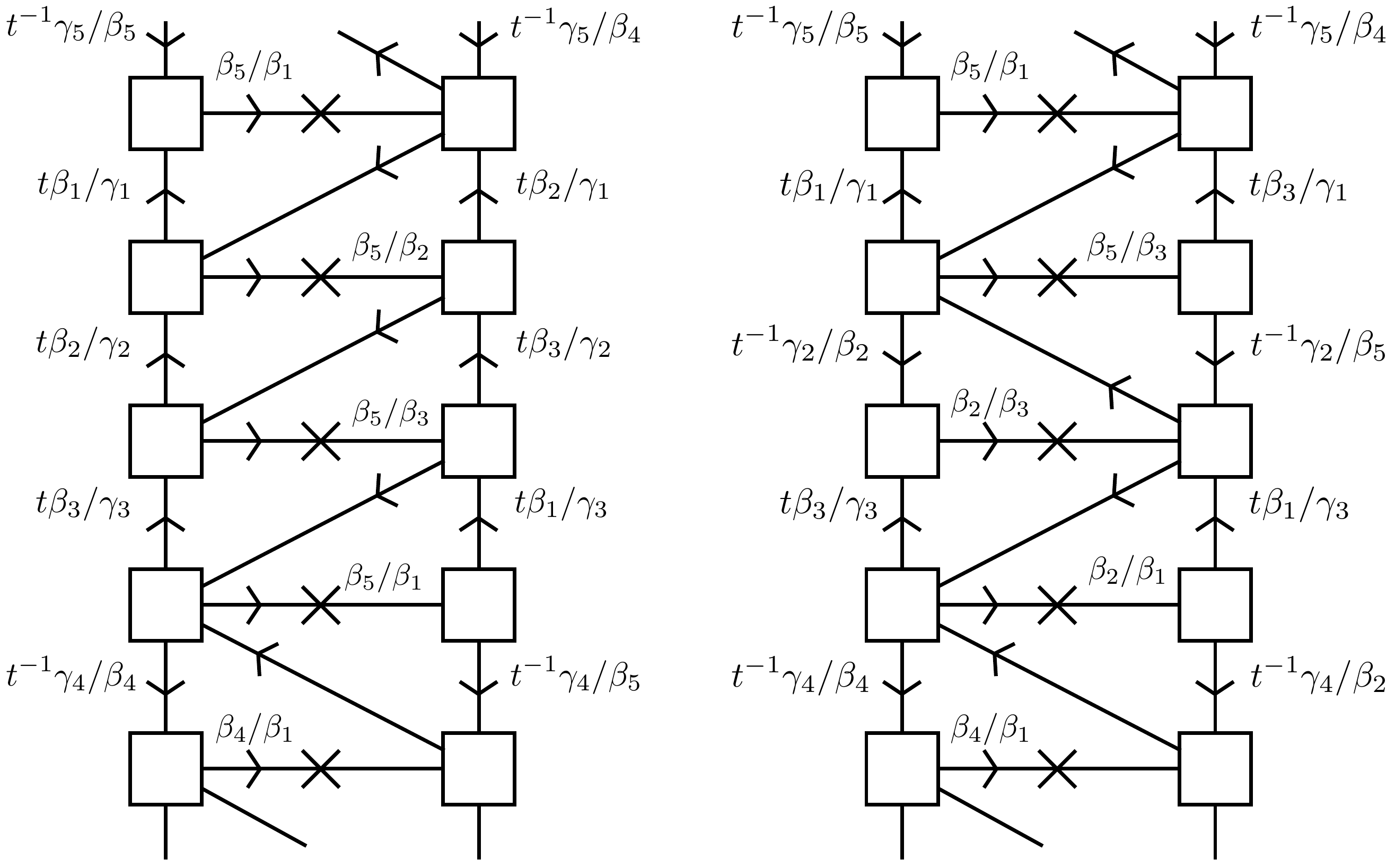}
\caption{Domain walls for $\mathcal{B}=\{+,+,+,-,-\}$ (left) and $\mathcal{B}=\{+,-,+,-,-\}$ (right). The $U(1)$ global charges of the 4d chiral fields denoted by their fugacities are fixed by the gauge-global mixed anomaly cancellation and superpotential terms. Here the permutation $\sigma_\beta=(1\;2\;3)(\;4\;5)$ and $\sigma_\gamma=\emptyset$ for the left tube. For the right tube $\sigma_\beta=(\;1\;3\;) (2\;5\;4\;)$ amd $\sigma_\gamma=\emptyset$.}\label{fig:an-walls-fugacity}
\end{center}
\end{figure}

The bulk contributions to the gauge-global mixed anomalies are not canceled by themselves, so the $U(1)_{\beta_i}$ and $U(1)_{\gamma_i}$ symmetries will be broken unless these anomalies are canceled by those from the 4d fields at the interface. It turns out that all the $U(1)$ flavor symmetry charges of the 4d chiral multiplets at the interface are uniquely fixed by requiring that all the Cartans of $SU(k)_\beta\times SU(k)_\gamma \times U(1)_t$ are gauge anomaly free, and that there are no additional flavor symmetries together with the superpotential constraints, with the exception of the two cases with the most symmetric boundary conditions, i.e. $s_i=+$ or $s_i=-$ for all $i$'s. We demand this property for for the domain walls realizing the 6d flux because the 6d flux compactified on a circle breaks no Cartans of the flavor symmetry. Under this requirement, for example, $U(1)_R$ charges for the 4d chiral multiplets $q_i$ and $\tilde{q}_i$ are fixed to be $0$ and $+1$ respectively. Two examples of domain walls in the $A_4$ quiver theory are presented in Figure \ref{fig:an-walls-fugacity}. Here the $U(1)$ charges of the 4d fields, which are determined by the this requirement, are denoted by the $U(1)^{k-1}_\beta\times U(1)^{k-1}_\gamma\times U(1)_t$ fugacities.

On the other hand, when $\mathcal{B}=(+,+,+,\cdots,+)$ or $\mathcal{B}=(-,-,-,\cdots,-)$ (so when $\mathcal{B}$ is the most symmetric), we find that there exists an additional $U(1)$ global symmetry apart from the bulk symmetry which does not arise from the circle reduction of the 6d theory with flux. Thus, we lose an interpretation for the most symmetric cases as a compactification of the six dimensional theory with flux. So we will not discuss the most symmetric boundary conditions from now on, however see the next section for a possible roundabout interpretation in four dimensions.

\subsection*{The relation between four and six dimensions}

We have constructed consistent domain wall configurations for 5d boundary condition $\mathcal{B}$'s. Let us now relate these domain walls in the 5d gauge theory to the 6d theory compactified on a 2d surface with flux.

We note that this domain wall permutes the $U(1)$ global symmetries of the 5d theory. More precisely, when we pass through it, the $U(1)_{\beta_i}$ symmetries acting on the hypermultiplets with the `$+$' boundary condition are cyclically permuted among themselves, and similarly the $U(1)_{\beta_j}$ symmetries on the hypermultiplets with `$-$' boundary condition are permuted.
We will label such permutations for $U(1)^{k-1}_\beta$ and $U(1)^{k-1}_\gamma$ by $\sigma_\beta$ and $\sigma_\gamma$ respectively. For a given $\mathcal{B}$, the $\sigma_\beta(\mathcal{B})$ is defined as a clockwise permutation of $U(1)_{\beta_i}$ symmetries with $s_i=+$ and a counterclockwise permutation of $U(1)_{\beta_j}$ symmetries with $s_j=-$. The permutation $\sigma_\gamma(\mathcal{B})$ is trivial for the above domain walls involving the 4d chiral fields $q_i$ with representations associated with the choice $\mathcal{T}=\beta$. As we will propose soon, these domain walls are associated to $SU(k)_\beta$ flux in 6d. We will construct another type of domain walls with non-trivial $\sigma_\gamma(\mathcal{B})$ below which come with 4d fields $q_i$ of other type with $\mathcal{T}=\gamma$.  Note however that the permutations will not specify the domain wall model in a unique way. This is because the permutations are invariant under cyclic permutations of $+$ and $-$, whereas the corresponding interface theories are different.

The definition of the permutations coming with the interfaces theory suffices for us  
to  make
 the basic statement about relation of the interface models and compactifications to four dimensions.
We conjecture the followings:
 \begin{framed}
\noindent {\bf Conjectures}	
\begin{enumerate}
	\item A flux domain wall with total flux $F_{\rm tot}$ in the 5d affine $G$ quiver theory on a circle realizes the 6d $(G,G)$ conformal matter theory with flux $F_{\rm tot}$ on a torus.
	\item When $\prod_{i=1}^l\sigma^{t_i}=1$, the flux domain wall with total flux $F_{\rm tot}$ in the 5d affine $G$ quiver theory on an interval realizes the 6d $(G,G)$ conformal matter theory with flux $F_{\rm tot}$ on a cylinder.
	% \item When we form a flux domain wall by gluing ${\rm lcm}(r,k-r)$ copies of a minimal flux domain wall and then combine such flux domain walls together to form a general domain wall configuration with flux $F_{\rm tot}=\sum_i {\rm lcm}(r_i,k-r_i)\cdot F_i$, this domain wall realizes the 6d $(A_k,A_k)$ conformal matter theory with flux $F_{\rm tot}$ on a cylinder.
\end{enumerate}
\end{framed}
Here, $G=A_k$ for A-type domain walls and we defined $\sigma^{t_i} = (\sigma_\beta(\mathcal{B}_i),\sigma_\gamma(\mathcal{B}_i))$ for $i$-th domain wall with boundary condition $\mathcal{B}_i$. We will propose the same conjectures also with $G=D_k,E_k$ for D-type and E-type flux domain walls which will be discussed below in detail.

The flux of the single domain wall, which we will call basic domain wall, is to be computed soon and the precise procedure to glue tubes together will be discussed. The total flux $F_{\rm tot}$ will be the sum of the contributions from each domain wall with the permutation of the symmetries properly considered. Therefore, even when we naively connect a domain wall to a copy of itself, as we are required to permute the symmetries, the flux we shall associate with the resulting domain wall is not twice the flux of the original one. Also when closing a tube on itself, to make a torus compactification, some symmetries may be broken. This then forces the flux to distribute accordingly, eliminating the flux from broken symmetries. As a result the flux associated with the closed surfaces may not be the same as the one associated with the tube when symmetries are broken upon closing the surface. 

For combinations of the domain walls which do not satisfy the condition, $\prod_{j=1}^n \sigma^{t_j}=1$, in particular the basic domain wall, we do not have a suggestion for the Riemann surface it is to be associated with. We merely use the basic walls as building blocks for constructing theories which we can identify with the compactifications. There are several reasons we do not make claims about the basic walls and we will discuss them here. First, we have not found an association of the flux to the basic domain wall such that the anomalies will agree with the six dimensional computation. This can be because either the walls not satisfying the condition do not correspond to compactifications or that there are subtleties with the computation of anomalies we miss. Another issue is that, as we will see soon, there is a natural way to associate flux to the basic blocks such that for surfaces satisfying the conditions given above, the anomalies of the 4d theories agree with the computations of anomalies from 6d. This flux however for a single wall is not properly quantized, which again hints that there is an issue with treating basic walls as arising in compactifications.  Here we should mention that improperly quantized fluxes for surfaces with punctures have occurred before  \cite{Razamat:2016dpl,Kim:2017toz}. While it is  important to resolve the fate of the basic tubes and the way they can be related to compactification, we will leave this for the future. Here we stress again that we only claim the statements appearing in the conjectures.\footnote{Let us  mention in which way  fractional fluxes can appear when one considers theories with punctures.
In 6d, we can turn on a  flux for the $SU(k)_\beta$ symmetry, like $F=\big(\overbrace{1/r,\cdots,1/r}^{r},-1/(k\!-\!r),\cdots,-1/(k\!-\!r)\big)$. This flux breaks the $SU(k)_\beta$ symmetry to $U(1)\times SU(r)\times SU(k\!-\!r)$. In this case, since the flux is fractional, we also need to turn on center fluxes in the subgroup $SU(r)\times SU(k\!-\!r)$.  These center fluxes lead to a cyclic ${\mathbb Z}_r\times {\mathbb Z}_{k-r}$ rotation on the $SU(r)\times SU(k\!-\!r)$ holonomies.  
  In the 5d reduction, the flux should  be realizable as a domain wall and the corresponding ${\mathbb Z}_r\times {\mathbb Z}_{k-r}$ actions become cyclic permutations of $U(1)^r\times U(1)^{k-r}\subset SU(k)_\beta$ symmetries as we move across the domain wall. The basic domain wall models we constructed behave in many ways like these tubes, for example they give same permutations, yet we do not  claim that they are the same models.
  
   }

%We stress that we will find that computation of the anomalies of the tubes does not support the assignment of flux for the tube but the conjectures stated do satisfy all the checks. That is the minimal blocks we have constructed {\textit{do not correspond}} to compactifications by themselves but models which one builds from them satisfying the conditions stated  in the conjectures{\it  do correspond }to flux  compactifications.

Below we will provide several evidences for these conjectures with examples by comparing anomalies of the 5d theory with flux domain walls against the expected anomalies of the 6d theory with the corresponding flux.

\subsection*{Gluing}

Let us explain how to connect two flux domain walls with boundary conditions $\mathcal{B}_1$ and $\mathcal{B}_2$ together. General domain walls can be constructed by repeating this gluing procedure. We consider the first domain wall with boundary condition $\mathcal{B}_1$ located at $x^4=t_1$ and then add the second domain wall with boundary condition $\mathcal{B}_2$ at $x^4=t_2$.  First, the vector multiplets in three chambers satisfy Neumann boundary condition, so the theory with the domain walls has $SU(N)^k_1 \times SU(N)^k_2\times SU(N)^k_3$ gauge symmetry. The hypermultiplets in the first and the third chambers will couple to the 4d chiral fields $q_i,\tilde{q}_i$ and $q_i',\tilde{q}_i'$ at two interfaces through cubic superpotentials of the form (\ref{eq:domainwall-W}). Now the 5d theory in the second chamber is put on a finite interval between $t_1$ and $t_2$. So at low energy the theory in the second chamber reduces to a 4d theory with $SU(N)^k_2$ gauge group. The chiral halves of the hypermultiplets satisfying Neumann boundary conditions at both ends reduce to 4d chiral multiplets. If a hypermultiplet in the second chamber satisfies opposite boundary conditions at the two ends, this hypermultiplet becomes massive and at low energy they are truncated. After integrating out the massive hypermultiplet, the cubic superpotentials involving this hypermultiplet turn into quartic superpotentials between the 4d chiral fields $q$ and $\tilde{q}$:
\begin{equation}\label{eq:domainwall-WW}
	\mathcal{W}' = \sum_{i=(+,-)}\left(q_{i+1} \tilde{q}_i \tilde{q}_i'q_{i+1}' \right)+\sum_{i=(-,+)} \left(q_{i} \tilde{q}_i \tilde{q}_i'q_{i}' \right) \ ,
\end{equation}
where $i=(s_1,s_2)$ runs over the massive hypermultiplets with boundary conditions $s_a$ at $t_a$. 

We shall consider various combinations of basic domain walls aligned along a spatial direction $x^4$. The gluing of two basic domain walls can naturally be generalized to the cases with multiple domain walls. In particular, when we identify the first and the last chambers, we will get a 5d system compactified on a circle along which a number of basic domain walls are distributed. Note that, when the first and the last chambers are identified, the hypermultiplets in the new chamber reduce to 4d chiral fields or are truncated in the same way as those in the second chamber in the two domain wall example above. Thus this system reduces to a 4d $\mathcal{N}=1$ quiver gauge theory at low energy. Following the above conjectures, we expect the resulting 4d theories implements torus compactifications of the 6d theory with fluxes.

\subsection*{Assignment of fluxes}

To derive an assignment of flux let us study the structure of the linear anomaly in six dimensions. We here will make the treatment general for $G$ type conformal matter. The fluxes we will discuss are for the Cartan of the $G\times G$ symmetry. For $A$ type we have an additional $U(1)$ symmetry but we do not construct models corresponding to flux for this symmetry.
From the anomaly polynomial in six dimensions we obtain that this anomaly in four dimensions is,

\be \label{sfdrtyet}
Tr G_i = n_i N \frac{h^\vee}{d_F} Q_i\,.
\ee Here $Q_i$ is the flux for the $U(1)$ subgroup $G_i$ in $G$ and $n_i$ is determined by the embedding of the $U(1)$ in $G$.  Here $N$ is the number of branes probing the singularity. We can absorb $N$ into the definition of $n_i$ however, in the way we will normalize the symmetries in all cases, $N$ will appear linearly in linear anomaly.
 On the other hand with a little thought, and we will discuss this in examples below, the only fields contributing to this anomaly in the field theory construction are the flip fields for non-minimal cases. It is thus natural to define the flux in the symmetry $G_i$ to be the sum of $G_i$ charges of the flip fields. The logic, assuming the theories built from the two punctured spheres and correspond to closed surfaces are the correct ones, and we conjecture they are, is as to follow. The gravity anomalies are proportional to the sum of charges of the flip fields for non-minimal cases
\be
Tr G_i =a^{(G,G)}_i   N
\sum_
{f} q_{i,f}\,.
\ee Here the sum is over flip fields and $a^{(G,\, G)}_i$ is a constant which depends on the symmetry and the type of conformal matter, we have that

\be
Q_i = a^{(G,\; G)}_i\frac{d_F}{n_i h^\vee} \sum_f q_{i,f}\,.
\ee That is the flux is the same as sum over charges  up to normalization which only depends on the compactification type and the symmetry. 
 Note that the anomaly scales as $N$ in six dimensions and the only fields giving a scaling with $N$ are the flip fields with other behaving quadratically. It is then that in case the models correspond to compactifications the anomalies only come from the flips. For all the cases we studied, we find a rather simple formula for the flux as
 \be\label{eq:domainwall-flux}
 	Q_i = \frac{1}{N h^\vee} \sum_f q_{i,f}\,,
 \ee
 in the orthogonal basis of the flavor symmetry $G\times G'$ which is the basis we will use  in this section for the flavor symmetries of ADE conformal matters.

The fluxes for minimal cases, on the other hand, are not solely determined by the charges of flip fields. Here we shall instead use the full linear anomalies, where the flux is chosen such that the linear anomalies of the 4d theories match those expected from 6d. 
For example, the flux of a basic domain wall can be determined by using the 4d tube theory with this domain wall. We compare the linear anomalies of this tube theory with those of the 6d theory on a tube involving both the geometric contributions and the puncture contributions which we will discuss in detail soon. Although we do not expect this tube theory matches the compactification of the 6d theory since $\sigma\neq 1$ for this case, we use this comparison to fix the flux of the basic domain wall.

We claim that with this identification of flux the anomalies for tori match between all the different computations both for minimal and non-minimal cases. This will be true for any closed Riemann surface if the total flux is integer in proper sense. If it is not then the anomaly only agrees for components of symmetry which have integer flux.  

%We propose that, when a basic domain wall is put on an interval, the sum of linear anomalies from 4d chiral fields and half the contribution of the 5d chiral fields with Neumann boundary condition in the first chamber, while not including the anomalies from 5d chiral fields in the second chamber, should be the same as the geometric contribution of the linear anomaly in (\ref{sfdrtyet}) of the 6d theory on a tube, and this determines the fluxes $Q_i$ for basic domain walls.

Our basic domain walls carry fluxes only on either $G_\beta$ or on $G_\gamma$ of the $G_\beta\times G_\gamma$ symmetry depending on the representations of the 4d chiral fields $q_i$ denoted by $\mathcal{T}=\beta,\gamma$, and the explicit form of $Q_i$ is fixed by the boundary condition $\mathcal{B}=\{\pm,\pm,\cdots,\pm\}$. So we will label the basic domain walls by $\mathcal{D}=\mathcal{B}_{\mathcal{T}}$. General flux domain walls carrying both $G_\beta$ and $G_\gamma$ fluxes can be built by joining flux domain walls of two  types $\mathcal{T}=\beta$ and $\mathcal{T}=\gamma$. 

%An interesting question is what is the range of fluxes we can construct using the tube models. In general we expect to have $2rank(G)$ independent choices with certain integrality conditions. The number of different constructions following the choices of the conditions at the interfaces matches the counting of independent variables. We thus do cover at least a sublattice of fluxes of maximal dimension.  For closed Riemann surface we could construct a model for any legal choice of flux. For surfaces with punctures we can construct theories for any integer flux. In principle the tubes, as already mentioned, allow fractional fluxes due to the possibility of center fluxes, see \cite{Kim:2017toz} for discussion of these. However, our construction does not produce such theories correctly and we will discuss this issue in detail.

We will now discuss examples of compactifications of different types of conformal matter.   We will discuss the prescription to associate theories to surfaces in more detail and give examples of various checks one can perform.

\subsection*{More general models and useful    examples}

\begin{figure}[htbp]
\begin{center}
\includegraphics[scale=0.4]{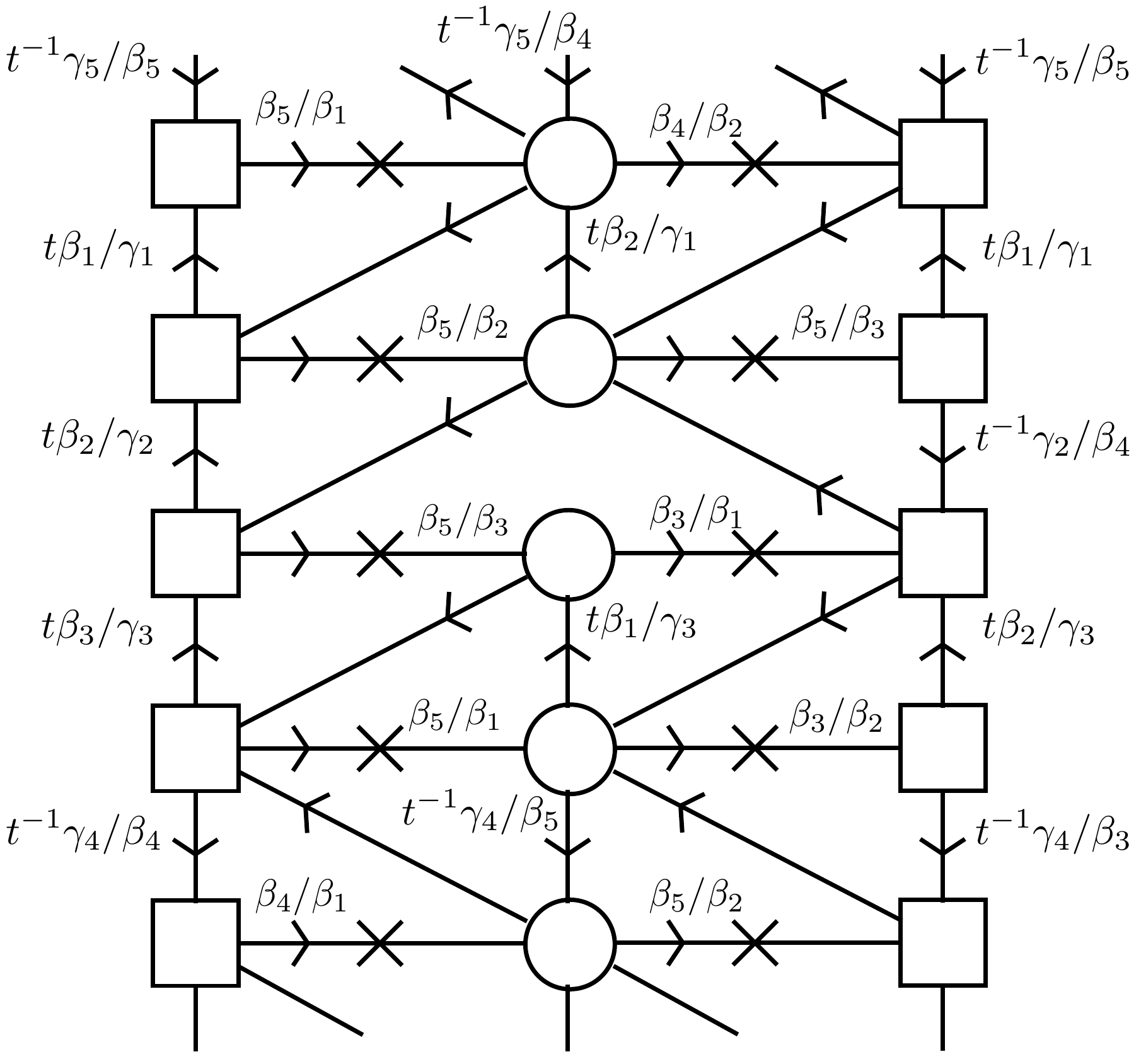}
\caption{Gluing two domain walls with $\mathcal{D}_1=\{+,+,+,-,-\}_\beta$ and $\mathcal{D}_2=\{+,-,+,-,-\}_\beta$. The total flux of the final domain wall is $F_{\rm tot}=(3/5,1/5,1/5,-1/5,-4/5)_\beta+(1/5,3/5,-1/5,-1/5,-2/5)_\beta=(4/5,4/5,0,-2/5,-6/5)_\beta$.}\label{fig:a4-2walls}
\end{center}
\end{figure}

For example, when we connect two domain walls in Figure \ref{fig:an-walls-fugacity}, we get a bigger domain wall with flux $F=2/5(2,2,0,-1,-3)$ for $SU(5)_\beta$ drawn in Figure \ref{fig:a4-2walls}.
Three or more domain walls can also be connected together by using the above gluing rules for each pair of adjacent domain walls. Also, by identifying two 5d theories in the first and the last chambers, we can construct the 5d $A_k$ quiver gauge theory on a circle with flux domain walls that corresponds to the 6d theory compactified on a torus with flux. 
% After the identification, the hypermultiplets in the new chamber reduce to 4d chiral fields or are truncated in the same way as those in the second chamber in the two domain wall example above.

\begin{figure}[htbp]
\begin{center}
\includegraphics[scale=0.35]{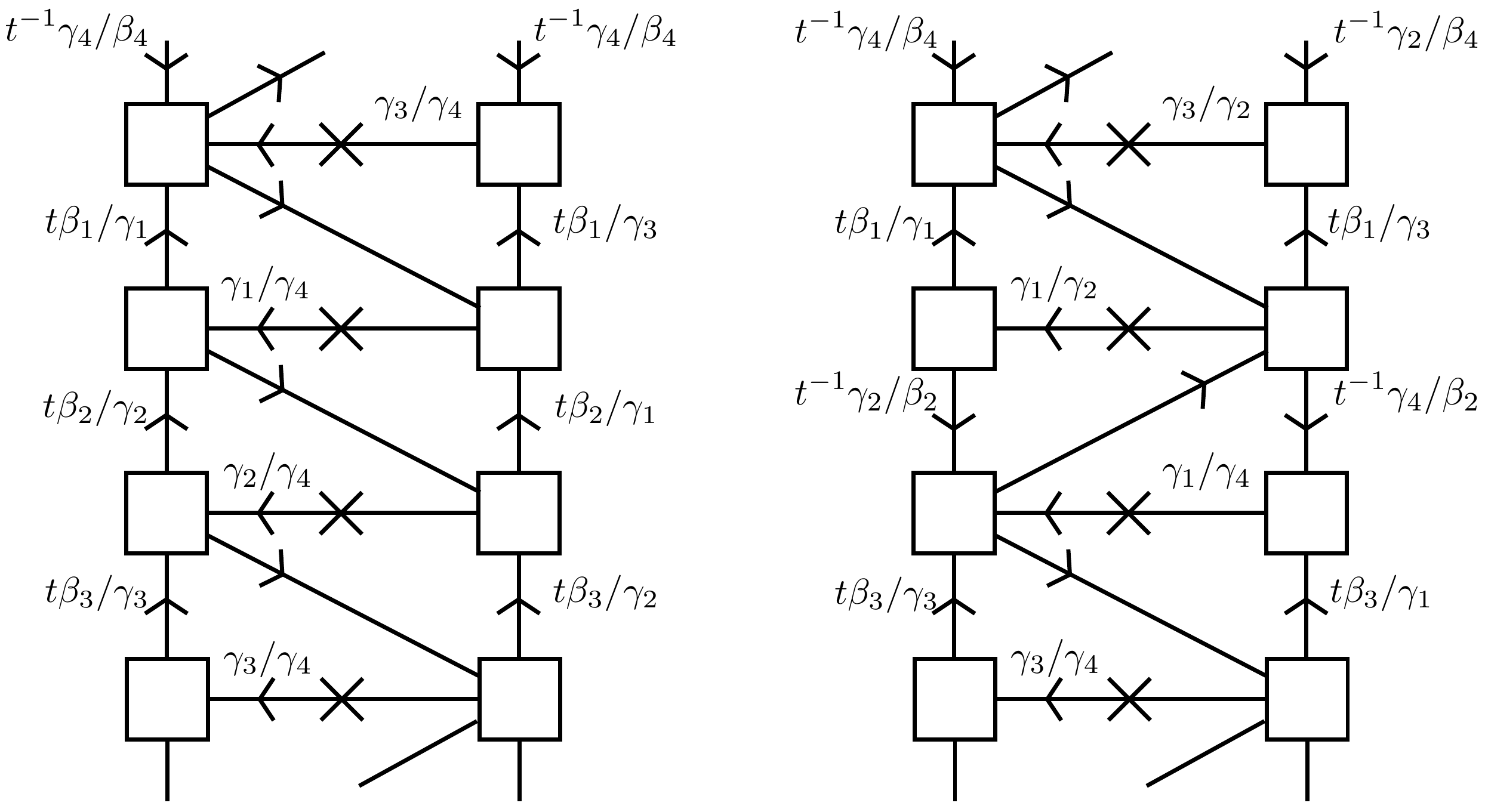}
\caption{Domain walls of type $\mathcal{T}=\gamma$ related to $SU(k)_\gamma$ fluxes. The left one is related to the flux $F=(-1/4,-1/4,-1/2,1)_\gamma$ and the right one is related to the flux $F=(-1/2,1/2,-1/2,1/2)_\gamma$ in the 6d $SU(k)_\gamma$ symmetry.}\label{fig:a3-walls}
\end{center}
\end{figure}

So far we discussed the domain walls of type $\mathcal{T}=\beta$ with fluxes only on the $SU(k)_\beta$ symmetry. We can construct the domain walls of type $\mathcal{T}=\gamma$ for $SU(k)_\gamma$ flux in a similar way. As discussed above, the main difference for a given boundary condition $\mathcal{B}$ is the representation of the 4d chiral fields $q_i$. We flip the representation of $q_i$ from $(\bar{\bf N},{\bf N})$ to $({\bf N},\bar{\bf N})$ of the $SU(N)_i\times SU(N)_i'$ gauge symmetry. It then follows that the interface hosts the following superpotentials: 
\begin{equation}\label{eq:domainwall-W2}
	\mathcal{W}_{x^4=0}=\sum_{i=+} \left( q_{i+1}\tilde{q}_i X_i + \tilde{q}_{i}q_iX_i'\right) +\sum_{i=-} \left( Y_i q_i \tilde{q}_{i}+Y_i'  \tilde{q}_i q_{i+1}\right) \ .
\end{equation}
The representations of the other chiral fields $\tilde{q}_i$ need to be chosen accordingly.
We also add flip chiral fields coupled to the baryonic operators of $q_i$'s. One can easily check that this domain wall configuration has no cubic gauge anomalies and also that all $U(1)$ charges  for the 4d fields are uniquely fixed with no additional abelian symmetry other than $U(1)_{\beta_i}\times U(1)_{\gamma_i}\times U(1)_t$ symmetries. Two examples  in the $A_3$ quiver gauge theory are depicted in Figure \ref{fig:a3-walls}. 

We define this type of domain walls with $\mathcal{B}_\gamma$ as the basic flux domain walls with flux  $F=(n_1,n_2,\cdots,n_k)$ for the $SU(k)_\gamma$ symmetry where $n_i=-1/r$ for $s_i=+$ or $n_i=1/(k-r)$ for $s_i=-$, and $r$ is the number of $+$ signs in $\mathcal{B}$. Note that this domain wall permutes cyclically $U(1)_\gamma^r$ and $U(1)^{k-r}_\gamma$ symmetries respectively. More precisely, the $\sigma_\gamma(\mathcal{B})$ is the counterclockwise permutation of $U(1)_{\gamma_i}$ symmetries with $s_i=+$ and the clockwise permutation of $U(1)_{\gamma_i}$ symmetries with $s_i=-$, and $\sigma_\beta(\mathcal{B})=1$.
 Following the conjectures above, we propose that a domain wall configuration constructed by these domain walls realize the flux compactification of the 6d theory when $\prod_i\sigma^{t_i}=1$ or when the system is compactified on a circle (so when the 6d theory is put on a torus).

\begin{figure}[htbp]
\begin{center}
\includegraphics[scale=0.35]{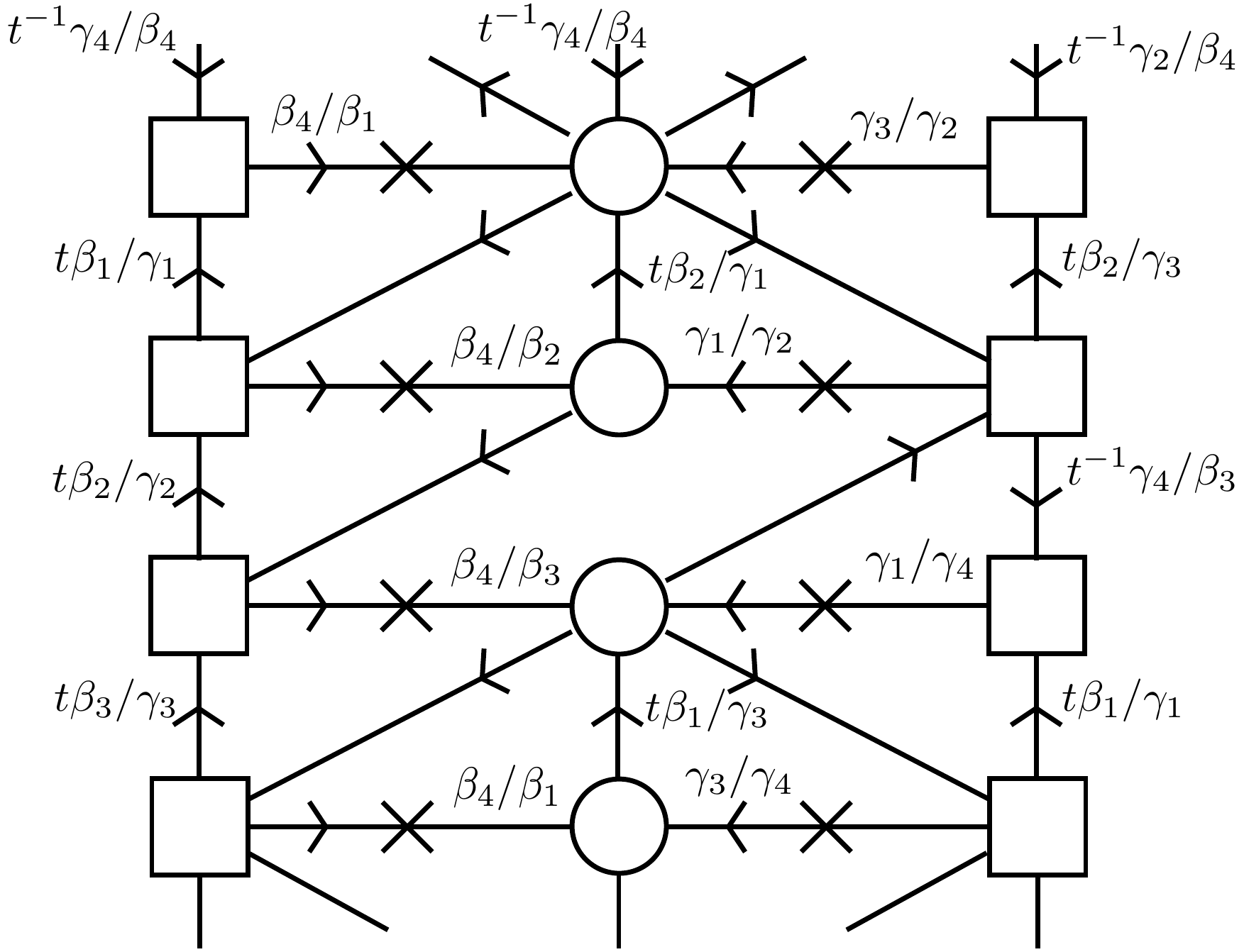}
\caption{A combination of a domain wall with $SU(k)_\beta$ flux and a domain wall with $SU(k)_\gamma$ flux in the affine $A_3$ quiver gauge theory. The total flux becomes $F_{\rm tot}=(1/2,1/4,1/4,-1)_\beta+(-1/2,1/2,-1/2,1/2)_\gamma$.  The top gauge group has four $N$ flavors, which means it is free in the IR. The way to think about this theory is to perform a Seiberg duality on the second node which will remove $N$ flavors from the first gauge groups. The resulting theory is free in the ultra violet. The theories we consider might have complicated dynamics when we flow to the IR. In all examples we consider there is a way to make sense of the models as complete in the ultra-violet.
}\label{fig:a3-2walls}
\end{center}
\end{figure}

For more general fluxes in both $SU(k)_\beta$ and $SU(k)_\gamma$ symmetries, we can simply combine the domain walls for $SU(k)_\beta$ flux with the domain walls of the second type for $SU(k)_\gamma$ flux.  Gluing these two different types of domain walls is straightforward. As the cases above, we will have a new chamber between two domain walls and at low energy the 5d theory in this chamber reduces to a 4d theory. The hypermultiplets with the same boundary conditions at the two ends leave 4d chiral fields coupled to the degrees of freedom at the interfaces and integrating out massive hypers with opposite boundary conditions at the two ends induces quartic superpotential couplings as discussed above. An example of gluing a flux domain wall of type $\mathcal{T}=\beta$ and another flux domain wall with $\mathcal{T}=\gamma$ in the $A_3$ quiver theory is given in Figure \ref{fig:a3-2walls}. Here, there is a quartic superpotential of the form $q_3\tilde{q}_2q_2'\tilde{q}_2'$ where $q,\tilde{q}$ are the 4d chiral fields in the first domain wall and $q',\tilde{q}'$ are the 4d fields in the second domain wall.

\subsection*{4d reduction and punctures}
Let us now compare our 5d domain wall configurations and the 6d theory with fluxes on Riemann surfaces. We first consider the 4d reduction of the 6d theory compactified on a tube (or a two punctured sphere). We can compute 't Hooft anomalies of the resulting 4d theories from the 6d anomaly polynomial by integrating it on a tube. In addition, there are anomaly inflow contributions from the 6d bulk theory toward two punctures. We will call the former as the geometric contribution and the latter as the inflow contribution \cite{Kim:2017toz,Kim:2018bpg}. By adding these two contributions, we can compute the total 't Hooft anomalies of the 4d compactification. The geometric contribution can be computed using the method studied in Section \ref{sec:sixdim}. The inflow contribution can be obtained from the 5d quiver gauge theories ending on a boundary. We will now explain how to compute this inflow contribution. See \cite{Kim:2017toz,Kim:2018bpg} for more discussions.

We can deform the 6d theory near a puncture as a long and thin tube ending on a boundary. Since we topologically twist the 6d theory on the 2d surface, this deformation has no effect in the 4d reduction. The 6d theory around the puncture at low energy reduces to the 5d affine quiver gauge theory ending on the boundary. This picture suggests a one-to-one correspondence between the type of punctures and the choice of boundary conditions in the 5d theory. Thus, for a given puncture on a Riemann surface, we can find the corresponding boundary condition. This boundary condition leads to additional 't Hooft anomalies in the 4d theory through the inflow mechanism.

So punctures on a Riemann surface are associated to boundary conditions in the 5d theory.  In this work, we will focus only on {\it maximal} punctures, which are defined by boundary conditions similar to those appearing in the domain walls, and without additional 4d degrees of freedom at the boundary. These are so named as they generalize the maximal punctures appearing in class $\mathcal{S}$ theories to the case of generic $ADE$ group. These types of punctures depend on a discrete parameter, called color, denoted by the permutations among the Cartans of $G\times G'$ symmetry. This additional degree of freedom comes from the option of performing $G\times G'$ Weyl transformations. The puncture with this boundary condition is defined as follows.

We first give Dirichlet boundary condition to the vector multiplets, so the gauge symmetries in the bulk 5d theory become 4d global symmetries at the boundary. This endows the puncture with an affine quiver type global symmetry in addition to the $G\times G'$ symmetry. The hypermultiplets satisfy the standard 1/2 BPS boundary condition $\mathcal{B}$ defined in (\ref{eq:bchyper}), and thus they give rise to 4d chiral multiplets charged under the affine quiver global symmetry of the puncture. When each hypermultiplet leaves a 4d chiral multiplet at the 4d boundary, we will call this type of punctures as maximal punctures for any 5d affine quiver gauge theory.

The boundary conditions for the 5d theory induce anomaly inflows toward the maximal puncture and thus the punctures in general carry non-trivial anomalies. These anomalies depend on the boundary condition $\mathcal{B}$ and can be considered as a defining property of the punctures. In particular, two or more maximal punctures for a 6d theory have the same type of affine quiver global symmetries, but, due to the permutations by flux, they can have different colors with respect to the Cartan of $G\times G'$ symmetry. This results in different 't Hooft anomalies of the punctures.

As we studied above, the anomaly inflows consist of matter contributions and the gauge kinetic term (or gauge-global mixed Chern-Simons term) contributions. The hypermultiplet contributions and the kinetic contributions are the same as before. As explained above, a chiral fermion from a 5d hypermultiplet with Neumann boundary condition induces half of the anomalies from a 4d chiral fermion with the same charges. Also, the gauge kinetic terms provide inflow contributions for mixed anomalies between the affine quiver global symmetry and subsets of $G\times G'$ associated to the instanton symmetry as (\ref{eq:Cartans-symmetry}). That is for the gauge group $G_i$
\begin{equation}
	Tr(U(1)G_i^2) = \frac{1}{2}Q_{i} \ ,
\end{equation}
where $Q_i$ is the $U(1)$ global charge of a unit instanton state.
% for the gauge group $G_i$ where $Q_i$ is the $U(1)$ global charge of a unit instanton state and
% \begin{equation}
% 	s_{SU(N)} = \frac{1}{2}, \quad s_{SO(k)} =1, \quad s_{E_6}=3, \quad s_{E_7}=6, \quad s_{E_8}=30 \ ,
% \end{equation}
% defined by $Tr_{\bf F}\,G^2 = s_G\, Tr\,G^2$.
On the other hand, the vector multiplets now satisfy Dirichlet boundary condition. So their inflow contributions are minus of those for the Neumann boundary condition, which we compute
\begin{equation}
	Tr(U(1)_R)=Tr(U(1)_R^3) = -\frac{d_{G_i}}{2}\ , \quad Tr(U(1)_RG_i^2) = -\frac{h^\vee}{2} \ ,
\end{equation}
from the vector multiplet of the gauge group $G_i$.
Collecting all these contributions, we can compute the anomalies assigned for a maximal puncture.

% For the affine $A_k$ quiver theory in this section, 

%   $(\sigma_\beta,\sigma_\gamma)\in S_k\times S_k/\mathbb{Z}_k$ for the $SU(k)_\beta\times SU(k)_\gamma$ symmetry. We will discuss more details about punctures in the next section. We will define {\it maximal} punctures of the 6d $(A_k,A_k)$ conformal matter theory as follows. We first give Dirichlet boundary condition for the $SU(N)^k$ vector multiplets at the puncture. This gives rise to $SU(N)^k$ global symmetry at the puncture. So a maximal puncture hosts $SU(N)^k$ global symmetry.

Let us discuss some more details of the punctures in the $(A_{k-1},A_{k-1})$ conformal matter theory. A maximal puncture in this theory supports $SU(N)^k$ global symmetry. The anomalies of this puncture can be computed as follows.
The vector multiplets induce anomaly inflows given by
\begin{equation}\label{eq:inflow-vector}
	Tr(U(1)_R) = Tr(U(1)^3_R) =- \frac{k(N^2-1)}{2} \ , \quad Tr(U(1)_R SU(N)_i^2) = -\frac{N}{2} \ {\rm for \ all } \ i \ .
\end{equation}
The $i$-th hypermultiplet provides the inflow contributions as
\begin{eqnarray}\label{eq:inflow-hyper}
	&& Tr(U(1)_t)=\sum_{i}\frac{s_iN^2}{2}\ , \quad Tr(U(1)_{\beta_i})=\frac{s_i N^2}{2} \ , \quad Tr(U(1)_{\gamma_i})=-\frac{s_i N^2}{2} \ , \nonumber \\
	&&Tr(U(1)_tSU(N)_i^2) =\frac{(s_i+s_{i-1})N}{4} \ , \quad Tr(U(1)_{\beta_i}SU(N)_i^2)=Tr(U(1)_{\beta_i}SU(N)_{i+1}^2)= \frac{s_i N}{4} \ , 
	% \quad Tr(U(1)_{\gamma_i}SU(N)_i^2)=  -\frac{s_i N}{4} \ , 
	\nonumber \\
	&& Tr(U(1)_{\gamma_i}SU(N)_{i}^2)= Tr(U(1)_{\gamma_i}SU(N)_{i+1}^2)=  -\frac{s_i N}{4} \ , \nonumber \\
	&& Tr(U(1)_aU(1)_bU(1)_c) = \sum_i\frac{s_iQ_aQ_bQ_cN^2}{2} \quad ({\rm with} \ a,b,c\in \{t,\beta_i,\gamma_i\} ) \ , \nonumber \\
	&& Tr(SU(N)_i^3) = (s_i-s_{i-1})\frac{N}{2} \ , 
\end{eqnarray}
where $s_i$ is the boundary condition and $Q_a$ denotes the $U(1)_a$ global charge of the $i$-th hypermultiplet.
Also the $SU(N)^k$ Yang-Mills terms provide additional contributions as
\begin{eqnarray}\label{eq:inflow-kinetic}
&&Tr(U(1)_{\beta_i} SU(N)_{i}^2) = Tr(U(1)_{\gamma_i} SU(N)_{i}^2) = \frac{N}{4} \ , \nonumber \\
	&&Tr(U(1)_{\beta_i} SU(N)_{i+1}^2) = Tr(U(1)_{\gamma_i} SU(N)_{i+1}^2) = -\frac{N}{4} \ .
\end{eqnarray}
Then the full anomaly inflow for a maximal puncture with $\mathcal{B}$ on a Riemann surface is given by a sum over these inflow contributions. We note that proper permutations $(\sigma_\beta,\sigma_\gamma)$ should be taken into account when there are two or more punctures.

Consider now the 5d affine $A_{k-1}$ quiver gauge theory with domain walls on an interval $0<I<L$. We impose maximal boundary conditions giving  maximal punctures at $x^4=0,L$.
At low energy $E \ll \frac{1}{L}$, this theory reduces to a 4d $\mathcal{N}=1$ quiver gauge theory. The resulting 4d theory will have $SU(N)^k\times SU(N)'{}^k$ global symmetries arising from the Dirichlet boundary conditions at $x^4=0,L$ and $U(1)^{k-1}_\beta\times U(1)^{k-1}_\gamma\times U(1)_t$ flavor symmetries from the hypermultiplets. We propose that this 4d theory, when $\prod_i\sigma^{t_i}=1$, corresponds to the 6d $(A_{k-1},A_{k-1})$ theory with fluxes on a tube with maximal punctures at both ends. 

%The global symmetry of the 4d quiver gauge theory agrees with that expected from the 6d theory. We map the $SU(N)^k_L\times SU(N)^k_R$ global symmetries to the symmetries assigned for the two maximal punctures and the $U(1)^{k-1}_\beta\times U(1)^{k-1}_\gamma$ symmetries to the Cartans of the 6d global symmetry $H\subset SU(k)_\beta\times SU(k)_\gamma$ where $H$ is the unbroken global symmetry under the 6d flux. We expect that the $U(1)^{k-1}_\beta\times U(1)^{k-1}_\gamma$ global symmetry of the 4d quiver theory enhances to $H$ by quantum effects.

\begin{figure}[htbp]
\begin{center}
\includegraphics[scale=0.35]{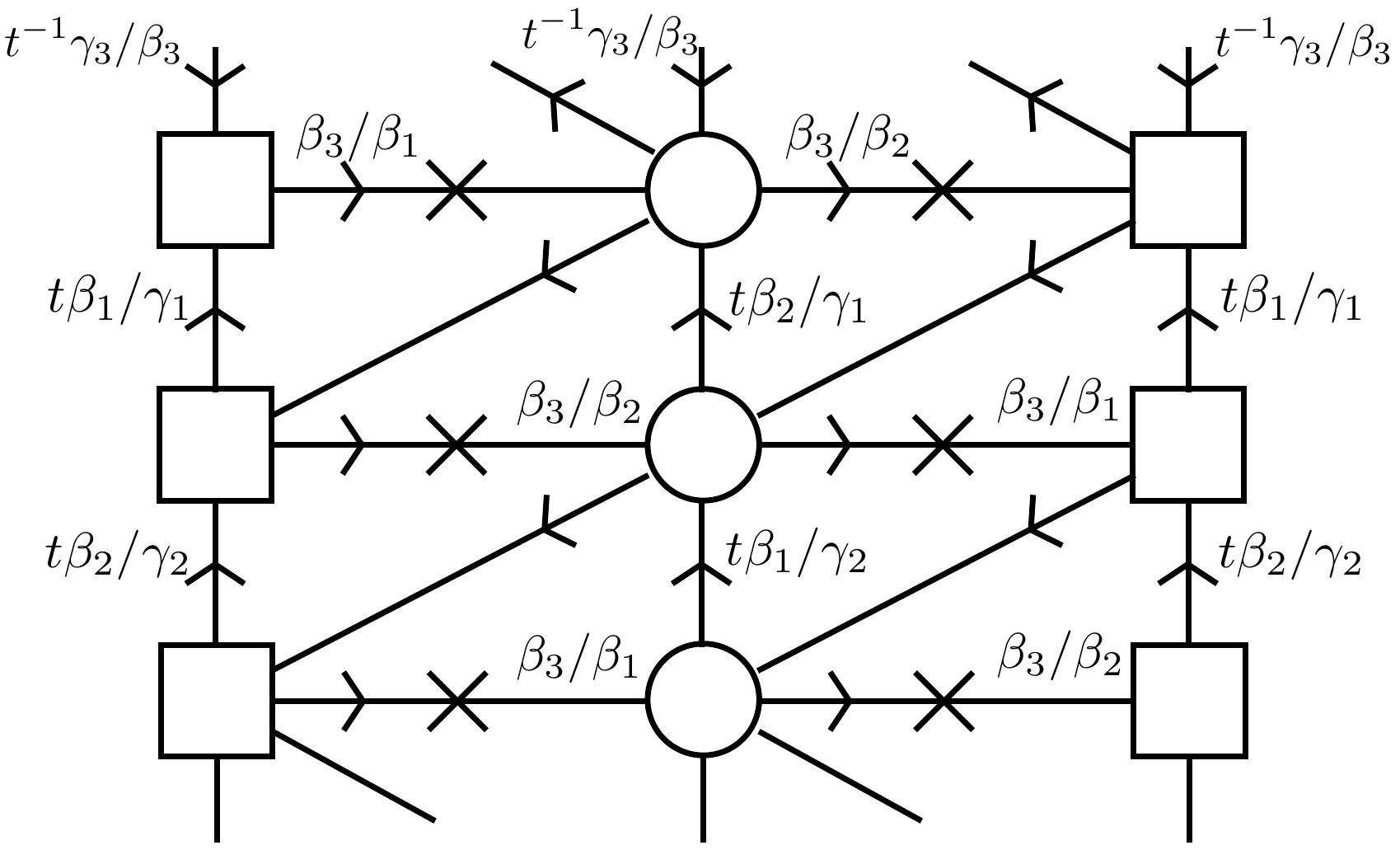}
\caption{Flux domain wall for $F_{\rm tot}=(1,1,-2)_\beta$ in the affine $A_2$ quiver theory. The permutations in the first and second domain walls are $\sigma^{t_1}=\sigma^{t_2}=(1 \ 2)$ and thus $\sigma^{t_1}\sigma^{t_2}=1$.}\label{fig:a2-walls}
\end{center}
\end{figure}

For example, we can engineer a 4d quiver gauge theory by connecting two basic domain walls of $\mathcal{D}=\{+,+,-\}_\beta$ in the 5d affine $A_2$ theory as drawn in Figure \ref{fig:a2-walls}. The fluxes associated with these domain walls are $F_1=(2/3,1/3,-1)_\beta, \, F_2=(1/3,2/3,-1)_\beta$, respectively, and their combination gives a flux of $F=(1,1,-2)_\beta$. Note that the combination of the two permutations becomes trivial, i.e. $\sigma^{t_1}\sigma^{t_2}=1$ where $\sigma^{t_1}=\sigma^{t_2}=(1 \ 2)_\beta$.
We thus propose that this 4d theory is the 6d $(A_2,A_2)$ conformal matter theory on a tube with two maximal punctures and flux $F=(1,1,-2)_\beta$. 
%
% This flux breaks the $SU(3)_\beta$ global symmetry to $SU(2)_\beta\times  U(1)_u$. Furthermore, the boundary conditions at two punctures break $SU(3)_\gamma$ symmetry down to $SU(2)_\gamma\times U(1)_v$. We expect that this 4d quiver theory will have enhanced non-abelian global symmetries $SU(2)_\beta \times SU(2)_\gamma$ in IR.

% From the 6d anomaly polynomial, we can read off the geometric contributions to the 4d anomalies which are given by
% \begin{eqnarray}
% 	&&Tr(U(1)_u) = 18N \ , \quad Tr(U(1)_u^3) = 6(9N^3 - 6N^2) \ , \quad Tr(U(1)_nU(1)_v^2)=36N^2 \\
% 	&&Tr(U(1)_uU(1)_R^2)=-18(N^2-1) \ , \quad Tr(U(1)_uU(1)_t^2) = 18N^2 \ , \quad  Tr(U(1)_u^2U(1)_t) = -18N^2 \ , \nonumber \\
% 	&&Tr(U(1)_tSU(2)_\beta^2) =3N^2 \ , \quad Tr(U(1)_uSU(2)_\beta^2) = 9N^3-6N^2 \ , \quad Tr(U(1)_uSU(2)_\gamma^2) = 3N^2 \ ,\nonumber
% \end{eqnarray}
% where $U(1)_R$ is the Cartan of the 6d $SU(2)_R$ symmetry.
%
Combining the geometric contribution, which is given by
\begin{eqnarray}
	&&Tr(U(1)_{\beta_{1,2}}) = 9N \ , \quad Tr(U(1)_R^2U(1)_{\beta_{1,2}}) = -27(N^2-1)  \ , \quad Tr(U(1)_t^2 U(1)_{\beta_{1,2}})=27N^2 \ , \nonumber \\
	&& Tr(U(1)_t U(1)_{\beta_{1,2}}^2) = -9N^2
	\ , \quad Tr(U(1)_t U(1)_{\beta_1}U(1)_{\beta_2}) = -36N^2 \nonumber \ , \\
	&& Tr(U(1)_{\beta_{1,2}}^3)= 9(3N^3\!-\!2N^2) \ , \quad Tr(U(1)_{\beta_1}^2U(1)_{\beta_2})=Tr(U(1)_{\beta_2}^2U(1)_{\beta_1}) =18(3N^3\!-\!2N^2) \ , \nonumber \\
	&& Tr(U(1)_{\beta_{1,2}}U(1)_{\gamma_a}U(1)_{\gamma_b})=18N^2 \quad {\rm for} \ a=1,2 \ ,
\end{eqnarray}
and the inflow contributions written in (\ref{eq:inflow-vector}), (\ref{eq:inflow-hyper}), (\ref{eq:inflow-kinetic}) with $(s_1,s_2,s_3)=(+,+,-)$ for the two punctures, we find the anomalies of the 6d theory on a tube perfectly agree with the 't Hooft anomalies of the 4d quiver gauge theory in Figure \ref{fig:a2-walls}.
Also, one can easily show that the 4d quiver gauge theory obtained by combining any number $n\in 2\mathbb{Z}$ of the basic domain walls for $\mathcal{D}=\{+,+,-\}_\beta$ has the same 't Hooft anomalies as those from the 6d theory with flux $F=n(1/2,1/2,-1)_\beta$ and two maximal punctures.

\begin{figure}[htbp]
\begin{center}
\includegraphics[scale=0.35]{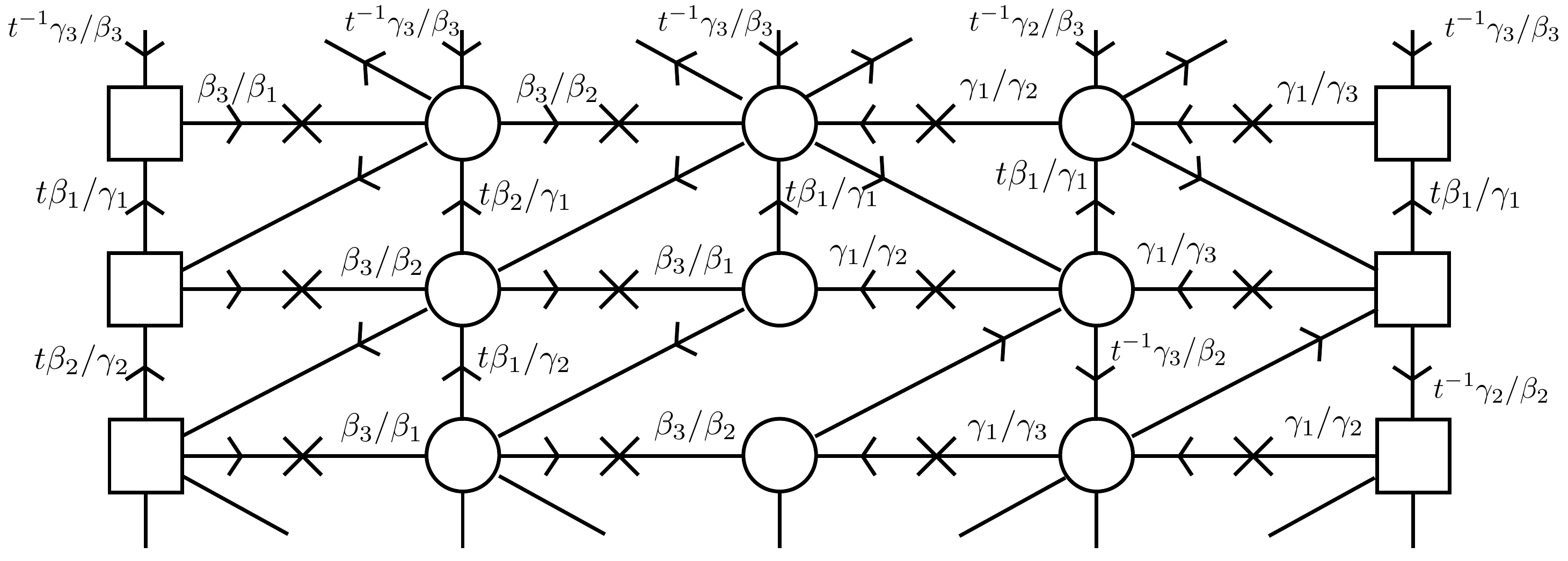}
\caption{A combination of four flux domain wall with $F_{\rm tot}=(1,1,-2)_\beta+(-2,1,1)_\gamma$ in the affine $A_2$ quiver theory. The permutations in four domain walls are $\sigma^{t_1}=\sigma^{t_2}=(1 \ 2)_\beta$ and $\sigma^{t_3}=\sigma^{t_4}=(2 \ 3)_\gamma$, and thus $\prod_{i=1}^4\sigma^{t_i}=1$.}\label{fig:a2-4walls}
\end{center}
\end{figure}

A more complicated example with $\mathcal{D}_1=\mathcal{D}_2=\{+,+,-\}_\beta$ and $\mathcal{D}_3=\mathcal{D}_4=\{+,-,-\}_\gamma$ is given in Figure \ref{fig:a2-4walls}. We expect that this 4d quiver gauge theory corresponds to the 6d $(A_2,A_2)$ theory on a tube with flux $F_{\rm tot}=(1,1,-2)_\beta+(-2,1,1)_\gamma$. In this case, two punctures amount to two different boundary conditions, $\mathcal{B}_1=\{+,+,-\}$ and $\mathcal{B}_2=\{+,-,-\}$ respectively. We checked that the 't Hooft anomalies of this 4d theory agree with the geometric and inflow results of the 6d theory with the $F_{\rm tot}$ and these two punctures.

We now consider gluing two boundaries of the 5d quiver theory on an interval in the presence of domain walls. From 5d perspective, this gluing can be simply considered as identifying the first and the last chambers without boundaries. Or we can also consider this as connecting two 5d theories in the first and the last chamber by a trivial interface between them. The 4d viewpoint of gluing two punctures will be presented in the next section.
At low energy after gluing two ends of the 5d theory, we will have a 4d quiver gauge theory. We conjecture that this 4d theory realizes a torus compactification the 6d $(A_{k-1},A_{k-1})$ conformal matter theory with flux. We expect that this conjecture, for the 6d theory on a torus, holds also for more general domain wall configurations with $\prod_i\sigma^{t_i}\neq 1$. When $\prod_i\sigma^{t_i}\neq 1$, the corresponding 6d theory has fractional flux with non-trivial center flux. This fractional flux breaks some subsets of $SU(k)_\beta \times SU(k)_\gamma$ symmetries. The same symmetry breaking occurs in the 5d quiver theory on a circle when the global symmetries in the first chamber and those in the last chamber are identified due to the non-trivial permutation. In the next section, we will see a number of examples of 4d quiver theories corresponding to the 6d theory on a torus with various fluxes and test them using superconformal indices and anomaly matchings.

We remark here that our domain wall construction fails to realize the compactifications of 6d conformal matters on a tube with fractional fluxes. The 4d theories we obtain using our domain walls on an interval with fractional fluxes have wrong 't Hooft anomalies against the expected anomalies of the compactification of the 6d theories.
This may imply that our flux domain wall is not the correct domain wall for 6d fractional fluxes. We may have missed some 4d degrees of freedom and associated superpotentials at the interfaces, but they disappear or decouple when we combine domain walls so that $\prod_i\sigma^{t_i}$ becomes trivial or when we locate the domain walls on a circle. Another possibility is that the 6d flux leaves non-trivial Chern-Simons terms for the global symmetries in the 5d reduction in the presence of flavor holonomies. These 5d Chern-Simons terms do not affect the dynamics of the 5d gauge theory, but they may induce additional inflow contributions toward the 4d boundaries. We leave further investigations on this mismatch to future research.

\subsection*{Generalization to D and E}
The same idea in this subsection will be used to build the D-type and E-type domain walls below. For these cases, the symmetry $G\times G'$ at the interface will be different, but, apart from this, all other ingredients will be essentially identical. All domain walls will be constructed by first specifying boundary conditions $\mathcal{B}=\{\pm,\pm,\cdots,\pm\}$ for the 5d hypermultiplets and then coupling them to the 4d chiral multiplets $q_i$ and $\tilde{q}_i$ and flip fields. The 4d bifundamental field $q_i$ is in either a $G_i\rightarrow G_i'$ representation or a $G_i\leftarrow G_i'$ representation and these two choices will be denoted by $\mathcal{T}=\beta$ or $\gamma$, respectively. All quiver nodes are connected to each other through the cubic superpotentials of the form in (\ref{eq:domainwall-W}). The representations of the other 4d fields $\tilde{q}_i$ are fixed by the boundary condition $\mathcal{B}$ and the superpotential terms accordingly. When the quiver node involves $U(1)$ gauge symmetries, we replace them by two fundamental hypermultiplets of the adjacent $SU(2)$ gauge nodes. In this case, we will add another cubic term like $\tilde{W}=XqX'$ between the chiral fields $X$ and $X'$ coming from the $SU(2)$ fundamentals and $SU(2)'$ fundamentals in the two chambers. 
Abelian charges of the 4d chiral multiplets are fixed by the gauge-global mixed anomaly cancellation and the superpotentials. In particular the 6d $U(1)_R$ charges of the 4d chiral multiplets $q_i$ and $\tilde{q}_i$ are always fixed to be $0$ and $+1$ respectively. The resulting domain walls labelled by $\mathcal{D}=\mathcal{B}_\mathcal{T}$ turn out to have no cubic gauge anomalies, therefore they can consistently couple to the 5d boundary conditions without introducing additional flavor symmetries.
We expect the same {\bf conjectures} hold for D- and E-type domain walls which we will discuss now.

\subsection{D-type domain walls}
Let us now turn to the construction of flux domain walls in the 5d reductions of 6d D-type conformal matter theories. The 5d theory without the domain walls is an affine $D_{k+3}$ quiver gauge theory with $SU(N)^2\times SU(2N)^{k}\times SU(N)^2$ gauge group. When $N=1$ the vertical lines at the edge of the quiver become free fields which form a mass term with the flip fields, as the $SU(1)$ gauge groups are empty.  In these cases 
the $SU(1)^2\times SU(1)^2$ 
gauge nodes at the two ends of the quiver can be replaced by four fundamental hypermultiplets for the first $SU(2)$ gauge node and another four fundamentals for the last $SU(2)$ gauge node. We will discuss the cases for $N=1$ separately at the end of this section.

The 6d global symmetry $SO(2k+6)_\beta\times SO(2k+6)_\gamma$ is broken by non-zero holonomies to $U(1)^{2k+6}=\prod_{i=1}^{k+3}U(1)_{\beta_i}\times U(1)_{\gamma_i}$ Cartans and the remaining abelian symmetries are mapped to certain combinations of the flavor symmetries acting on the bifundamental hypers and the topological instanton symmetries in the 5d gauge theory. In our notation, the bifundamental hypermultiplets $\Phi_i=(X_i,Y_i)$ carry the $U(1)^{2k+6}$ charges $Q_i$ as follows:
\begin{eqnarray}\label{eq:DD-bifundamental-charge}
	&&\Phi_1 \ : \ (Q_{\beta_1},Q_{\beta_{k+2}},Q_{\gamma_1},Q_{\gamma_{k+2}}) = \frac{1}{2}(1,1,-1,-1) \ , \nonumber \\
	&& \Phi_2 \ : \ (Q_{\beta_1},Q_{\beta_{k+2}},Q_{\gamma_1},Q_{\gamma_{k+2}}) = \frac{1}{2}(1,-1,-1,1) \ , \nonumber \\
	&& \Phi_i \ : \ (Q_{\beta_{i-1}},Q_{\gamma_{i-1}}) = \frac{1}{2}(1,-1) \quad {\rm for} \quad 3\le i \le k+1 \ , \nonumber \\
	&& \Phi_{k+2} \ : \  (Q_{\beta_{k+1}},Q_{\beta_{k+3}},Q_{\gamma_{k+1}},Q_{\gamma_{k+3}}) = \frac{1}{2}(1,1,-1,-1) \ , \nonumber \\
	&&\Phi_{k+3} \ : \  (Q_{\beta_{k+1}},Q_{\beta_{k+3}},Q_{\gamma_{k+1}},Q_{\gamma_{k+3}}) = \frac{1}{2}(1,-1,-1,1) \ . 
\end{eqnarray}
Here, the fields $\Phi_1,\Phi_2,\Phi_{k+2},\Phi_{k+3}$ are in the fundamental represntations of four $SU(N)$ gauge groups, which we will denote by $SU(N)_{1,2,3,4}$, respectively.

The flux in the 6d theory is expected to be realized as a certain domain wall configuration in this 5d theory. We will first propose basic domain walls and then construct general flux domain walls by gluing a series of basic domain walls in the appropriate manner. As discussed the domain wall construction in the D-type quiver theory is similar to that of the A-type theory. The domain wall comes with a 4d interface between two 5d affine $D_{k+3}$ quiver gauge theories, and 4d degrees of freedom and superpotentials at the interface linking boundary conditions of two 5d theories on both sides of the wall.

The 1/2 BPS boundary condition at the interface ($x^4=0$) is the same as that in the A-type domain wall dicussed before. The vector multiplets of the $SU(N)^2\times SU(2N)^{k}\times SU(N)^2$ gauge group satisfy the Neumann boundary condition defined in equation (\ref{eq:Neumann}). Thus we will have $(SU(N)^2\times SU(2N)^{k}\times SU(N)^2)^2$ gauge symmetries at the interface. The $i$-th bifundamental hypermultiplet satisfies the boundary condition in equation (\ref{eq:bchyper}) labelled by a sign $s_i=\pm$. Thus the boundary condition of the 5d theory at the interface is defined by a vector $\mathcal{B}=\{s_1,s_2,\cdots,s_{k+3}\}$ with $s_i=\pm$. Basic domain walls have the same boundary condition for two 5d theories on both sides.

At the interface, we introduce additional 4d chiral multiplets $q_i$ and $\tilde{q}_i$ coupled to the 5d boundary conditions through cubic superpotentials. Like the A-type cases, the 4d chiral field $q_i$ is a bifundamental field between $H_{i}\times H_{i}'$, where $H_{i}$ and $H_{i}'$ represent the $i$-th gauge group in the quiver diagram in the first and the second chamber respectively, and $\tilde{q}_i$ is a bifundamental field between either $H_{i}\times H_{i+1}'$ or $H_{i+1}\times H_{i}'$ which is determined by the 4d cubic superpotentials. For a given boundary condition $\mathcal{B}$, we can construct two types of basic domain walls, which we call as $\mathcal{T}=\beta$ and $\mathcal{T}=\gamma$, related to the fluxes on $SO(2k+6)_\beta$ and the fluxes on $SO(2k+6)_\gamma$ respectively. 

\begin{figure}[htbp]
\begin{center}
\includegraphics[scale=0.50]{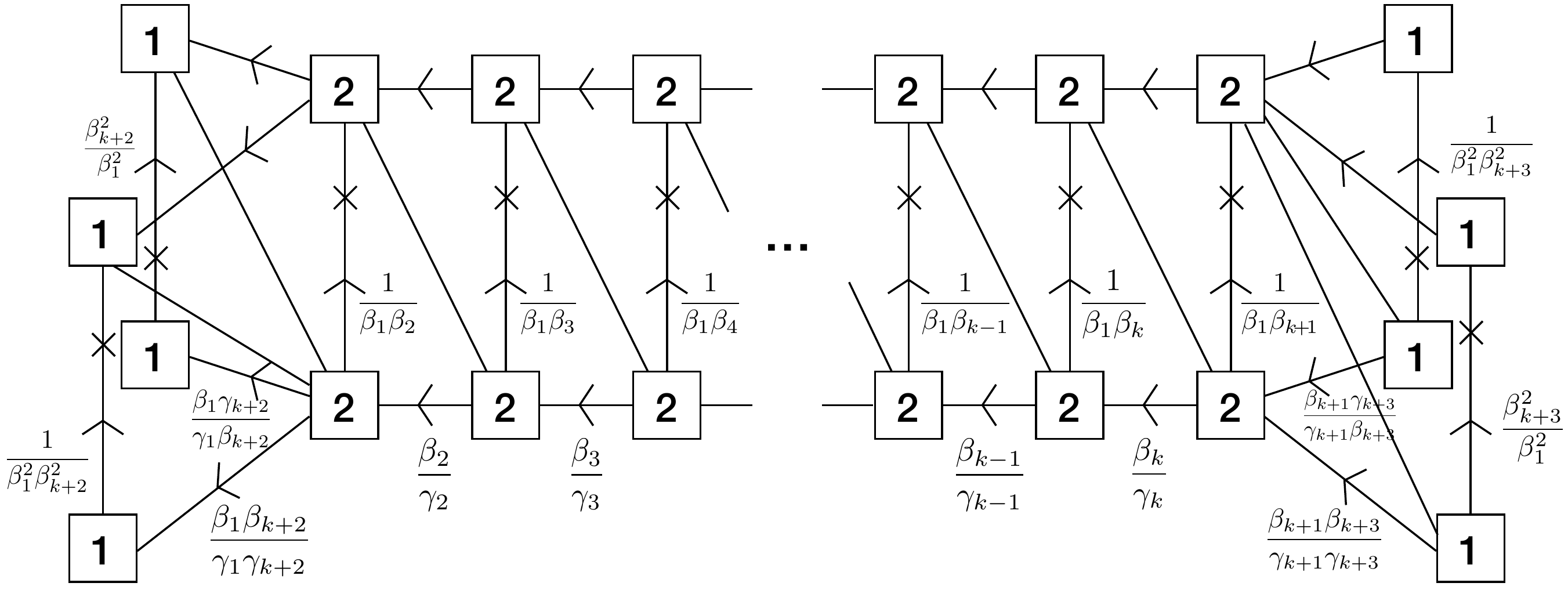}
\caption{Basic domain wall for $\mathcal{D}=\{+,+,\cdots,+\}_\beta$ in the affine $D_{k+3}$ quiver theory. The boxes with $1$ denote the 5d $SU(N)_{i=1,2,3,4}$ gauge nodes and the boxes with $2$ denote the 5d $SU(2N)_{i=1,\cdots,k}$ gauge nodes.}\label{fig:dn-fugacities}
\end{center}
\end{figure}

Let us first consider the basic domain walls for $SO(2k+6)_\beta$ fluxes. The simplest boundary condition is $\mathcal{B}=\{+^{k+3}\}$ where all $X_i$ in $\Phi_i$ satisfy Neumann boundary condition. In this case we propose a basic domain wall as dipicted in Figure \ref{fig:dn-fugacities}. The 5d chiral fields $X_i$ with Neumann boundary condition in the bottom (or first) chamber are denoted by the horizontal arrows in the bottom forming an affine $D_{k+3}$ diagram with boxes of $1$ and $2$.  Similarly, the horizontal arrows in the top forming another affine $D_{k+3}$ diagram correspond to another 5d chiral multiplets $X_i'$ with Neumann boundary condition in the top (or second) chamber. These 5d chiral multiplets $X_i$ and $X_i'$ couple to additional 4d chiral multiplets $q$ and $\tilde{q}$ living at the interface represented by the vertical arrows and diagonal arrows, respectively, connecting the top and the bottom affine quiver diagrams. There is a cubic superpotential for each triangle in the Figure \ref{fig:dn-fugacities}. Also, the baryonic operators from the 4d chiral fields $q$ couple to the flip fields denoted by $\times$.

The system with a basic domain wall inserted between two 5d affine $D_{k+3}$ quiver theories with boundary condition $\mathcal{B}=\{+^{k+3}\}$ is consistent in a sense that it has no gauge anomalies. We will show this now. First, we compute the anomaly inflows toward the 4d interface from the boundary conditions of the 5d theory. As we discussed above, there are two inflow contributions: one from the 5d Yang-Mills terms and another one from the matter multiplets with Neumann boundary condition. We first compute the contributions from the YMs terms. The Cartans of the $SO(2k+6)_\beta\times SO(2k+6)_\gamma$ symmetry are related to the instanton and the baryon charges as given in (\ref{eq:Cartans-symmetry}) with the affine $D_{k+3}$ Cartan matrix $A_{ij}$. This relation and the charge assignment for the hypermultiplets in (\ref{eq:DD-bifundamental-charge}) tells us that the gauge kinetic terms induce anomaly inflows given by
\begin{eqnarray}\label{eq:D-inflow-instanton}
	&&Tr(U(1)_{\beta_{1},\gamma_1}SU(N)_{1,2}^2)=\frac{N}{2} \ , \quad Tr(U(1)_{\beta_1,\gamma_1}SU(2N)^2_1) = -\frac{N}{2} \ ,\nonumber \\
	&&Tr(U(1)_{\beta_{k+2},\gamma_{k+2}}SU(N)_1^2)=\frac{N}{2} \ , \quad Tr(U(1)_{\beta_{k+2},\gamma_{k+2}}SU(N)_2^2)=-\frac{N}{2}\nonumber \\
	&&Tr(U(1)_{\beta_{i},\gamma_i}SU(2N)_{i-1}^2)=\frac{N}{2} \ , \quad Tr(U(1)_{\beta_{i},\gamma_i}SU(2N)_i^2)=-\frac{N}{2} \ , \nonumber \\
	&&Tr(U(1)_{\beta_{k+1},\gamma_{k+1}}SU(N)_{3,4}^2)=-\frac{N}{2} \ , \quad Tr(U(1)_{\beta_{k+1},\gamma_{k+1}}SU(2N)^2_k) = \frac{N}{2} \ ,\nonumber \\
	&&Tr(U(1)_{\beta_{k+3},\gamma_{k+3}}SU(N)_3^2)= -\frac{N}{2} \ , \quad Tr(U(1)_{\beta_{k+3},\gamma_{k+3}}SU(N)_4^2)=\frac{N}{2} \ ,
\end{eqnarray}
with $2\le i\le k$. 
There are also matter contributions to the anomaly inflows.
The vector multiplets with Neumann boundary condition contribute to the anomaly inflow as
\begin{eqnarray}\label{eq:D-inflow-vector}
	&&Tr(U(1)_R)=Tr(U(1)_R^3) =2(k+1)N^2-k/2-2 \ , \nonumber \\
	&&Tr(U(1)_R SU(N)^2_{1,2,3,4})=\frac{N}{2} \ , \quad Tr(U(1)_R SU(2N)^2_i)=N \quad {\rm for} \quad 1\le i\le k \ .
\end{eqnarray}
The hypermultiplet contributions depend on the boundary condition $\mathcal{B}$. For the boundary condition $\mathcal{B}=\{+^{k+3}\}$, the anomaly inflow contributions from the hypermultiplets are given by
\begin{eqnarray}\label{eq:D-inflow-hyper}
	&&Tr(U(1)_{\beta_i}) = N^2 \ , \quad Tr(U(1)_{\gamma_i})= -N^2 \quad {\rm for} \quad 1\le i \le k+1 \  ,  \\
	&&Tr(U(1)_{\beta_1,\beta_{k+2}}SU(N)^2_{1})=-Tr(U(1)_{\gamma_1,\gamma_{k+2}}SU(N)^2_{1})=\frac{N}{4} \ , \nonumber \\
	&&Tr(U(1)_{\beta_1,\gamma_{k+2}}SU(N)^2_{2})=-Tr(U(1)_{\gamma_1,\beta_{k+2}}SU(N)^2_{2})=\frac{N}{4} \ , \nonumber \\
	&&Tr(U(1)_{\beta_{i},\beta_{i+1}}SU(2N)^2_i) = - Tr(U(1)_{\gamma_i,\gamma_{i+1}}SU(2N)^2_i)= \frac{N}{4} \quad {\rm for} \quad 1\le i \le k \ , \nonumber \\
	&&Tr(U(1)_{\beta_{k+1},\beta_{k+3}}SU(N)^2_{3})=-Tr(U(1)_{\gamma_{k+1},\gamma_{k+3}}SU(N)^2_{3})=\frac{N}{4} \ , \nonumber \\
	&&Tr(U(1)_{\beta_{k+1},\gamma_{k+3}}SU(N)^2_{4})=-Tr(U(1)_{\gamma_{k+1},\beta_{k+3}}SU(N)^2_{4})=\frac{N}{4} \ , \nonumber \\
	&& Tr(U(1)_aU(1)_bU(1)_c) = \sum_i \!\frac{Q_{a,i}Q_{b,i}Q_{c,i}n_i}{2} \ , \quad Tr(SU(N)^3_{1,2}) \!=\! N \ , \quad Tr(SU(N)^3_{3,4}) \!=\! -N \ , \nonumber
\end{eqnarray}
where $Q_{a,i}$ and $n_i$ denote the $U(1)_a$ flavor charge and the number of the $i$-th hypermultiplet respectively. The total anomaly inflows from the 5d theory with the boundary condition $\mathcal{B}=\{+^{k+3}\}$ are sum of these three contributions in (\ref{eq:D-inflow-instanton}), (\ref{eq:D-inflow-vector}), (\ref{eq:D-inflow-hyper}). Anomaly inflows for other cases with different boundary conditions can be computed in the same way.

% The inflow contributions
% \begin{eqnarray}
% && Tr(U(1)_{\beta_1,\beta_{k+2}}SU(N)_1^2) = \frac{s_1N}{2} \ , \quad Tr(U(1)_{\beta_1,\beta_{k+2}}SU(2N)_1^2) = \frac{s_1N}{4} \ , \nonumber \\
% && Tr(U(1)_{\gamma_1,\gamma_{k+2}}SU(N)_1^2) = -\frac{s_1N}{2} \ , \quad Tr(U(1)_{\gamma_1,\gamma_{k+2}}SU(2N)_1^2) = -\frac{s_1N}{4} \ , \nonumber \\
% && Tr(U(1)_{\beta_1,\gamma_{k+2}}SU(N)_2^2) = \frac{s_2N}{2} \ , \quad Tr(U(1)_{\beta_1,\gamma_{k+2}}SU(2N)_1^2) = \frac{s_2N}{4} \ , \nonumber \\
% && Tr(U(1)_{\gamma_1,\beta_{k+2}}SU(N)_2^2) = -\frac{s_2N}{2} \ , \quad Tr(U(1)_{\gamma_1,\beta_{k+2}}SU(2N)_1^2) = -\frac{s_2N}{4} \ , \nonumber \\
% && Tr(U(1)_{\beta_i}SU(2N)_{i-1}^2)=Tr(U(1)_{\beta_i}SU(2N)_{i}^2) = \frac{s_{i+1}N}{2} \ , \nonumber \\
% && Tr(U(1)_{\gamma_i}SU(2N)_{i-1}^2)=Tr(U(1)_{\gamma_i}SU(2N)_{i}^2) = -\frac{s_{i+1}N}{2} \ , \nonumber \\
% && Tr(U(1)_{\beta_{k+1},\beta_{k+3}}SU(N)_3^2) = \frac{s_{k+2}N}{2} \ , \quad Tr(U(1)_{\beta_{k+1},\beta_{k+3}}SU(2N)_k^2) = \frac{s_{k+2}N}{4} \ , \nonumber \\
% && Tr(U(1)_{\gamma_{k+1},\gamma_{k+3}}SU(N)_3^2) = -\frac{s_{k+2}N}{2} \ , \quad Tr(U(1)_{\gamma_{k+1},\gamma_{k+3}}SU(2N)_k^2) = -\frac{s_{k+2}N}{4} \ , \nonumber \\
% && Tr(U(1)_{\beta_{k+1},\gamma_{k+3}}SU(N)_4^2) = \frac{s_{k+3}N}{2} \ , \quad Tr(U(1)_{\beta_{k+1},\gamma_{k+3}}SU(2N)_k^2) = \frac{s_{k+3}N}{4} \ , \nonumber \\
% && Tr(U(1)_{\gamma_{k+1},\beta_{k+3}}SU(N)_4^2) = -\frac{s_{k+3}N}{2} \ , \quad Tr(U(1)_{\gamma_{k+1},\beta_{k+3}}SU(2N)_k^2) = -\frac{s_{k+3}N}{4} \ , \qquad
% \end{eqnarray}

We shall check the gauge anomaly cancellation at the interface. There are cubic gauge anomalies coming from the anomaly inflow in (\ref{eq:D-inflow-hyper}) and they cancel out beautifully by the cubic anomalies from the 4d chiral multiplets $q$ and $\tilde{q}$ given in Figure \ref{fig:dn-fugacities}. Also, the gauge-global anomaly cancellation for the 6d global symmetries as well as the conditions from the 4d superpotentials uniquely fix all the charges of the additional 4d degrees of freedom inserted at the interface which we find as drawn in Figure \ref{fig:dn-fugacities}.  For convenience, we scaled the fugacities as $\beta_i\rightarrow \beta_i^2$ and $\gamma_i\rightarrow \gamma_i^2$ in the quiver diagrams for D-type domain walls in this section.
This domain wall configuration is thus consistent with no  gauge anomaly and no additional global symmetry.

As a consequence, we constructed a consistent domain wall configuration interpolating two 5d affine $D_{k+3}$ quiver gauge theories. Note that the $U(1)_{\beta_1}, U(1)_{\beta_2}, \cdots, U(1)_{\beta_{k+1}}$ symmetries are cyclically permutted, and $U(1)_{\beta_{k+2}}$ and $U(1)_{\beta_{k+3}}$ symmetries are flipped as we move across the domain wall. Namely,
\begin{equation}
	\sigma(\beta_1,\beta_2,\cdots,\beta_k,\beta_{k+1}, \beta_{k+2},\beta_{k+3}) \ 
	\rightarrow (\beta_2,\beta_3,\cdots,\beta_{k+1},\beta_1, 1/\beta_{k+2},1/\beta_{k+3}) \ ,
\end{equation}
in terms of fugacities $\beta_i$.
 % while permutting and flipping flavor symmetries accordingly.
All other basic domain walls for other $\mathcal{D}_{\mathcal{T}}$ can be similarly constructed.

\begin{figure}
\begin{center}
\includegraphics[scale=0.4]{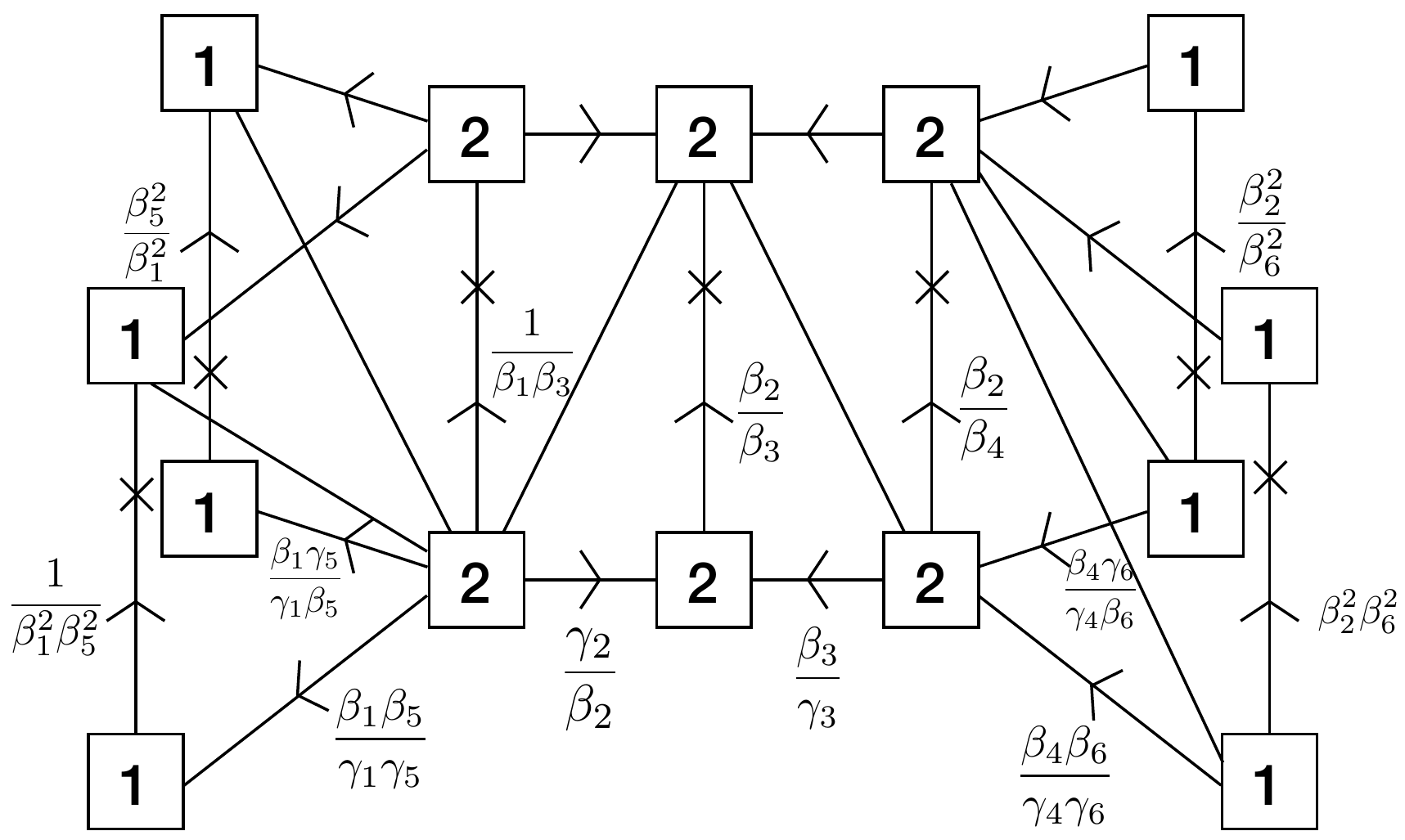}\qquad
\includegraphics[scale=0.4]{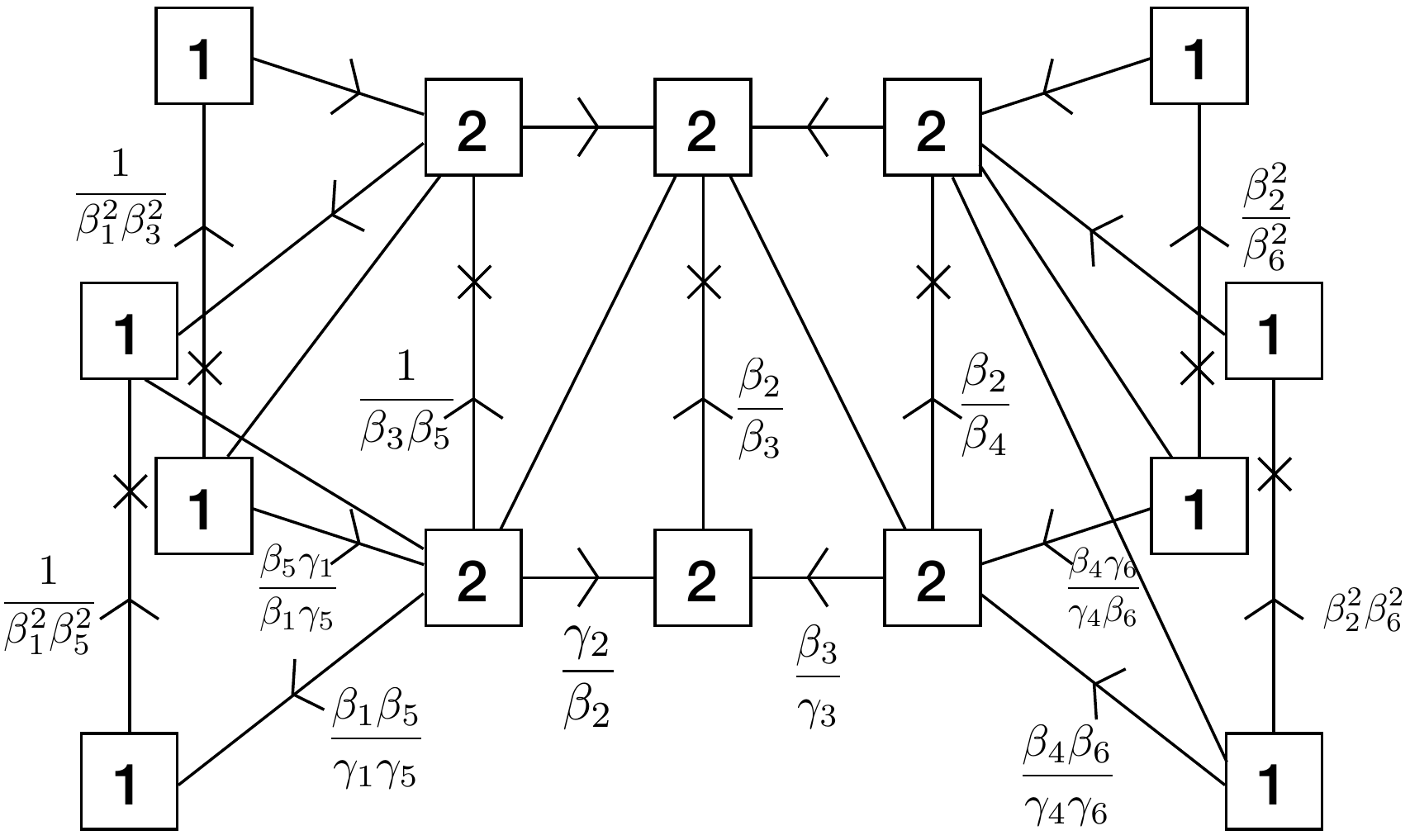}
\caption{Basic domain walls in the affine $D_6$ quiver theory. The left wall is for $\mathcal{D}_1=\{++|-+|++\}_\beta$ and the right wall is for $\mathcal{D}_2=\{+-|-+|++\}_\beta$. 
}
\label{fig:D6-domainwalls}
\end{center}
\end{figure}

\begin{figure}[htbp]
\begin{center}
\includegraphics[scale=0.50]{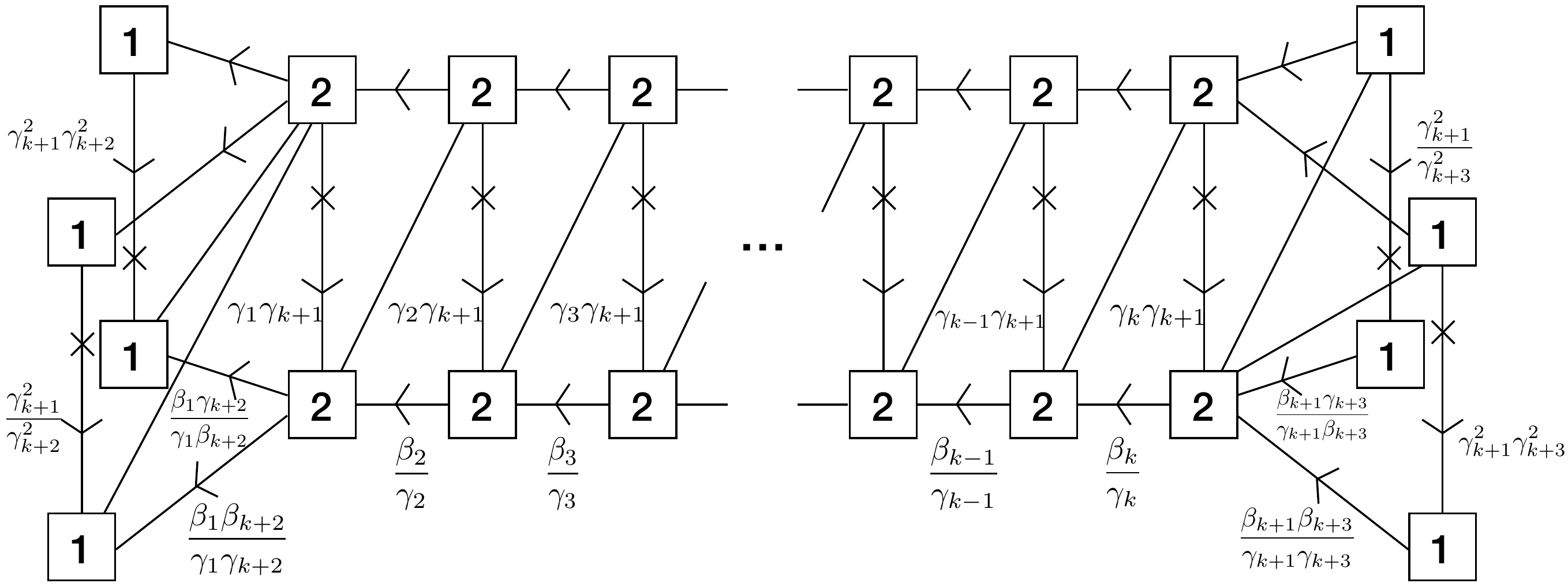}
\caption{Basic domain wall for $\mathcal{D}=\{+,+,\cdots,+\}_\gamma$ in the affine $D_{k+3}$ quiver theory. }\label{fig:dn-fugacities2}
\end{center}
\end{figure}

For example, two other basic domain walls in the affine $D_6$ quiver gauge theory are given in Figure \ref{fig:D6-domainwalls}. The left quiver diagram corresponds to the basic domain wall for $\mathcal{D}_1=\{++|-+|++\}_\beta$ where the first two and the last two signs denote the boundary conditions for the $SU(N)_{1,2}\times SU(2N)_1$ and the $SU(N)_{3,4}\times SU(2N)_3$ bifundamental hypers respectively. The $U(1)_\beta$ global symmetries are permutted by this domain wall as 
\begin{equation}
	\sigma(\beta_1,\beta_2,\beta_3,\beta_4,\beta_5,\beta_6) \ \rightarrow \ (\beta_3,1/\beta_1,\beta_4,1/\beta_2,1/\beta_5,1/\beta_6) \ ,
	% \beta_1\rightarrow \beta_3\, , \ \beta_2\rightarrow 1/\beta_1\,,\ \beta_3\rightarrow \beta_4 \,, \ \beta_4 \rightarrow 1/\beta_2 \,, \ \beta_5\rightarrow 1/\beta_5 \,, \ \beta_6\rightarrow 1/\beta_6 \ ,
\end{equation}
in terms of the fugacities $\beta_i$ for the $U(1)_{\beta_i}$. On the other hand, the right quiver diagram corresponds to the domain wall for $\mathcal{D}_2=\{+-|-+|++\}_\beta$ and it permutes the $U(1)_\beta$ symmetries as
\begin{equation}
	\sigma(\beta_1,\beta_2,\beta_3,\beta_4,\beta_5,\beta_6)\ \rightarrow \ (1/\beta_1,1/\beta_5,\beta_4,1/\beta_2,\beta_3,1/\beta_6)\ .
	% \beta_1 \rightarrow 1/\beta_1 \,, \ \beta_2 \rightarrow 1/\beta_5 \,, \ \beta_3 \rightarrow \beta_4 \,, \ \beta_4 \rightarrow 1/\beta_2 \,, \ \beta_5 \rightarrow \beta_3 \,, \ \beta_6 \rightarrow 1/\beta_6 \ .
\end{equation}
Another example is depicted in Figure \ref{fig:dn-fugacities2}. This domain wall is for $\mathcal{D}=\{+,+,\cdots,+\}_\gamma$ associated to the flux on $SO(2k+6)_\gamma$.

We will now relate the domain walls constructed by connecting multiple basic domain walls with fluxes in the 6d $(D_{k+3},D_{k+3})$ conformal matter theory. We first need to identify fluxes for the basic domain walls. We will employ the flux assignement given in (\ref{eq:domainwall-flux}). We find that the flux $Q_i$ on a $U(1)_i$ global symmetry is given by
\begin{equation}
	Q_i = \frac{1}{h^\vee_{D_{2k+6}}N}\sum_f q_{i,f}= \frac{1}{(2k+4)N}\sum_f q_{i,f} \ ,
\end{equation}
where $q_{i,f}$ denotes the $U(1)_i$ charge for the $f$-th flip field. For instance, the basic domain wall of $\mathcal{D}=\{+^{k+3}\}_\beta$ drawn in Figure \ref{fig:dn-fugacities} corresponds to the flux $\left(\frac{k+4}{2k+4},(\frac{1}{2k+4})^k,0,0\right)_\beta$ in $SO(2k+6)_\beta$ symmetry. Similarly, the two basic domain walls in Figure \ref{fig:D6-domainwalls} for $\mathcal{D}=\{++|-+|++\}_\beta$ and  $\mathcal{D}=\{+-|-+|++\}_\beta$ correspond to the fluxes $\frac{1}{10}(3,-4,2,1,0,0)_\beta$ and $\frac{1}{10}(2,-4,3,1,2,0)_\beta$, respectively, in $SO(12)_\beta$. When we join multiple basic domain walls, the total flux $F_{\rm tot}$ of the final domain wall configuration is simply the sum of the fluxes on all basic domain walls.

For the D-type flux domain walls, we will propose {\bf Conjectures} in Section \ref{sec:a-type-walls} with $G=D_{k+3}$.  
The simplest exercise is to combine $k+1$ (or $2k+2)$ basic domain walls of the same type for odd $k$ (or even $k$). When we put this 5d domain wall configuration on a finite interval, it corresponds to the 6d theory on a tube with integer fluxes. For example, we can consider the 5d theory on an interval with $k+1$ copies of the basic domain wall of $\mathcal{D}=\{+,+,\cdots,+\}_\beta$ drawn in Figure \ref{fig:dn-fugacities} which gives rise to an integer flux $F_{\rm tot}=(1^{k+1},0,0)_\beta$ for odd $k$. Choosing the maximal boundary condition, this theory reduces to a 4d $\mathcal{N}=1$ quiver gauge theory at low energy. We claim this 4d theory realizes the 6d $(D_{k+3},D_{k+3})$ conformal matter theory on a tube with flux $F_{\rm tot}=(1^{k+1},0,0)_\beta$ and maximal punctures at the two ends. 
% We can check this by comparing 't Hooft anomalies. For this, we first need to compute the expected 't Hooft anomalies from the 6d theory. The geometric contribution to the anomalies can be easily computed by integrating the anomaly polynomial of the 6d conformal matter theory with flux $F_{\rm tot}$ on a tube. There are also inflow contributions at each puncture from the boundary condition of the 5d theory. The inflow contributions from the gauge kinetic terms and the hypermultiplets with $\mathcal{B}=\{+^{k+3}\}$ were already computed in (\ref{eq:inflow-kinetic}), (\ref{eq:inflow-hyper}). The anomaly inflows from the vector multiplets with Dirichlet boundary condition can be obtained by flipping overall signs from the result for those with Neumann boundary condition given in (\ref{eq:inflow-vector}).
%Combining the geometric and the inflow contributions, one can compute the total 't Hooft anomalies for the 6d theory on a tube with flux $F_{\rm tot}$.
When $k$ is even, we can combine $2k+2$ basic domain walls of type $\mathcal{D}=\{+,+,\cdots,+\}_\beta$, and this theory on an interval gives rise to the 4d quiver theory corresponding to the 6d theory on a tube with flux $F_{\rm tot}=(2^{k+1},0,0)_\beta$.
We have checked for several $k$'s that the 't Hooft anomalies of the 4d quiver theory perfectly agree with those from the 6d anomaly polynomial and anomaly inflow at the two punctures.

Similarly, when we combine 4 copies of the basic domain walls in Figure \ref{fig:D6-domainwalls} on a tube, we will obtain the 4d quiver gauge theories corresponding to the 6d conformal matter theory on a tube with fluxes $F=(1,-1,1,1,0,0)_\beta$ for the left type and $F=(0,-1,1,1,1,0)_\beta$ for the right type. We checked these theories by comparing their 't Hooft anomalies against expected anomalies from the 6d theory.

We can also consider domain wall configurations on a circle which realize the 6d conformal matter theories on a torus with flux. The simplest example is to glue two ends of the tube theory from the $k+1$ copies of the basic domain wall in Figure \ref{fig:dn-fugacities}. 
% This gluing can be done by identifying two affine $D_{k+3}$ global symmetries and corresponding two sets of bifundamental chiral multiplets at two ends. 
Indeed, the resulting 4d theory has the expect 't Hooft anomalies for the torus theory with flux $F=(1^{k+1},0,0)_\beta$.

\begin{figure}[htbp]
\begin{center}
\includegraphics[scale=0.40]{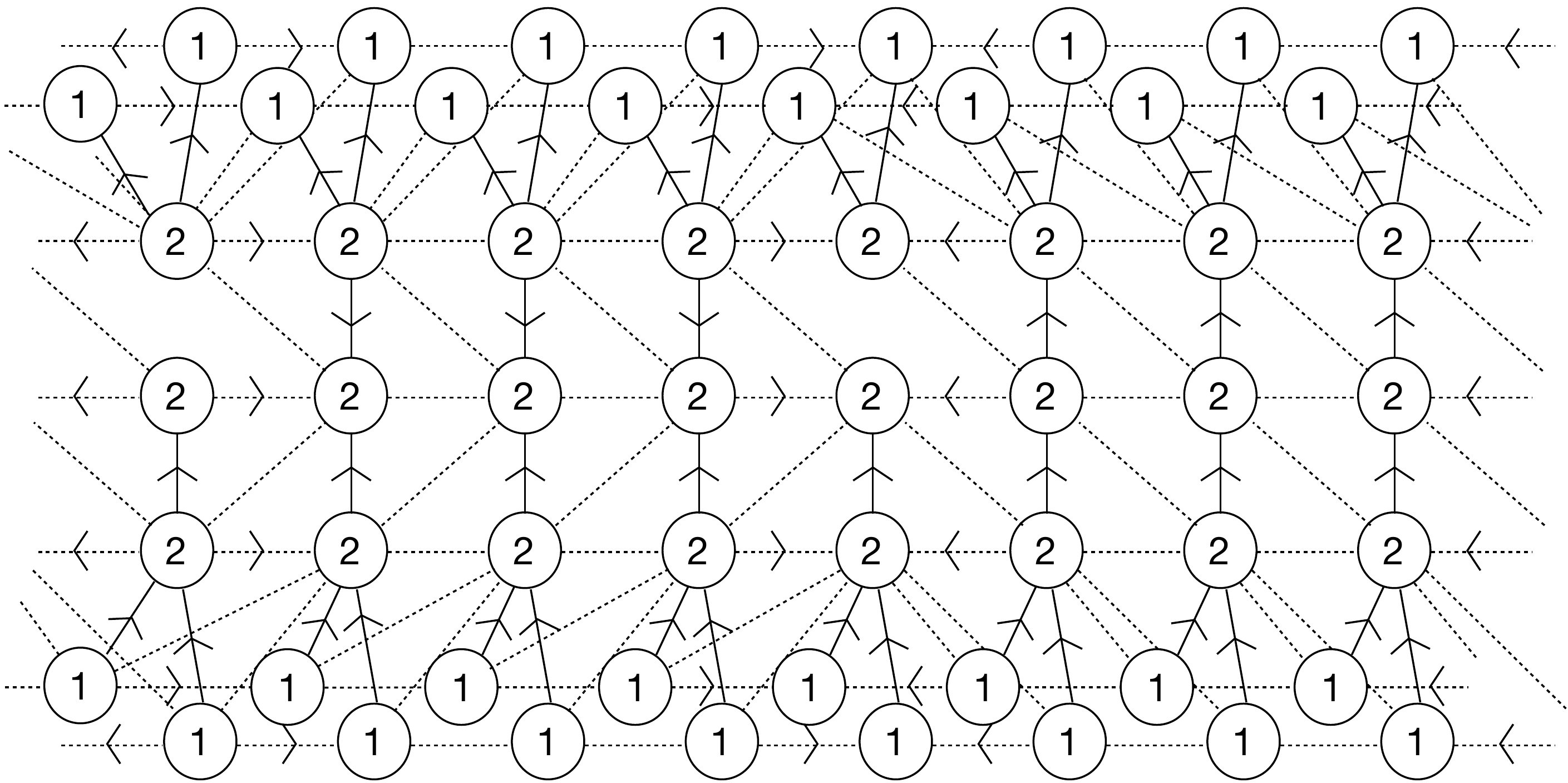}
\caption{Flux domain wall configuration with flux $F_{\rm tot}=(1,-1,1,1,0,0)_\beta+(1,1,1,1,0,0)_\gamma$ in the affine $D_6$ quiver theory on a circle. The horizontal direction is the circle direction and both ends are identified. The 4d chiral fields are denoted by dashed lines. There is a flip field on each horizontal arrow.}\label{fig:d6-torus}
\end{center}
\end{figure}

More general flux domain walls can be constructed by considering more complicated combinations of basic domain walls. An example is given in Figure \ref{fig:d6-torus}. Here, we combined four copies of a domain wall with $F_1=\frac{1}{10}(3,-4,2,1,0,0)_\beta$  and another four copies of a domain wall with $F_2=\frac{1}{10}(6,1,1,1,0,0)_\gamma$. So the total flux of the final domain wall configuration is $F_{\rm tot}=(1,-1,1,1,0,0)_\beta+(1,1,1,1,0,0)_\gamma$. When we compactify this 5d theory with domain walls on a circle, we will obtain at low energy the 4d quiver theory given in Figure \ref{fig:d6-torus}. This 4d theory corresponds to the 6d $(D_6,D_6)$ conformal matter theory on a torus with flux $F_{\rm tot}$.
We have checked that this 4d theory has the correct 't Hooft anomalies for being the torus theory with flux $F_{\rm tot}$.

\begin{figure}[htbp]
\begin{center}
\includegraphics[scale=0.35]{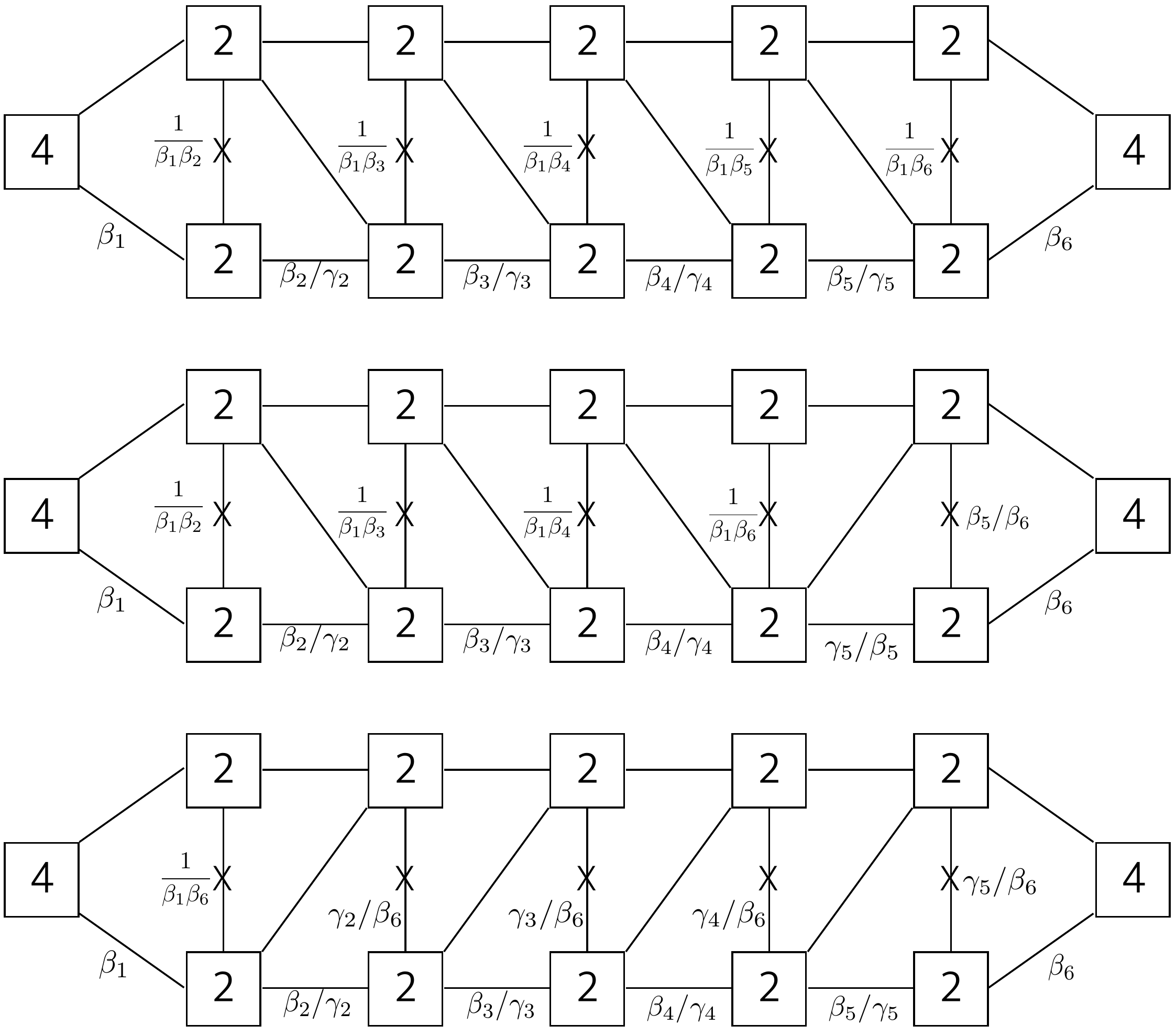}
\caption{Basic domain walls with the boundary conditions $\mathcal{B}_1=\{+^4|+^4|+^4\}$ and $\mathcal{B}_2=\{+^4|+^3,-|+^4\}$ and $\mathcal{B}_3=\{+^4|+^4|+^4\}$, respectively, in the first chamber. The first and the third domain walls have the same boundary condition in the first chamber, but they have different 4d chiral fields $q_i$ and superpotentials.}\label{fig:d8-tube}
\end{center}
\end{figure}

Let us now discuss the domain walls for the minimal $(D_{k+3},D_{k+3})$ conformal matter theories with fluxes. The construction of the flux domain wall in this theory is almost parallel to that in the non-minimal conformal matter theory with $N>1$.
The 5d theory is a linear quiver gauge theory with $SU(2)^k$ gauge groups and the first and last gauge nodes have 4 fundamental hypermultiplets. As explained already, the basic domain walls can be constructed by 4d interfaces with 4d chiral multiplets $q_i,\tilde{q}_i$ and flip fields coupled to 5d boundary conditions on both sides through cubic superpotentials. The new feature of the minimal conformal matter here is that the symmetry is enhanced to $D_{2k+6}$ (and for $k=1$ to $E_8$). This means that we have a larger Weyl symmetry, and thus domain walls can be engineered which manifest this by permuting $\gamma$ and $\beta$ symmetries. This is related to the fact that in the end of the quiver the $SU(1)$ symmetries are empty which gives rise to combining the $U(1)$ symmetries under which bifundamentals at the end of the quivers are charged into two $SU(4)$ global symmetries. We can write the domain wall as in Figure \ref{fig:dn-fugacities} or as Figure \ref{fig:d8-tube}. Note that these differ by choices of boundary conditions. The latter exists only for the minimal case and is more natural here so we will use it. However, the procedure of reading off the fluxes from the flip fields only applies to the former. 

%and should be done with the ll the flip fields even if they become massive .

The 5d boundary condition is labeled by a sign vector $\mathcal{B}=\{s_1,\cdots,s_4|s_5,\cdots, s_{k\!+\!3}|s_{k\!+\!4},\cdots,s_{k\!+\!7}\}$ where the $i$-th element $s_i$ denotes the boundary condition of the $i$-th hypermultiplet. In this $\mathcal{B}$, the first four elements are for the fundamental hypermultiplets of the first gauge node and the last four are for the fundamentals of the last gauge node. This condition breaks the $SO(8)\times SO(8)$ global symmetries rotating the fundamental flavors at the ends of the quiver to $U(1)^2\times SU(4)\times SU(4)$ symmetry. At the interface, the chiral halves of $k-1$ bifundamental hypermultiplets, chosen by $s_i$, satisfy Neumann boundary conditions and they couple to the 4d chiral fields $q_i$ and $\tilde{q}_i$ through cubic superpotentials, which we have seen in the non-minimal cases. Note that, since the gauge groups are now all pseudo-real, we can freely choose the chiral field $\tilde{q}_i$ to be in either $SU(2)_i\times SU(2)_{i+1}'$ or $SU(2)_{i+1}\times SU(2)_i'$ and these choices yield different domain walls.
 % We will identify the boundary conditions for the bifundamental hypers on both sides as $s_i=s_i'$ for $5\le i\le k+3$ where the primed signs denote the boundary conditions in the second chamber. With this choice, only the flavor symmetries with non-trivial flux are permutted by the domain wall.
Also, we will introduce new cubic superpotentials at the interface for the chiral halves $M_i$ and $M'_i$ of the fundamental hypers at the two ends of the quiver as $\tilde{\mathcal{W}}= \sum_{i=1}^4M_iq_1M_i'+\sum_{i=k+4}^{k+7}M_iq_kM_i'$. This identifies the boundary conditions as $s_i=-s_i$ for $i=1,\cdots,4,k+4,\cdots,k+7$. 

%The $SU(2)^k\times SU(2)'{}^k$ gauge anomalies should be cancelled at the interface. 
Since the gauge groups are $SU(2)$, cubic gauge anomalies are absent at the interface.
The gauge-global mixed anomalies from the boundary conditions of the 5d theory in the first chamber are the followings. First, the anomaly inflows from the Yang-Mills kinetic terms are given by
\begin{eqnarray}
	&Tr(U(1)_{\beta_{i}}SU(2)_{i-1}^2)=Tr(U(1)_{\gamma_{i}}SU(2)_{i-1}^2)=\frac{1}{2} \ , \nonumber \\
	&Tr(U(1)_{\beta_i}SU(2)_{i}^2)=Tr(U(1)_{\gamma_{i}}SU(2)_{i}^2)=-\frac{1}{2} \ ,
\end{eqnarray}
with $2\le i \le k$. Here, $\prod_{i=1}^{k+1}U(1)_{\beta_i}\times \prod_{i=2}^{k}U(1)_{\gamma_i}$ are the abelian global symmetries of the 5d theory. Then the hypermultiplets with $\mathcal{B}=\{+^{k+7}\}$ induce the inflow contributions as
\begin{eqnarray}
	&Tr(U(1)_{\beta_1}SU(2)_1^2) = Tr(U(1)_{\beta_{k+1}}SU(2)_k^2)= \frac{1}{2} \ , \nonumber \\
	& Tr(U(1)_{\beta_i}SU(2)_{i-1}^2) = Tr(U(1)_{\beta_i}SU(2)_{i}^2)=\frac{1}{4} \ , \nonumber \\
	&Tr(U(1)_{\gamma_i}SU(2)_{i-1}^2) = Tr(U(1)_{\gamma_i}SU(2)_{i}^2)=-\frac{1}{4} \ ,
\end{eqnarray}
with $2\le i \le k$. The requirement of these anomaly cancellation uniquely fixes all flavor charges of the 4d chiral fields. Also, the 4d fields $q_i$ and $\tilde{q}_i$ should have R-charges $0$ and $+1$ to cancel the gauge-R mixed anomaly from the 5d vector multiplets.  The same is true for the other boundary conditions $\mathcal{B}$. So these basic domain walls can be consistently inserted into the 5d system.

Three different basic domain walls in the affine $D_8$ quiver theory are depicted in Figure \ref{fig:d8-tube}. The fluxes for these domain walls can be determined from the linear anomalies as we have outlined previously. The corresponding fluxes are
\begin{eqnarray}
	&&F_1= \frac{1}{14}(7,2,1,1,1,2,0,0,0,0) \ , \nonumber \\
	&&F_2 = \frac{1}{14}(5,2,1,1,-2,3,0,0,0,0) \ , \nonumber \\
	&&F_3 = \frac{1}{14}(2,0,0,0,0,7,-1,-1,-1,-2) \ ,
\end{eqnarray}
in the $U(1)_\beta^6\times U(1)_\gamma^4$ abelian symmetries for the three domain walls respectively. For all these cases, the fluxes for $\beta_7$, $\beta_8$, $\gamma_1$, and $\gamma_{6-8}$ are zero and we have not written them for brevity. 
These domain walls permute the flavor symmetries with non-zero fluxes as follows:
\begin{eqnarray}
	&&\sigma(\beta_1,\beta_2,\beta_3,\beta_4,\beta_5,\beta_6) \ \rightarrow \ (\beta_2,\beta_3,\beta_4,\beta_5,\beta_6,\beta_1) \ , \nonumber \\
	&&\sigma(\beta_1,\beta_2,\beta_3,\beta_4,\beta_5,\beta_6) \ \rightarrow \ (\beta_2,\beta_3,\beta_4,\beta_6,-\beta_1,-\beta_5)  \ , \nonumber \\
	&&\sigma(\beta_1,\beta_6,\gamma_2,\gamma_3,\gamma_4,\gamma_5) \ \rightarrow \ (\beta_6,-\gamma_5,-\beta_1,\gamma_2,\gamma_3,\gamma_4) \ .
\end{eqnarray}

We claim that the {\bf Conjectures} above hold for the flux domain walls constructed by joining these basic domain walls in the minimal D-type conformal matter theories.
For example, we can connect 6 copies of the first domain wall in Figure \ref{fig:d8-tube} and then put this 5d theory on an interval with maximal boundary conditions at the two ends. 
% Here, the maximal boundary condition means that the $SU(2)^k$ vector multiplets satisfy Dirichlet boundary condition and the hypermultiplets satisfy the same boundary conditions at two ends in the first and the last chamber. This boundary condition is associated with a maximal puncture in the Riemann surface. 
Then we conjecture that the resulting 4d quiver gauge theory at low energy corresponds to the minimal $(D_{8},D_{8})$ conformal matter theory with flux $F_{\rm tot}=(1^6,0^4)$ on a tube with two maximal punctures.
Indeed, the 't Hooft anomalies of this 4d quiver theory agree with the expected results obtained from the 6d anomaly polynomial on the tube together with the inflow contributions at the two punctures. We will see more examples for these conjectures for tube theories and torus theories in the next section.

\subsection{E-type domain walls}

Now we turn to the flux domain walls in the 5d affine E-type quiver gauge theories.
We can construct these domain walls by applying the same idea used for the A- and D-type domain wall systems presented in the previous subsections.  

%, and these symmetries are expected to enhance to $E_6\times E_6 \times U(1)$ symmetry by non-perturbative instanton effects. Here the last $U(1)$ corresponds to the KK momentum.

Let us start by discussing the domain walls in the affine $E_6$ quiver gauge theory for general $N$ M5-branes.  This quiver theory has $U(1)^6$ flavor symmetry of the bifundamental hypers and $U(1)^7$ instanton symmetry. There are basic domain walls arising from a single interface interpolating between two 5d theories with 1/2 BPS boundary conditions given in (\ref{eq:Neumann}) and (\ref{eq:bchyper}). These domain walls are labeled by $\mathcal{D}=\mathcal{B}_{\mathcal{T}}$ where $\mathcal{B}=\{s_1,s_2,\cdots,s_6\}$ denotes the boundary conditions of six hypermultiplets and $\mathcal{T}$ denotes the representations of the 4d chiral fields $q_i$.
At the interface, we have $\prod_{i=1}^6SU(l_iN) \times SU(l_iN)'$ gauge symmetry with $\{l_i\}=\{1,2,3,2,1,2,1\}$ coming from the 5d vector multiplets with Neumann boundary condition in two sides. The 4d interface includes 4d chiral fields $q_i$ and $\tilde{q}_i$ and the cubic superpotentials of the form in (\ref{eq:domainwall-W}) whose explicit expressions are fixed by the domain wall data $\mathcal{B}$ and $\mathcal{T}$.

\begin{figure}[htbp]
\begin{center}
\includegraphics[scale=0.34]{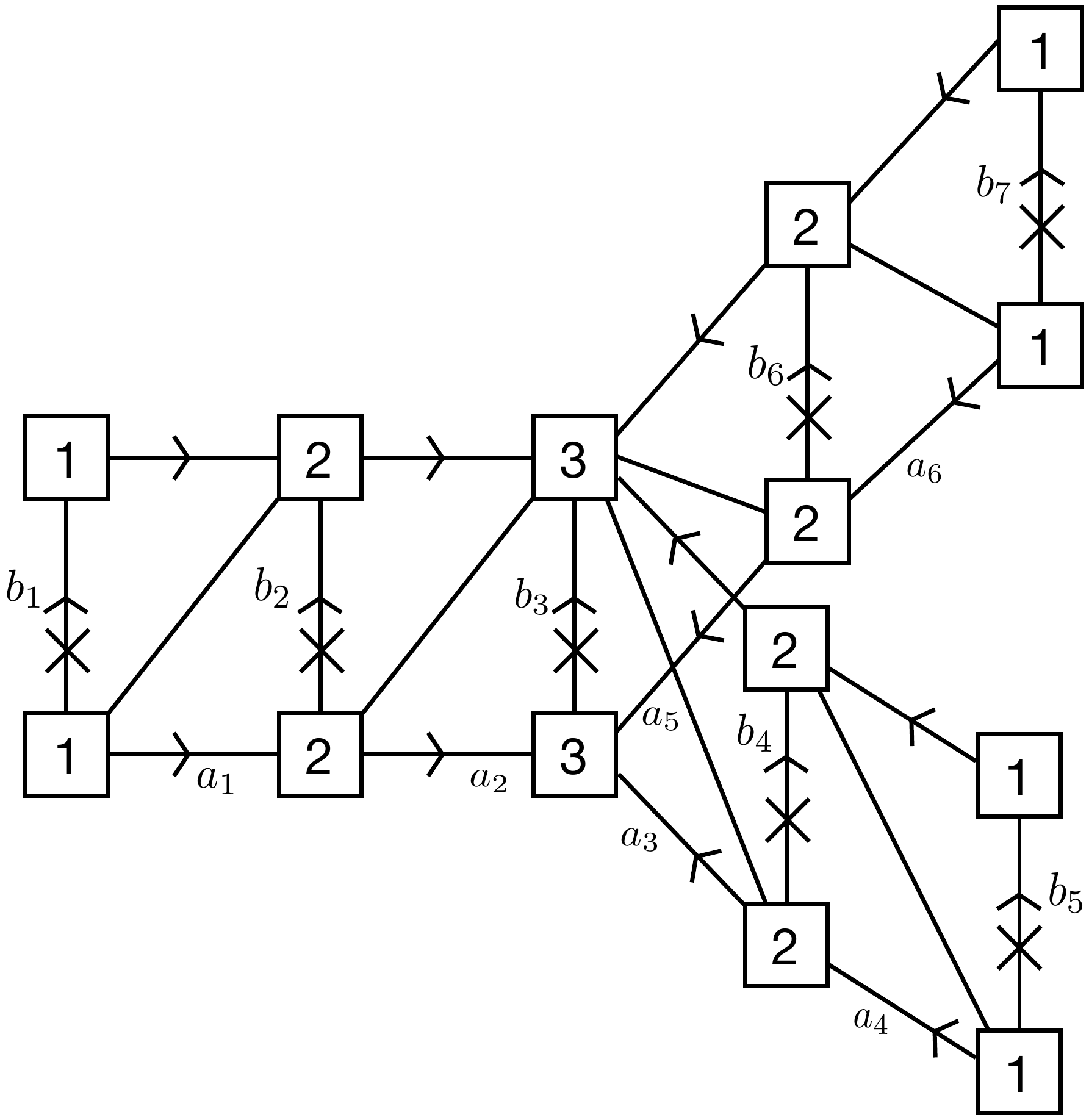}\qquad \qquad \includegraphics[scale=0.34]{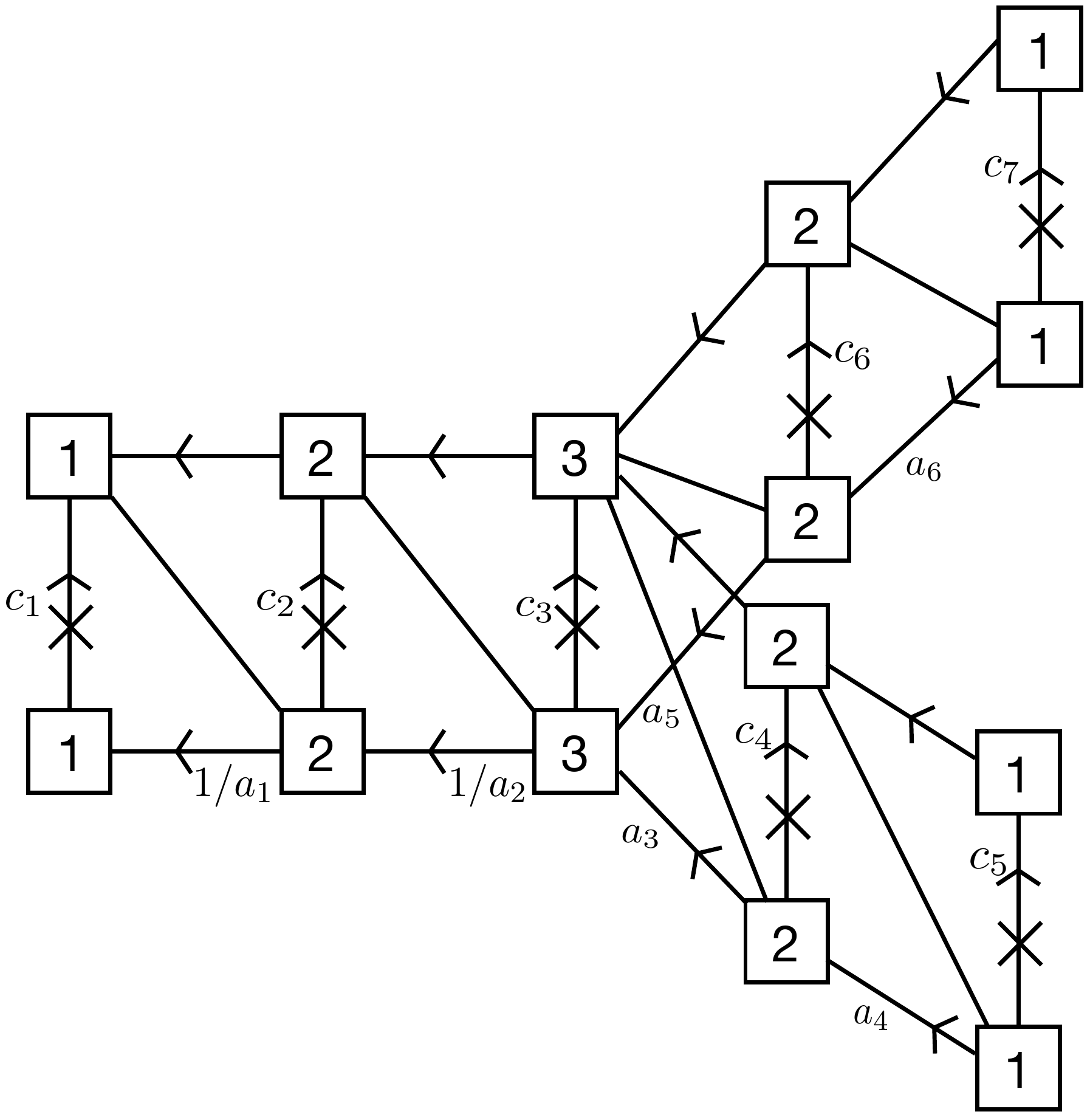}
\caption{Basic domain walls in the affine $E_6$ quiver theory with $\mathcal{D}_1=\{+^6\}_\beta$ and $\mathcal{D}_2=\{-,-,+^4\}_\beta$, respectively. The integer numbers $l_i$ in the boxes denote the $SU(l_iN)$ gauge symmetries at the interface.}\label{fig:E6-tubess}
\end{center}
\end{figure}

We present two basic domain walls for fluxes on the $U(1)^6_\beta$ of the first $E_6$ global symmetry in Figure \ref{fig:E6-tubess}.  The fugacities $a_i,b_i,c_i$ are related to the fugacities $\beta_i,\gamma_i$ for the Cartans $\prod_{i=1}^6 U(1)_{\beta_i}\times U(1)_{\gamma_i}$ of  $E_6\times E_6$.  Regarding the mass parameters of the baryon symmetries given in (\ref{eq:coupling-baryon}), the fugacities $a_i$ for the chiral halves of the 5d hypermultiplets are given by
\begin{eqnarray}\label{eq:E6-hyper-fugacities}
	&&a_1 = (\beta_2\gamma_1/\beta_1\gamma_2)^{1/2}, \quad a_2 = (\beta_3^2\gamma_1\gamma_2/\beta_1\beta_2\gamma_3^2)^{1/6} , \quad a_3 = (\beta_1\beta_2\beta_3\beta_5^3\beta_6\gamma_4^3/\beta_4^3\gamma_1\gamma_2\gamma_3\gamma_5^3\gamma_6)^{1/12} , \nonumber \\
	&&a_4 = (\beta_1\beta_2\beta_3\beta_4\beta_6\gamma_5/\beta_5\gamma_1\gamma_2\gamma_3\gamma_4\gamma_6)^{1/4} , \quad a_5=(\beta_1\beta_2\beta_3\gamma_4^3\gamma_5^3\gamma_6/\beta_4^3\beta_5^3\beta_6\gamma_1\gamma_2\gamma_3)^{1/12} , \nonumber \\
	&&a_6 = (\beta_1\beta_2\beta_3\beta_4\beta_5\gamma_6/\beta_6\gamma_1\gamma_2\gamma_3\gamma_4\gamma_5)^{1/4} \ .
\end{eqnarray}
Here, we have chosen the Cartans $U(1)_{\beta_i}$ and $U(1)_{\gamma_i}$ in the orthogonal basis of $E_6$ such  that the fundamentals of $SO(10)\subset E_6$ carry unit charges under these Cartans. Specifically, using the $SO(10)\times U(1)$ subgroup of $E_6$, the $\beta_{1-5}$ fugacities parametrize the $SO(10)$ and $\beta_{6}$ parametrize the $U(1)$, where the latter is normalized such the the $\bold{10}$ of $SO(10)$, appearing in the decomposition of the $\bold{27}$, has charge $1$. The $U(1)_{\gamma_i}$ symmetries use the same basis.
%  the $E_6$ mass parameters $m_i$ in (\ref{eq:coupling-baryon}) are related to the mass parameters $e_i$ in the orthogonal basis as
% % \begin{equation}
% % 	C_2(E_6)_{\bf F} = 6 \sum_{i=1}^5 C_1(U(1)_{\beta_i})^2 + 2 C_1(U(1)_{\beta_6})^2 \ .
% % \end{equation}
% \begin{eqnarray}
% 	&&m_1 =e_1\!-\!\frac{e_6}{3} , \quad m_2 = e_1\!+\!e_2\! -\!\frac{2e_6}{3}
% 	\ , \quad m_3 = e_1\!+\!e_2\!+\!e_3\!-\!e_6 , \quad m_4 = \frac{e_1\!+\!e_2\!+\!e_3\!+\!e_4\!-\!e_5}{2}\!-\!\frac{5e_6}{6} , \nonumber \\
% 	&&m_5 = -\frac{2e_6}{3}, \quad m_6 = \frac{e_1+e_2+e_3+e_4+e_5-e_6}{2} , \quad m_7=0 \ .
% \end{eqnarray}
% Also, the cancellation of the gauge-global mixed anomaly inflows regarding the 5d gauge kinetic terms with (\ref{eq:coupling-baryon}) 
Also, the anomaly free condition for the Cartans $U(1)_{\beta_i}$ and $U(1)_{\gamma_i}$
and the 4d superpotential constraints fully determine all fugacities for the 4d chiral fields  as
\begin{eqnarray}
	&&b_1= \beta_4/\beta_1 , \quad b_2 = (\beta_4/\beta_2)^{1/2} , \quad b_3 = (\beta_4/\beta_3)^{1/3} , \quad b_4 = (\beta_5/\beta_3)^{1/2} , \nonumber \\
	&& b_5 = (\beta_1\beta_2\beta_4\beta_5\beta_6/\beta_3)^{1/2} , \quad b_6 = (\beta_3\beta_5)^{-1/2} , \quad b_7 = (\beta_1\beta_2\beta_4/\beta_3\beta_5\beta_6)^{1/2}  \ , 
\end{eqnarray}
for the first domain wall and 
\begin{eqnarray}
	&&c_1 = \beta_2/\beta_1 , \quad c_2 = (\beta_3/\beta_1)^{1/2}, \quad c_3 = (\beta_4/\beta_1)^{1/3}, \quad c_4 =(\beta_5/\beta_1)^{1/2} , \nonumber \\
	&&c_5 =(\beta_2\beta_3\beta_4\beta_5\beta_6/\beta_1)^{1/2}, \quad c_6=(\beta_1\beta_5)^{-1/2} , \quad c_7 = (\beta_2\beta_3\beta_4/\beta_1\beta_5\beta_6)^{1/2} \ ,
\end{eqnarray}
for the second domain wall. One can check that these domain walls, when coupled to the 5d boundary conditions, have no cubic gauge anomalies and in total $U(1)^6_\beta\times U(1)^6_\gamma\subset E_6\times E_6$ anomaly-free abelian global symmetries. We can construct all the other basic domain walls in the same way by choosing different domain wall data $\mathcal{B}=\{s_i\}$ and $\mathcal{T}=\beta$ or $\gamma$. 
% All the abelian charges of the other chiral fields are fixed by the cubic superpotentials for the triagles in the quiver diagrams. With the abelian charge assignments in these two quiver diagrams, it is easy to check that the anomaly inflows from the 5d boundary conditions are all cancelled by the 't Hooft anomalies from the 4d degrees of freedom inserted at the interface. 

% \begin{eqnarray}
% 	&&Tr(SU(N)_1^3) = -Ns_1 , \quad Tr(SU(2N)_1^3)=\frac{N}{2}(s_1-3s_2),\quad Tr(SU(3N)^3)=N(s_2+s_3+s_5) , \nonumber \\
% 	&&Tr(SU(2N)_2^3) = \frac{N}{2}(s_4-3s_3) , \quad Tr(SU(N)_2^3) = -Ns_4,\quad Tr(SU(2N)_3^3) = \frac{N}{2}(s_6-3s_5) , \quad Tr(SU(N)_3^3) = -s_6 \nonumber \\
% 	&&Tr(U(1)_{\beta_1}SU(2N)_1^2)=-Tr(U(1)_{\beta_2}SU(2N)_1^2) = -\frac{N}{4}, \quad Tr(U(1)_{\beta_3}SU(3N)^2)=-Tr(U(1)_{\beta_4}SU(3N)^2)=\frac{N}{2}
% \end{eqnarray}

The global symmetries are permuted by the domain walls. The first domain wall with $\mathcal{D}_1=\{+^6\}_\beta$ in Figure \ref{fig:E6-tubess} permutes the $U(1)_{\beta_i}$ symmetries as
\begin{eqnarray}
	&&\beta_1 \rightarrow (\beta_3\beta_4/\beta_1\beta_2)^{1/2} , \quad \beta_2 \rightarrow (\beta_1\beta_3/\beta_2\beta_4)^{1/2}, \quad \beta_3 \rightarrow(\beta_2\beta_3/\beta_1\beta_4)^{1/2}, \nonumber \\
	&& \beta_4 \rightarrow (\beta_1\beta_2\beta_3\beta_4)^{-1/2}, \quad
	\beta_5\rightarrow (\beta_6/\beta_5)^{1/2},\quad \beta_6\rightarrow(\beta_5^3\beta_6)^{-1/2} \ ,
\end{eqnarray}
in terms of the $U(1)_{\beta_i}$ fugacities, and the second domain wall with $\mathcal{D}_1=\{-,-,+^4\}_\beta$ permutes the symmetries as
\begin{eqnarray}
	&&\beta_1 \rightarrow (\beta_1\beta_2/\beta_3\beta_4)^{1/2} , \quad \beta_2 \rightarrow (\beta_1\beta_3/\beta_2\beta_4)^{1/2}, \quad \beta_3 \rightarrow(\beta_1\beta_4/\beta_2\beta_3)^{1/2}, \nonumber \\
	&& \beta_4 \rightarrow (\beta_1\beta_2\beta_3\beta_4)^{-1/2}, \quad
	\beta_5\rightarrow (\beta_6/\beta_5)^{1/2},\quad \beta_6\rightarrow(\beta_5^3\beta_6)^{-1/2} \ .
\end{eqnarray}

We propose the {\bf Conjectures} in Section \ref{sec:a-type-walls} hold for the flux domain walls in the affine $E_6$ quiver theory engineered by connecting these basic domain walls.
The flux assignment for each basic domain wall is given by (\ref{eq:domainwall-flux}) with $h^\vee=12$.
So the first domain wall in Figure \ref{fig:E6-tubess} corresponds to the 6d flux $F=(0,0,\frac{1}{3},-\frac{1}{3},0,0)_\beta$ and the second domain wall is mapped to the 6d flux $F=(\frac{1}{2},-\frac{1}{6},-\frac{1}{6},-\frac{1}{6},0,0)_\beta$.

When we connect 6 copies of the first domain wall, the resulting domain wall configuration has flux $F=(0,1,2,-1,0,0)_\beta$ with $\prod_{i=1}^6\sigma^{t_i}=1$ after carefully taking into account the above permutations. This flux is the minimal integral flux breaking $E_6\rightarrow SU(3)\times SU(3)\times SU(2)\times U(1)$. 
The {\bf Conjectures} predict that 4d reductions of this configuration by putting it on a circle or an interval with maximal boundary conditions give rise to the 6d $(E_6,E_6)$ conformal matter theory of $N$ M5-branes carrying the same flux $F$ compactified on a torus or a tube with two maximal punctures.
Indeed, we checked these 4d theories obtained from the 5d flux domain walls have the same 't Hooft anomalies as those computed from the compactification of the 6d anomaly polynomial and the anomaly inflows for the maximal punctures. For example, the 4d theory on a torus has the central charges as
\begin{equation}
	a = \frac{2N\sqrt{3}(9N-4)^{3/2}}{\sqrt{N(3N-1)}} \ , \quad 
	c = \frac{N(18N-7)\sqrt{3(9N-4)}}{\sqrt{N(3N-1)}} \ ,
\end{equation}
which are precisely the central charges of the 4d quiver theory obtained from the 5d theory with domain walls of the flux $F=(0,1,2,-1,0,0)_\beta$  on a circle.
We have performed similar computations using other combinations of basic domain walls and the results agree with our conjectures.
Note that the six dimensional computation is agnostic about some of  the fields becoming free and thus for comparison we do not decouple the free fields. The above values are not the superconformal anomalies as we do not take into account the accidental $U(1)$ symmetries coming from free fields. As we claim to identify the symmetries correctly in six and four dimensions, the above computation is a simple non-trivial check of matching 't Hooft anomalies between the six dimensional and four dimensional computations.

The flux domain walls for the minimal $E_6$ conformal matter theory when $N=1$ can be  constructed as follows. The 5d theory is a quiver gauge theory of $SU(2)^3\times SU(3)$ gauge groups with two fundamental hypers for each $SU(2)$ gauge node. Let us define a basic domain wall labeled by $\mathcal{D}=\mathcal{B}_{\mathcal{T}}$ as a 4d interface connecting two 5d quiver theories with boundary condition $\mathcal{B}=\{s_1,s_2,\cdots,s_6\}$. Here we assume that two $SU(2)$ fundamental hypermultiplets for an $SU(2)$ gauge node have the same boundary conditions for $\mathcal{T}=\beta$ and the opposite boundary conditions for $\mathcal{T}=\gamma$. The domain wall adds four 4d chiral multiplets $q_i$ and $\tilde{q}_i$ as explained.
% The domain wall adds four 4d chiral multiplets $q_i$ in the bifundamental of $G_i\times G'_i$ (with $G_i,G_i'=SU(2)$ or $SU(3)$) and three chiral multiplets $\tilde{q}_i$ whose representations are fixed by cubic interactions of the form (\ref{eq:domainwall-W}). There exist cubic superpotentials between the chiral fields $q_{i}$ for $i=1,2,3$ and  a set of the 5d hypers with Neumann boundary condition in the fundamental representations of the $SU(2)_i$ and $SU(2)_i'$ in two sides.
The index $\mathcal{T}=\beta$ or $\gamma$ denotes the representation of the 4d field $q_4$ either in $(\bar{\bf 3},{\bf3})$ or $({\bf3},\bar{\bf 3})$ of the $SU(3)\times SU(3)'$ gauge symmetry.

\begin{figure}[htbp]
\begin{center}
\includegraphics[scale=0.38]{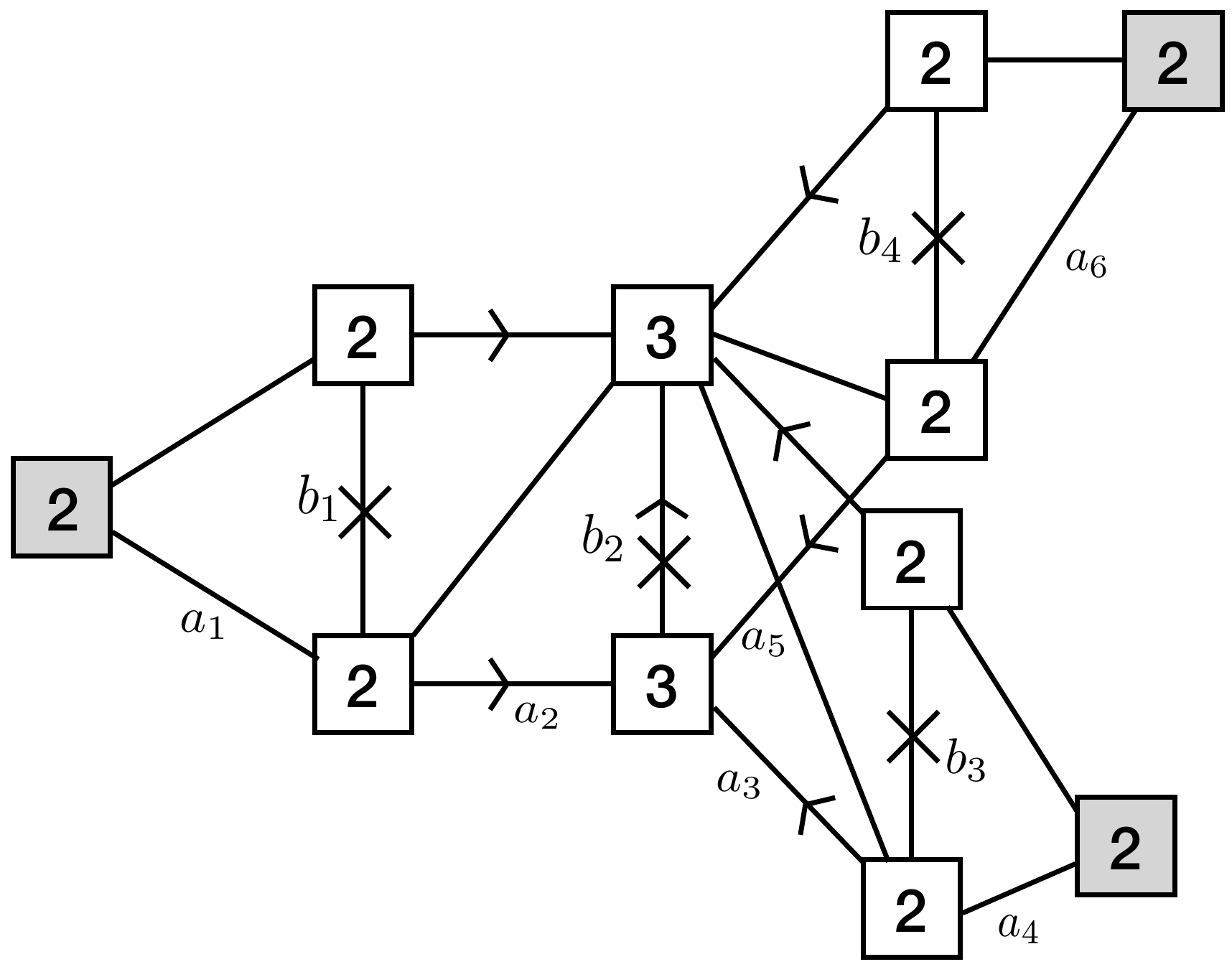}\qquad \qquad \includegraphics[scale=0.38]{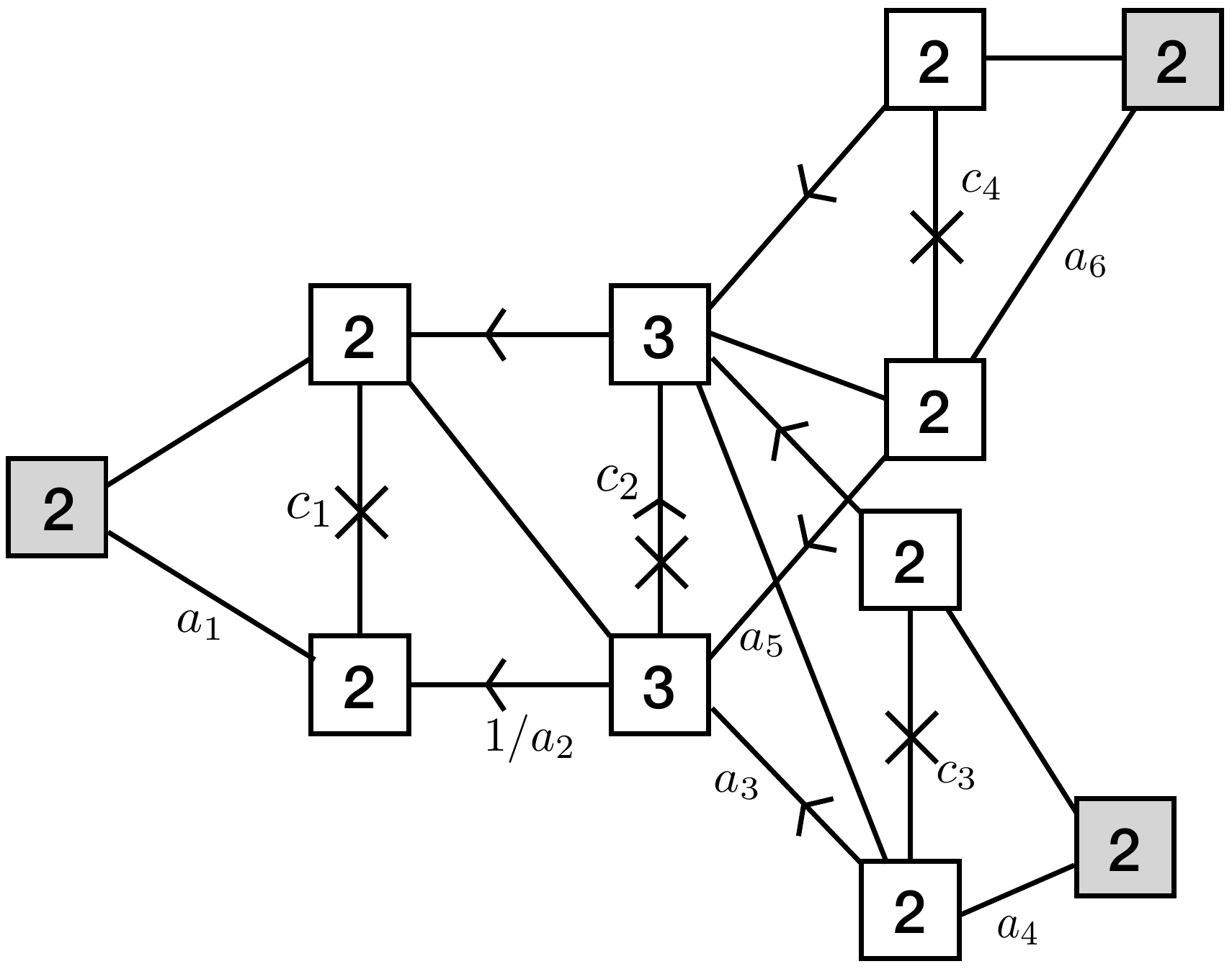}
\caption{Basic domain walls in the minimal $E_6$ theory with $\mathcal{D}_1=\{+^6\}_\beta$ and $\mathcal{D}_2=\{+,-,+^4\}_\beta$, respectively. The shaded boxes stand for $SU(2)^3$ flavor nodes in the bulk 5d theory.}\label{fig:E6-minimal-tubes}
\end{center}
\end{figure}

The quiver diagrams for two examples are depicted in Figure \ref{fig:E6-minimal-tubes}. The abelian fugacities for the 5d hypermultiplets in the first (or bottom) chamber are given in (\ref{eq:E6-hyper-fugacities}). Other abelian charges for the 4d fields are fixed by the gauge-global anomaly cancellation and the cubic superpotential couplings. We find
\begin{equation}
	b_1 = (\beta_4/\beta_2)^{1/2} , \quad b_2 = (\beta_4/\beta_3)^{1/3}, \quad
	b_3 = \left(\beta_5/\beta_3\right)^{1/2}, \quad b_4 = \left(1/\beta_3\beta_5\right)^{1/2} \ ,
\end{equation}
for the first quiver diagram and
\begin{eqnarray}
	c_1= (\beta_3/\beta_2)^{1/2},\quad c_2=(\beta_4/\beta_2)^{1/3},\quad
	c_3=(\beta_5/\beta_2)^{1/2},\quad c_4=(1/\beta_2\beta_5)^{1/2} \ ,
\end{eqnarray}
for the second diagram. The first and the second domain walls correspond to the fluxes $F_1=(0,\frac{1}{8},\frac{1}{3},-\frac{5}{24},0,0)_\beta$ and $F_2=(0,\frac{11}{12},-\frac{8}{24},-\frac{1}{12},0,0)_\beta$ respectively. Other basic domain walls can be similarly constructed and generic flux domain walls can be obtained from various combinations of these basic domain walls.

We briefly comment on the other possible choices of the boundary conditions for the $SU(2)_{i=1,2,3}$ fundamental hypermultiplets. As mentioned, the above basic domain walls choose the same (or the opposite) boundary conditions for each pair of two $SU(2)_i$ fundamental hypers when $\mathcal{T}=\beta$ (or $\mathcal{T}=\gamma$). However, we can for example consider a domain wall with opposite boundary conditions for two $SU(2)_1$ fundamental hypers while keeping other boundary conditions and 4d chiral fields the same as those drawn in the first quiver in Figure \ref{fig:E6-minimal-tubes}. In this case, we have a new domain wall with $\beta_2/\beta_1$ and $\gamma_2/\gamma_1$ exchanged. Similarly, other choices of boundary conditions for the $SU(2)_{i=2,3}$ fundamentals can lead to other types of domain walls which we can obtain by exchanging some $\beta_i$ and $\gamma_i$'s.

\subsection*{ $E_7$ and $E_8$}
Lastly, let us discuss the flux domain walls in the affine $E_7$ and $E_8$ quiver gauge theories. The basic domain walls in these theories can be built by a single interface supporting two copies of affine $E_7$ or affine $E_8$ quiver gauge symmetries coupled to 5d boundary conditions, e.g. $\mathcal{B}=\{s_1,\cdots ,s_7\}$ for $E_7$ or $\mathcal{B}=\{s_1,\cdots,s_8\}$ for $E_8$, and 4d chiral fields $q_i$, $\tilde{q}_i$ through the cubic superpotentials of the form (\ref{eq:domainwall-W}).
We denote these basic domain walls by $\mathcal{D}=\mathcal{B}_{\mathcal{T}}$. There exists a unique domain wall system for each $\mathcal{D}$.
Under the {\bf Conjecture} stated above, we expect that the flux domain wall systems constructed by gluing the basic domain walls can realize the compactification of the 6d $E_7$ and $E_8$ conformal matter theories with flux on a torus or a tube.

\begin{figure}[htbp]
\begin{center}
\includegraphics[scale=0.27]{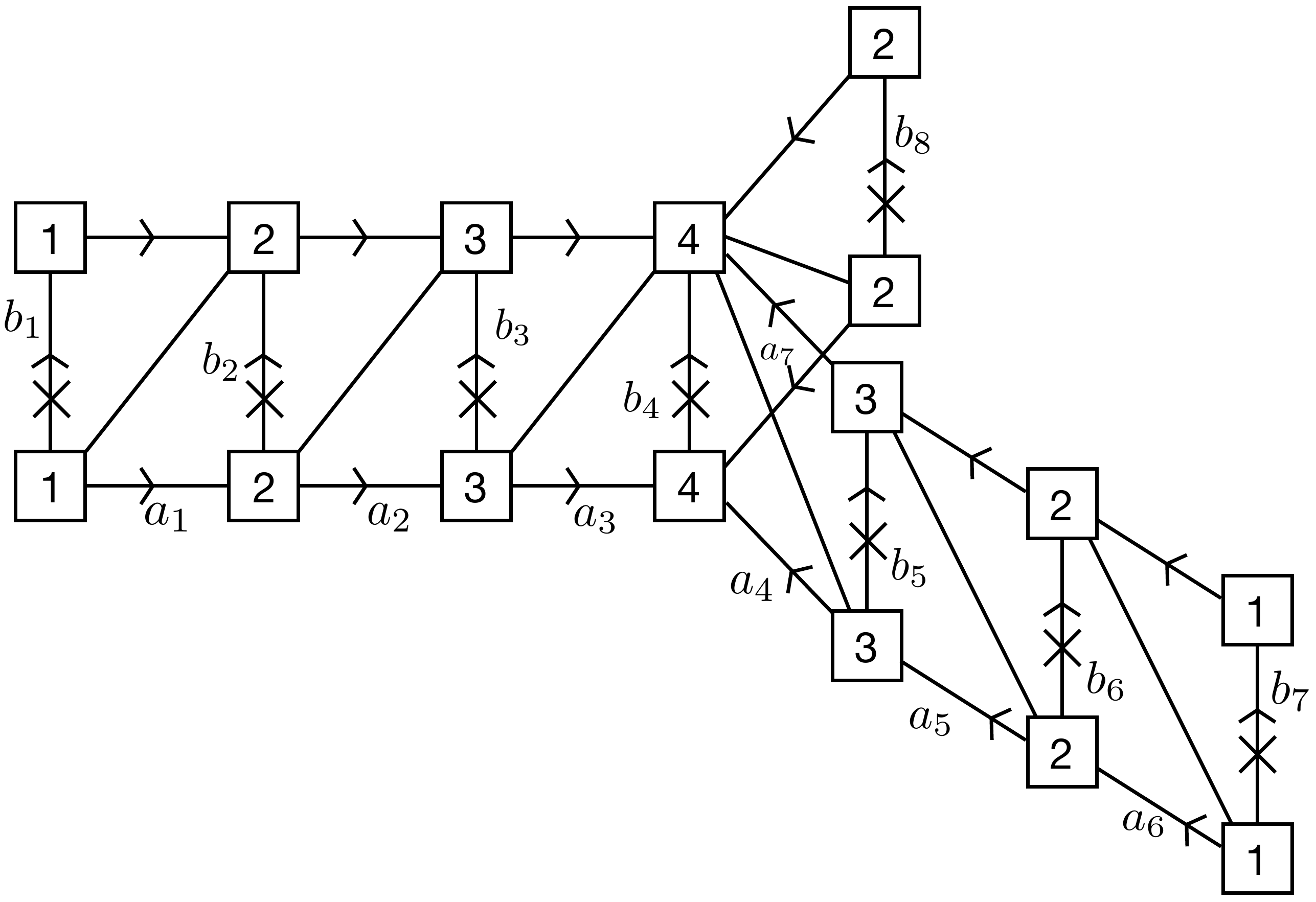}\qquad  \includegraphics[scale=0.27]{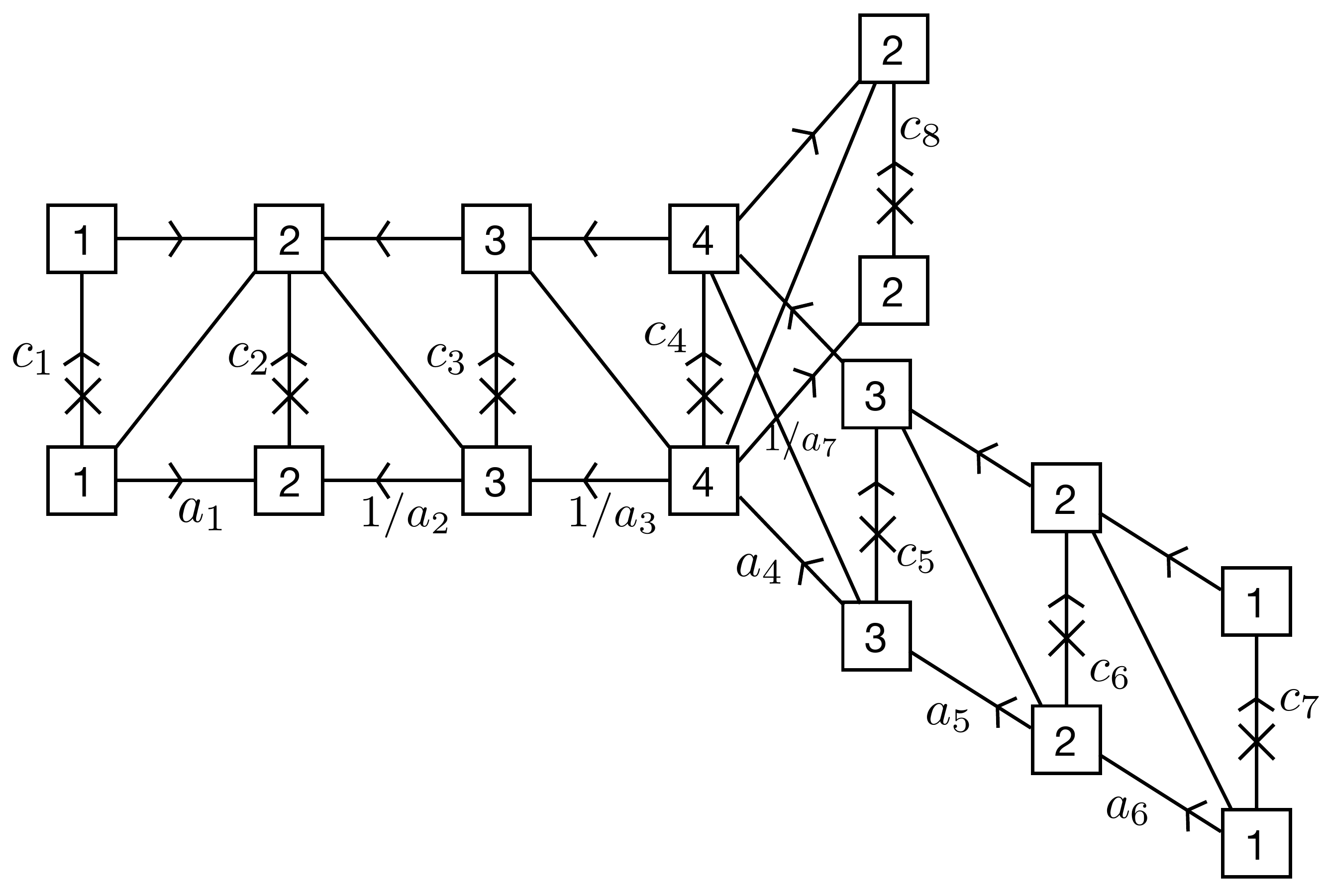}
\caption{Basic domain walls in the affine $E_7$ quiver theory with $\mathcal{D}_1=\{+^7\}_\beta$ and $\mathcal{D}_2=\{+,-,-,+,+,+,-\}_\beta$, respectively. The integer numbers $l_i$ in the boxes denote the $SU(l_iN)$ gauge symmetries at the interface.}\label{fig:E7-tubes}
\end{center}
\end{figure}

Two basic domain walls in the affine $E_7$ quiver theory are given in Figure \ref{fig:E7-tubes}. From (\ref{eq:coupling-baryon}), the fugacities $a_i$ for the 5d hypermultiplets are given by
\begin{eqnarray}\label{eq:fugacities-a-E7}
	&&a_1 = \left(\frac{\beta_2\gamma_1}{\beta_1\gamma_2}\right)^{\!\!1/2}, \quad a_2=\left(\frac{\beta_3^2\gamma_1\gamma_2}{\beta_1\beta_2\gamma_3^2}\right)^{\!\!1/6}, \quad a_3 = \left(\frac{\beta_4^3\gamma_1\gamma_2\gamma_3}{\beta_1\beta_2\beta_3\gamma_4^3}\right)^{\!\!1/2},
	\quad a_4 =\left(\frac{\beta_1\beta_2\beta_3\beta_4\beta_6^2\gamma_5^2}{\beta_5^2\gamma_1\gamma_2\gamma_3\gamma_4\gamma_6^2}\right)^{\!\!\frac{1}{12}}, \nonumber \\
	&&a_5 = \left(\frac{\beta_1\beta_2\beta_3\beta_4\beta_5\gamma_6}{\beta_6\gamma_1\gamma_2\gamma_3\gamma_4\gamma_5}\right)^{1/6}, \quad a_6=\left(\frac{\gamma_7}{\beta_7}\right)^{1/2}, \quad a_7 = \left(\frac{\gamma_5\gamma_6}{\beta_5\beta_6}\right)^{1/4} \ ,
\end{eqnarray}
with the fugacities $\beta_i,\gamma_j$ for $E_7\times E_7$ symmetry in the orthogonal basis where the fundamentals of $SU(2)\times SO(12)\subset E_7$ carry unit charges of $U(1)_{\beta_i}$ or $U(1)_{\gamma_i}$ symmetries. Here $\beta_7$ and $\gamma_7$ are for the $SU(2)$ and the rest for the $SO(12)$.
% Note that the mass parameters $m_i$ in (\ref{eq:coupling-baryon}) are mapped to the parameters $e_i$ in the orthogonal basis as
% \begin{eqnarray}
% 	&&m_1 = e_1\!-\!\frac{e_7}{2}, \quad m_2 = e_1\!+\!e_2\!-\!e_7, \quad m_3=e_1\!+\!e_2\!+\!e_3\!-\!\frac{3e_7}{2}, \quad
% 	m_4 = e_1\!+\!e_2\!+\!e_3\!+\!e_4\!-\!2e_7, \nonumber \\
% 	&&m_5 = \frac{e_1\!+\!e_2\!+\!e_3\!+\!e_4\!+\!e_5\!-\!e_6\!-\!3e_7}{2}, \quad m_6 = -e_7, \quad m_8 = \frac{e_1\!+\!e_2\!+\!e_3\!+\!e_4\!+\!e_5\!+\!e_6\!-\!2e_7}{2} \ \ \qquad .
%  \end{eqnarray}
The interface introduces no other anomaly-free abelian symmetry. The abelian charges for the 4d fields are uniquely fixed by the gauge-global mixed anomaly cancellation.
 % which involves the gauge kinetic term contributions with the gauge couplings in (\ref{eq:coupling-baryon}), together with superpotential constraints. 
 We find
\begin{eqnarray}
	&&b_1 = \frac{\beta_5}{\beta_1}, \quad b_2 = \left(\frac{\beta_5}{\beta_2}\right)^{1/2}, \quad b_3 = \left(\frac{\beta_5}{\beta_3}\right)^{1/3}, \quad
	b_4 = \left(\frac{\beta_5}{\beta_4}\right)^{1/4}, \quad b_5=\left(\frac{\beta_6}{\beta_4}\right)^{1/3}, \nonumber \\
	&&b_6 = \left(\frac{\beta_1\beta_2\beta_3\beta_5\beta_6\beta_7}{\beta_4}\right)^{1/4}, \quad
	b_7 = \left(\frac{\beta_1\beta_2\beta_3\beta_5\beta_6}{\beta_4\beta_7}\right)^{1/2}, \quad 
	b_8 = \left(\frac{1}{\beta_4\beta_6}\right)^{1/2} \ ,
\end{eqnarray}
for the first domain wall with $\mathcal{D}_1=\{+^7\}_\beta$ and
\begin{eqnarray}
	&&c_1 = \frac{\beta_3}{\beta_1}, \quad c_2 =\left(\frac{\beta_3}{\beta_2}\right)^{1/2},\quad
	c_3 = \left(\frac{\beta_4}{\beta_2}\right)^{1/3}, \quad c_4 = \left(\frac{1}{\beta_2\beta_6}\right)^{1/4}, \quad c_5 = \left(\frac{1}{\beta_2\beta_5}\right)^{1/3}, \nonumber \\
	&& c_6 = \left(\frac{\beta_1\beta_3\beta_4\beta_7}{\beta_2\beta_5\beta_6}\right)^{1/4},\quad c_7 = \left(\frac{\beta_1\beta_3\beta_4}{\beta_2\beta_5\beta_6\beta_7}\right)^{1/2}, \quad c_8 = \left(\frac{1}{\beta_5\beta_6}\right)^{1/2} \ ,
\end{eqnarray}
for the second domain wall with $\mathcal{D}_2=\{+,-,-,+,+,+,-\}_\beta$. The cubic gauge anomalies are also absent and therefore these domain walls can consistently couple to the 5d affine $E_7$ quiver gauge theory. Using (\ref{eq:domainwall-flux}) with $h^\vee=18$, the fluxes are $F_1=(0,0,0,\frac{2}{9},-\frac{5}{18},-\frac{1}{18},0)_\beta$ for the first domain wall and $F_2=(0,\frac{5}{18},-\frac{1}{6},-\frac{1}{9},\frac{1}{6},\frac{1}{6},0)_\beta$ for the second domain wall.
Gluing 12 copies of the first basic domain wall with $\mathcal{D}_1=\{+^7\}_\beta$ leads to a flux domain wall with $F=(-\frac{1}{2},\frac{1}{2},\frac{3}{2},\frac{5}{2},-\frac{3}{2},-\frac{1}{2},-1)_\beta$ corresponding to the 6d theory on a circle with a unit flux breaking $E_7\rightarrow SU(4)\times SU(3)\times SU(2)\times U(1)$. The circle reduction of this 5d theory with the flux domain wall yields a 4d quiver gauge theory at low energy and the resulting 4d theory has the central charges
\begin{equation}
	a=\frac{72N(3N-1)^{3/2}}{\sqrt{N(4N-1)}}\ , \quad
	c=\frac{9N(24N-7)\sqrt{3N-1}}{\sqrt{N(4N-1)}} \ ,
\end{equation}
which precisely coincide with the expected central charges of the $E_7$ conformal matter theory on a torus with flux $F$. We also checked that other 't Hooft anomalies of this 4d theory match the anomalies obtained by integrating the 6d anomaly polynomial in the presence of the flux $F$.
One can similarly construct other basic domain walls by choosing different $\mathcal{B}$ and $\mathcal{T}$ and generic flux domain walls from other combinations of the basic domain walls.

\begin{figure}[htbp]
\begin{center}
\includegraphics[scale=0.35]{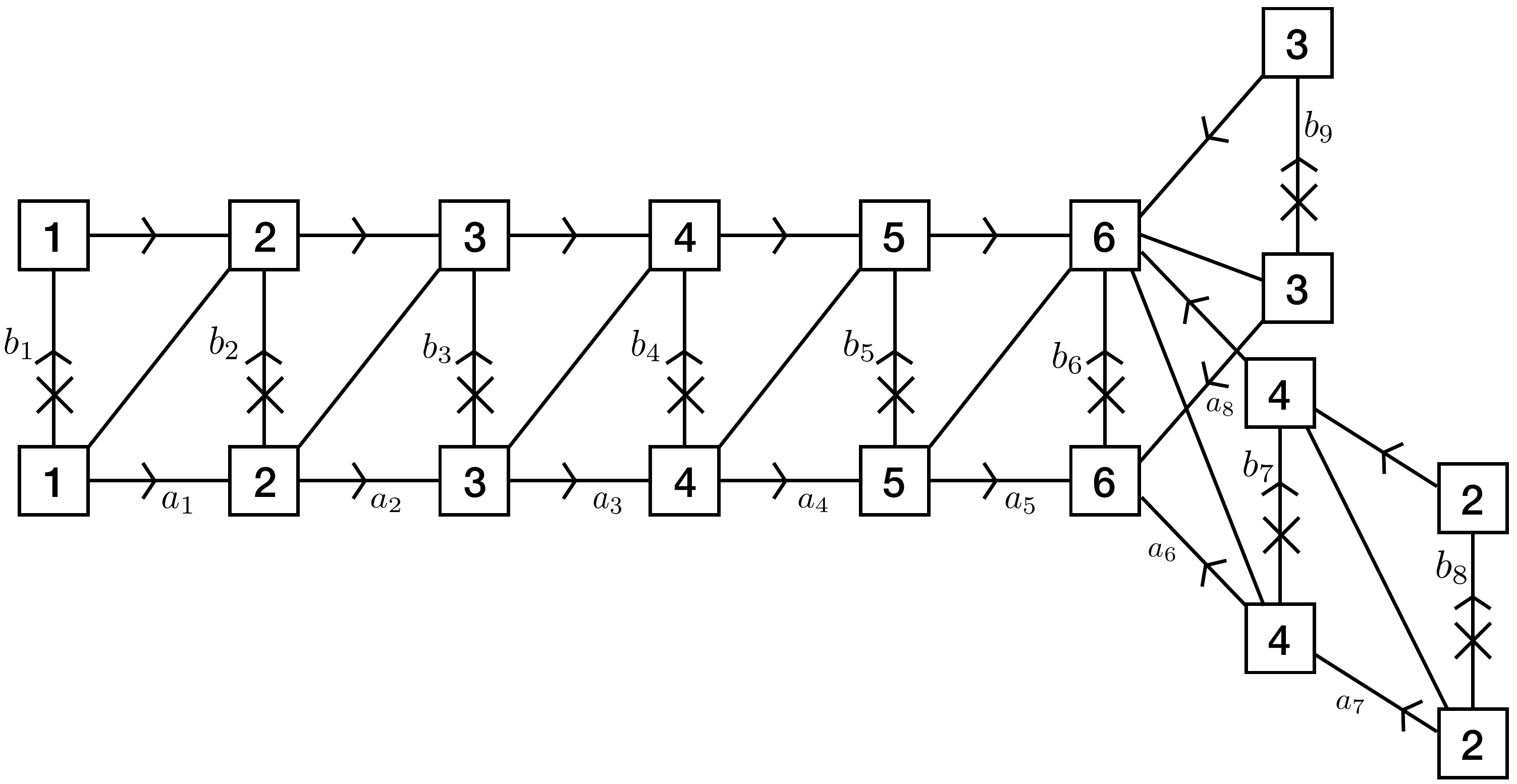}
\caption{Basic domain wall in the affine $E_8$ quiver theory with $\mathcal{D}=\{+^8\}_\beta$. The integer numbers $l_i$ in the boxes denote the $SU(l_iN)$ gauge symmetries at the interface.}\label{fig:E8-tubes}
\end{center}
\end{figure}

In Figure \ref{fig:E8-tubes}, we have a basic domain wall for $\mathcal{D}=\{+^8\}_\beta$ in the affine $E_8$ quiver gauge theory. One can readily check that all cubic gauge anomalies are absent when it is inserted between two 5d affine $E_8$ quiver theories. The 5d hypermultiplet fugacities $a_i$ are
\begin{eqnarray}\label{eq:fugacities-a-E8}
	&&a_1=\left(\tfrac{\beta_1\gamma_8}{\beta_8\gamma_1}\right)^{\!\!1/2}, \quad
	a_2=\left(\tfrac{\beta_2^2\gamma_1\gamma_8}{\beta_1\beta_8\gamma_2^2}\right)^{\!\!1/6}, \quad
	a_3=\left(\tfrac{\beta_3^3\gamma_1\gamma_2\gamma_8}{\beta_1\beta_2\beta_8\gamma_3^3}\right)^{\!\!1/12}, \quad
	a_4 = \left(\tfrac{\beta_4^4\gamma_1\gamma_2\gamma_3\gamma_8}{\beta_1\beta_2\beta_3\beta_8\gamma_4^4}\right)^{\!\!1/20}, \\
	&&a_5=\left(\tfrac{\beta_5^5\gamma_1\gamma_2\gamma_3\gamma_4\gamma_8}{\beta_1\beta_2\beta_3\beta_4\beta_8\gamma_5^5}\right)^{\!\!1/30},\quad 
	a_6=\left(\tfrac{\beta_1\beta_2\beta_3\beta_4\beta_5\beta_7^3\beta_8\gamma_6^3}{\beta_6^3\gamma_1\gamma_2\gamma_3\gamma_4\gamma_5\gamma_7^3\gamma_8}\right)^{\!\!1/24}, \quad
	a_7 = \left(\tfrac{\gamma_7\prod_{i\neq7}\beta_i}{\beta_7\prod_{i\neq7}\gamma_i}\right)^{\!\!1/8}, \quad
	a_8 =\left(\tfrac{\gamma_6\gamma_7}{\beta_6\beta_7}\right)^{\!\!1/6} \nonumber \ ,
\end{eqnarray}
in the orthogonal bases where the fundamentals of $SO(16)\subset E_8$ carry  charge $\pm1$ under $U(1)_{\beta_i}$ or $U(1)_{\gamma_i}$. By demanding the gauge-global mixed anomaly cancellation and the superpotential constraints, we fix the $U(1)$ charges of the 4d chiral fields as
\begin{eqnarray}
	&&b_1=\frac{\beta_6}{\beta_8}, \quad b_2=\left(\frac{\beta_6}{\beta_1}\right)^{1/2}, \quad
	b_3 = \left(\frac{\beta_6}{\beta_2}\right)^{1/3}, \quad
	b_4 = \left(\frac{\beta_6}{\beta_3}\right)^{1/4} , \quad
	b_5 = \left(\frac{\beta_6}{\beta_4}\right)^{1/5} , \nonumber \\
	&&b_6 = \left(\frac{\beta_6}{\beta_5}\right)^{1/6},\quad
	b_7 = \left(\frac{\beta_7}{\beta_5}\right)^{1/4},\quad
	b_8 = \left(\frac{\prod_{i\neq 5}\beta_i}{\beta_5}\right)^{1/4}, \quad b_9 = \left(\frac{1}{\beta_5\beta_7}\right)^{1/3} \ .
\end{eqnarray}
From (\ref{eq:domainwall-flux}) with $h^\vee=30$, one reads the flux $F=(\frac{1}{60},\frac{1}{60},\frac{1}{60},\frac{1}{60},\frac{7}{60},-\frac{13}{60},-\frac{1}{60},\frac{1}{60})_\beta$ for this basic domain wall. The other basic domain walls can be similarly constructed.

We suggest that the 5d affine $E_8$ quiver theory with 30 copies of the basic domain wall in Figure \ref{fig:E8-tubes} realizes the 6d $(E_8,E_8)$ conformal matter theory with flux $F=(-\frac{1}{2},\frac{1}{2},\frac{3}{2},\frac{5}{2},\frac{7}{2},-\frac{5}{2},-\frac{1}{2},-\frac{3}{2})_\beta$ on a circle. The flux $F$ is the minimal flux breaking one $E_8$ global symmetry to $SU(5)\times SU(3)\times U(1)$. A circle reduction of this 5d domain wall configuration leaves a 4d quiver gauge theory corresponding to the 6d theory with flux $F$ on a torus. Indeed, the central charges of this 4d theory
\begin{equation}
	a=\frac{50\sqrt{3}N(9N-2)^{3/2}}{\sqrt{N(6N-1)}}\ , \quad c=\frac{25N(36N-7)\sqrt{N(27N-6)}}{2\sqrt{N(6N-1)}} 
\end{equation}
agree with those computed by integrating the 6d anomaly polynomial with $F$. Also, all other anomalies in this 4d theory coincide with the anomalies of the 6d theory with $F$ on a torus.

Now consider the domain walls in the minimal affine $E_7$ and $E_8$ quiver gauge theories. When $N=1$, the $U(1)$ gauge nodes in the quiver diagram are replaced by two fundamental hypermultiplets charged under the adjacent $SU(2)$ gauge nodes. Domain walls in these theories have almost the same form of those in the non-minimal cases. The differences are the boundary conditions of the $SU(2)$ fundamental hypers and 4d chiral multiplets coupled to these 5d fundamental fields. The other parts are the same.

\begin{figure}[htbp]
\begin{center}
\includegraphics[scale=0.33]{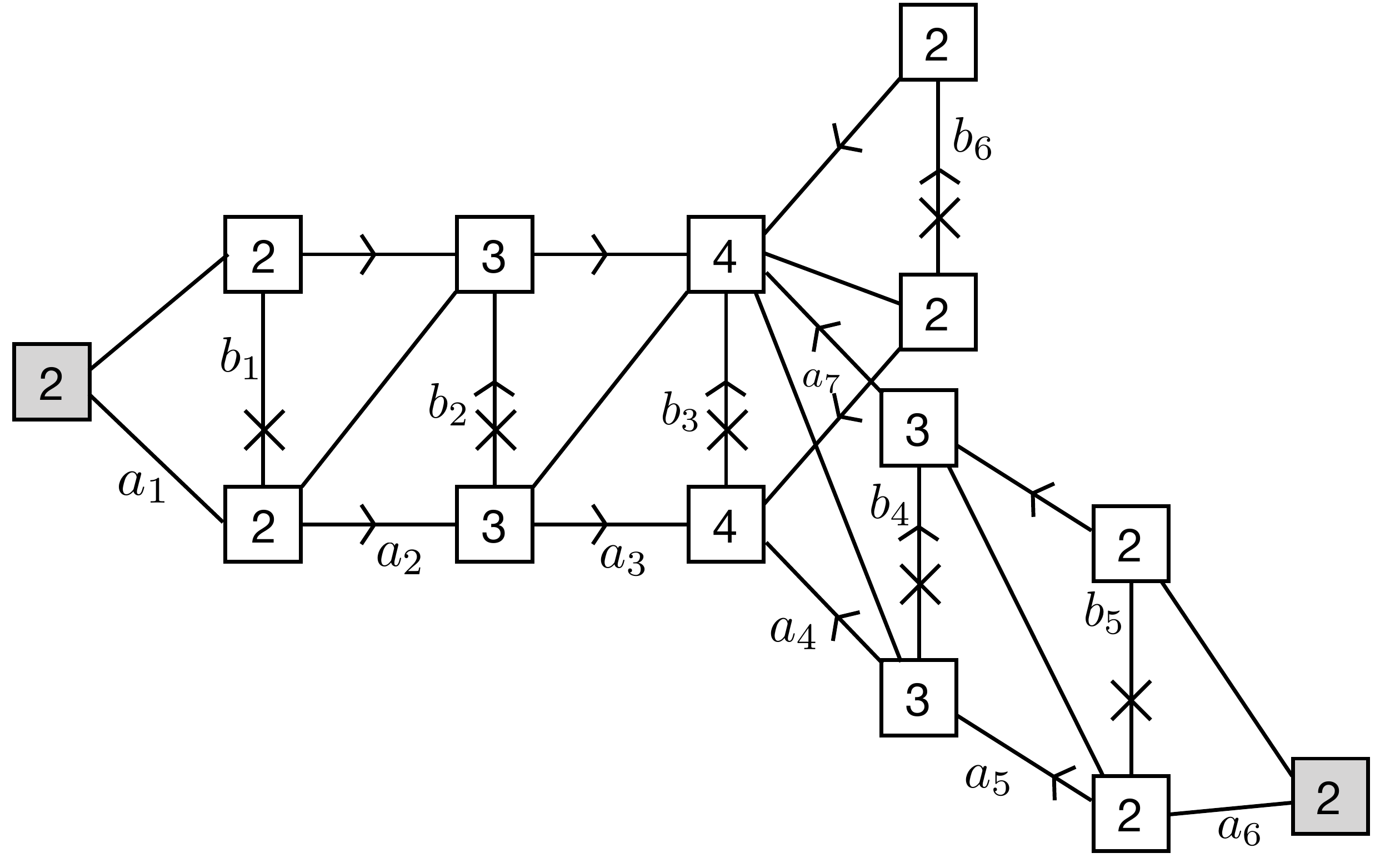}
\caption{Basic domain walls in the minimal $E_7$ theory with $\mathcal{D}_1=\{+^7\}_\beta$.}\label{fig:E7-minimal-tubes}
\end{center}
\end{figure}

For $E_7$, the basic domain walls are defined by $\mathcal{D}=\mathcal{B}_{\mathcal{T}}$ with $\mathcal{B}=\{s_1,s_2,\cdots,s_7\}$. We choose two fundamentals for an $SU(2)$ gauge node to have the same (or the opposite) boundary conditions for $\mathcal{T}=\beta$ (or $\mathcal{T}=\gamma$). As the minimal $E_6$ cases above, the interface connects these boundary conditions of the $SU(2)$ fundamentals in two sides by using the cubic superpotentials including the 4d bifundamental chirals $q_i$ charged under $SU(2)_i\times SU(2)_i'$. These also hold for the $E_8$ cases below with $\mathcal{B}=\{s_1,s_2,\cdots s_8\}$.

One example of $E_7$ is drawn in Figure \ref{fig:E7-minimal-tubes} with fugacities $a_i$ in (\ref{eq:fugacities-a-E7}) and
\begin{eqnarray}
&&b_1 = (\beta_5/\beta_2)^{1/2}, \quad b_2 = (\beta_5/\beta_3)^{1/3}, \quad b_3 = (\beta_5/\beta_4)^{1/4},
 \nonumber \\
&& b_4 = (\beta_6/\beta_4)^{1/3}, \quad b_5 = (\beta_1\beta_2\beta_3\beta_5\beta_6\beta_7/\beta_4)^{1/4}, \quad b_6 = (1/\beta_4\beta_6)^{1/2} \ ,
\end{eqnarray}
which are again determined by the gauge-global anomaly cancellation and the superpotential constraints.
This quiver diagram describes a basic domain wall with $\mathcal{D}_1=\{+^7\}_\beta$ and it has the flux $F=\frac{1}{72}(-3,3,1,15,-17,-3,-3)_\beta$ in the orthogonal basis of $E_7$.

\begin{figure}[htbp]
\begin{center}
\includegraphics[scale=0.33]{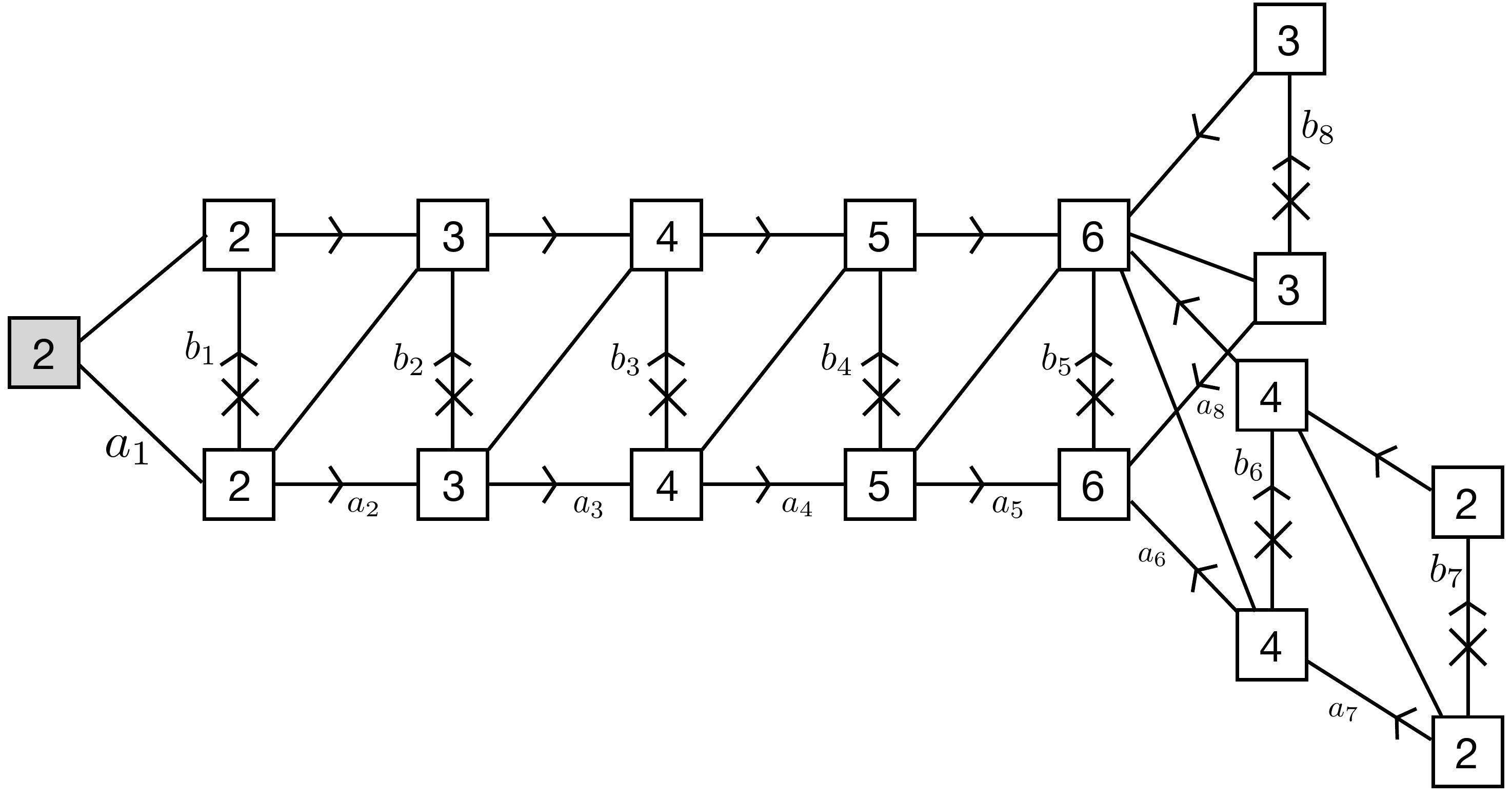}
\caption{Basic domain walls in the minimal $E_8$ theory with $\mathcal{D}_1=\{+^8\}_\beta$.}\label{fig:E8-minimal-tubes}
\end{center}
\end{figure}

The $E_8$ basic domain wall for $\mathcal{D}_1=\{+^8\}_\beta$ is drawn in Figure \ref{fig:E8-minimal-tubes}. Here, the 5d fugacities $a_i$ are written in (\ref{eq:fugacities-a-E8}) and 4d fugacities $b_i$ are given by
\begin{eqnarray}
	&&b_1 = (\beta_6/\beta_1)^{1/2}, \quad b_2 = (\beta_6/\beta_2)^{1/3}, \quad b_3 = (\beta_6/\beta_3)^{1/4}, \quad b_4 = (\beta_6/\beta_4)^{1/5} \nonumber \\
	&&b_5 = (\beta_6/\beta_5)^{1/6}, \quad b_6 = (\beta_7/\beta_5)^{1/4}, \quad b_7 = (\prod_{i\neq 5}\beta_i/\beta_5)^{1/4}, \quad b_8 =(1/\beta_5\beta_7)^{1/3} \ .
\end{eqnarray}
This domain wall corresponds to the flux $F=\frac{1}{60}(2,1,1,1,7,-12,-1,-1)_\beta$. We will see more examples and tests for our flux domain wall conjectures by reducing them to 4d in the next section.

\section{Four dimensions}

We have obtained a conjecture for the theories corresponding to compactifications of ADE conformal matter on a torus with  flux for the flavor symmetry.
The way we construct the models is by gluing together building blocks which formally correspond to spheres with two maximal punctures and have some flux. 
In this section we will subject these conjectures to various tests directly in four dimensions. There are two types of checks we can perform, we can compare anomalies and check for enhancements of symmetry.
The checks of anomalies we have already discussed so here we will give examples of enhancement of symmetry as well as some dynamical interesting features such as dualities.  We will also connect the results to other constructions appearing in  the literature .

Note that the two punctured spheres in general have the non abelian flavor symmetry of the models in six dimensions broken down to abelian factors both due to presence of punctures and the flux. 
The tube theories we have defined also in general, with the exception of minimal $D$ and $E$ conformal matter,  have only abelian flavor  symmetries which are not associated to the punctures. However, when combining the theories to form a torus and selecting the combination of the two punctured spheres so that the flux is non generic, the symmetry can be enhanced at some loci on the conformal manifold. Such an enhancement is highly non obvious from the four dimensional perspective. One can check such enhancements of symmetry by studying different supersymmetric partition functions, and in particular the supersymmetric index \cite{Romelsberger:2005eg,Kinney:2005ej,Beem:2012yn,Dolan:2008qi}. As the models involved have many gauge group factors, the computations, though straightforward, are computationally intense. We will thus restrict in what follows to verifying the claims in some simple examples in which the computations can be performed more easily.

\subsection*{$S$ gluing $\Phi$ gluing and color for punctures}

We have derived theories we naturally associate with tubes from our five dimensional discussion. It is natural from the four dimensional point of view to define slightly modified tubes. This will not change the theories we associate to closed Riemann surface as we will define the gluing procedure to achieve that, however this will make the discussion more uniform with the existing literature. The difference is with the chiral fields one couples to operators charged under the puncture symmetries.
 In particular, there are different types of punctures with the different choices denoted by color, sign, and orientation \cite{Gaiotto:2015usa,Razamat:2016dpl}. As punctures break some of the $G\times G$ symmetry these choices specify what is exactly the preserved symmetry group and what are the anomalies associated to the puncture. Moreover, each puncture comes with a set of operators $M_i$ which are charged under the puncture symmetry. 
These operators generalize the moment maps of class ${\cal S}$, which is the $(A_1,A_1)$ conformal matter. Different punctures give rise to different charges under the Cartan of $G\times G$ for $M_i$. We glue punctures of same color and sign and opposite orientation by gauging the puncture symmetry and adding a field $\Phi_i$ in conjugate representations to $M_i$, and coupling them through a superpotential,

\be
W^\Phi_{gluing}=M_i \Phi_i -\widetilde M_i   \Phi_i\,.
\ee Here $M_i$ are the operators of one of the punctures and $\widetilde M_i$ of the other. We denote this gluing as $\Phi$ gluing \cite{Gaiotto:2015usa,Razamat:2016dpl}. We also can glue punctures of same color, same orientation, and opposite sign. Punctures of opposite sign have operators $M_i$ and $\widetilde M_i$ in conjugate representations. Thus we turn on the superpotential,

\begin{figure}[htbp]
\begin{center}
\includegraphics[scale=0.8]{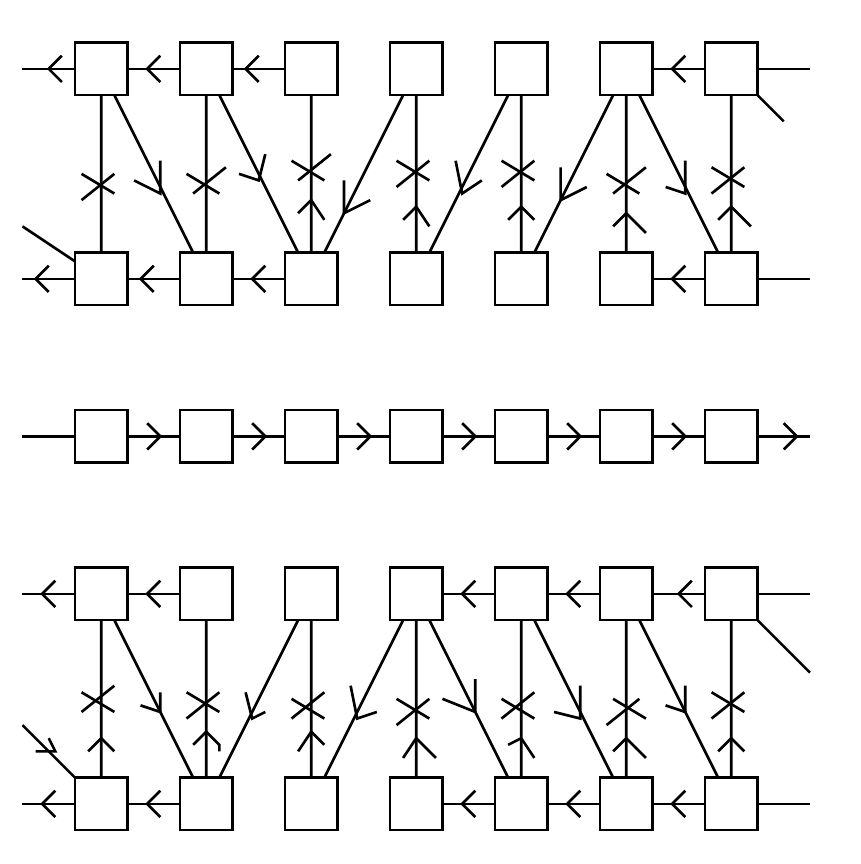}
\caption{This is $\Phi$ gluing for $A$ theories. The two punctures are of the same sign, opposite orientation, and different colors. The puncture symmetry is $SU(N)^k$. The operators $M_i$ are charged under two factors of $SU(N)$ symmetry. For $+$ boundary conditions they are bifundamental fields and for $-$ they are bi-linear operators. For simplicity we have written arrows on some of the lines and the orientation of the rest is determined by the superpotentials we turn for each face of the quiver.
 In the middle we have the fields we add when gluing, $\Phi$, in the bifundamental representation of two of the $SU(N)^k$ symmetries.
  }\label{deffphu}
\end{center}
\end{figure}

\be 
W^S_{gluing} = \widetilde M_i M_i\,.
\ee This gluing will be denoted as $S$ gluing \cite{Gaiotto:2015usa,Hanany:2015pfa,Razamat:2016dpl}.

\begin{figure}
\begin{center}
\includegraphics[scale=0.8]{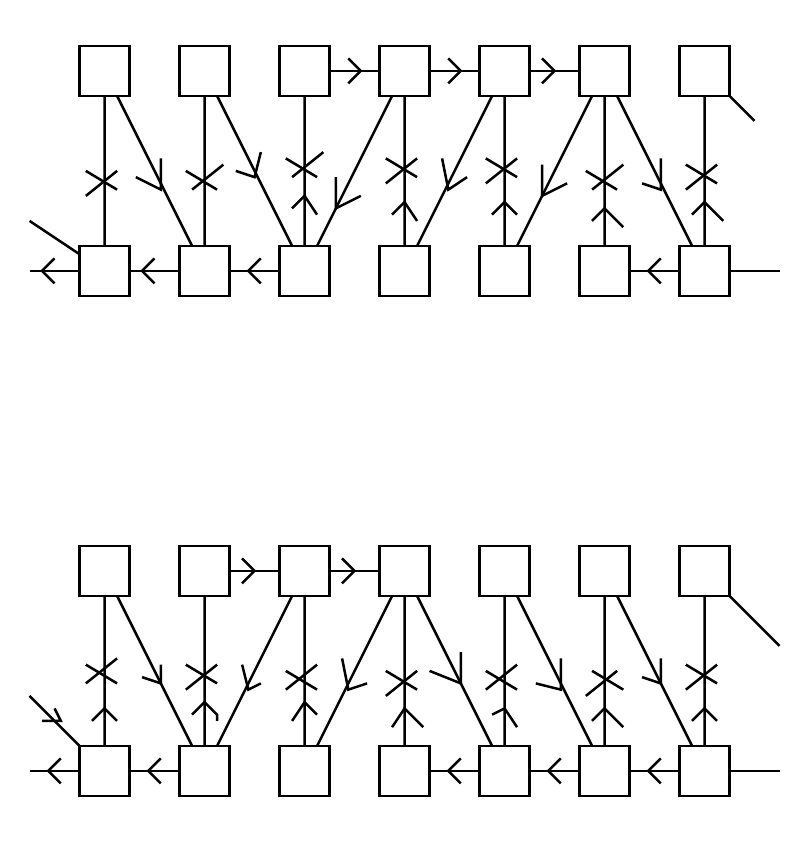}
\caption{This is $S$ gluing for $A$ theories. The two punctures have opposite sign, same orientation, and different color. We glue the left and right punctures with S gluing turning on $W^S_{gluing}$ superpotential.}\label{deffphuy}
\end{center}
\end{figure}

For the theories we have defined, to follow the same pattern of gluing as above, we need to add bifundamental fields between symmetry factors on the same side of the duality wall
for the tubes only for $+$ boundary conditions and not for minus and call these positive punctures. We can also add the lines for $-$ boundary conditions and not for $+$ and call those negative punctures. 
With this definition of tubes we  glue them with $\Phi$ and $S$ gluing. See Figures \ref{deffphu} for $\Phi$ gluing and \ref{deffphuy} for $S$ gluing. 
For simplicity we will add fields such that the left puncture is of one sign and the right of opposite sign and build surfaces  using $S$ gluing. In this way some of the properties will become simpler.
 
\

\subsection{Examples of $A$}

We start with the case of A-type conformal matter. 
 The case of a single M5 brane, the minimal matter, leads to free models so we will not consider it. The cases of $N>1$ were discussed recently in a variety of papers \cite{Gaiotto:2015usa,Hanany:2015pfa,Franco:2015jna,Razamat:2016dpl}. The compactifications on the torus were considered in \cite{Bah:2017gph}. There, the torus with no punctures was constructed by first gluing together theories corresponding to spheres with two maximal and one minimal puncture. Such theories are given by the Wess-Zumino type of models. Then the minimal punctures were closed by turning on vacuum expectation values to certain operators. In particular, it was claimed that in such a manner one can produce models corresponding to flux which is multiple of a $1/k$. Nevertheless, our discussion has something to add even for this case. The theories we get from this construction generally have flux that is a multiple of $1/r$ for $r$ an integer obeying $0<r<k$, and thus give theories that are not accessible from the existing construction\footnote{To be more precise, the theories in question have fractional fluxes and require a central flux element for their consistency, see section 5 in \cite{Bah:2017gph} and appendix C in \cite{Kim:2017toz}. The theories so far constructed in the literature embed the central flux element for one $SU(k)$ group in the other $SU(k)$ group, while the theories considered here embed it in the unbroken part of the same $SU(k)$ group.}. The two construction overlap for theories with integer fluxes for which this provides a different systematic construction of the theories.
Let us then discuss some of the general properties of the compactifications and analyze several concrete instances in detail.

\subsubsection*{The two puncture spheres}

\begin{figure}
\begin{center}
\includegraphics[scale=0.9]{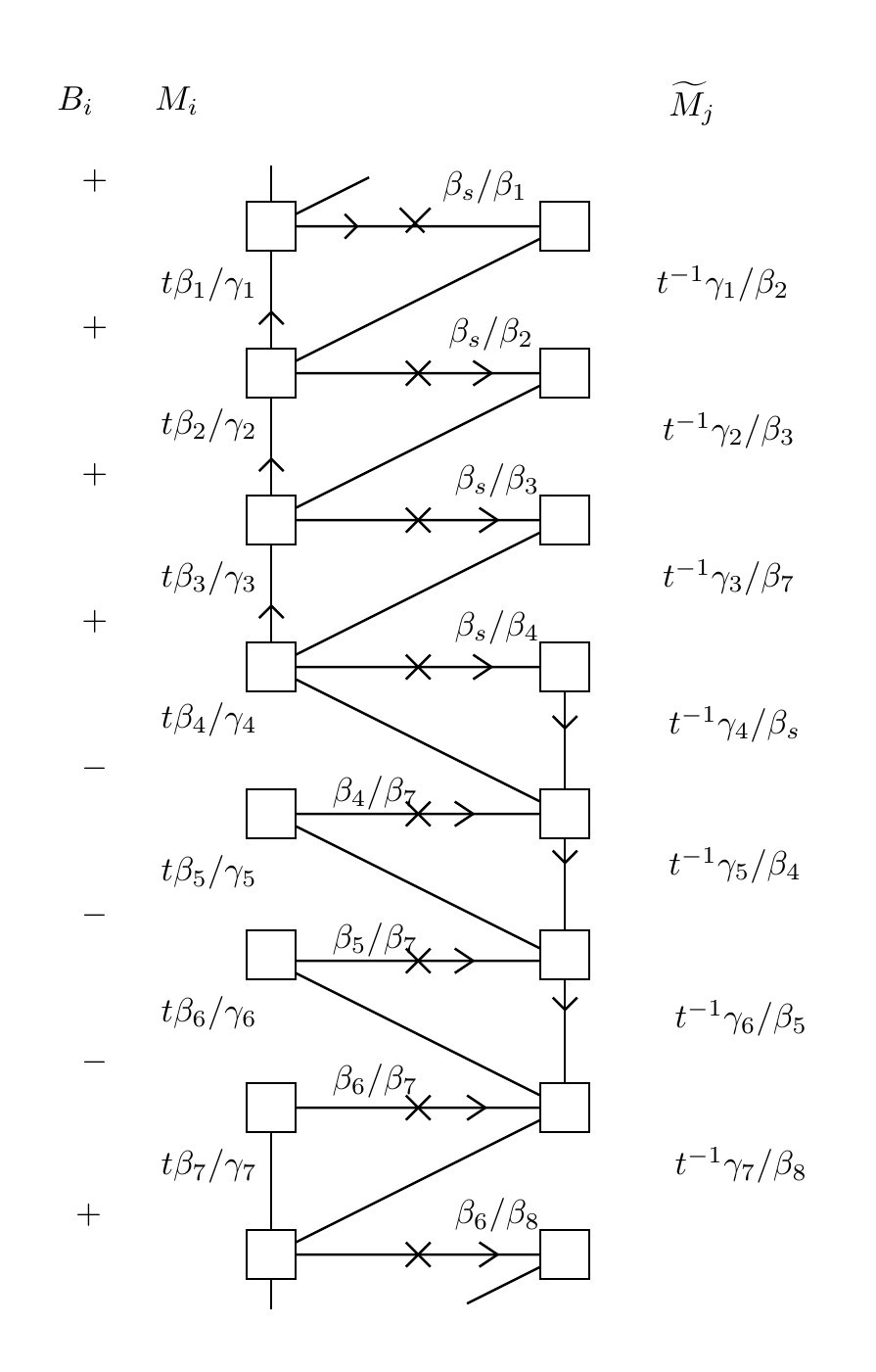}
\caption{Example of a tube in the four dimensional language. Note this differs from what we have defined before by horizontal fields which are chiral fields associated to punctures.
 The signs $B_i$ correspond to boundary conditions. We denote the charge of operators $M_i$ and $\widetilde M_j$ for the two punctures as manifest in their fugacities. The charges of bifundamental fields are denoted in their fugacities. The missing fugacities can be derived by demanding superpotential terms for every face of the quiver. The R charges of flipped fields are zero, flip fields are two, and fields which are not flipped are one. The permutation is $(\dots 1 \; 2\; 3\; 7\; 8\, \dots) \;\, (\,\dots 6\; 5\; 4\; s\,\dots)$.
}\label{tubexst}
\end{center}
\end{figure}

The color of a puncture is defined as follows. The symmetry group in the $A$ case is $SU(k)\times SU(k)\times U(1)$. The puncture symmetry is $\prod_{i=1}^kSU(N)_i$. We have a cyclic order of the $SU(N)$ groups  coming from the affine Dynkin diagram of type $A$. 
We have operators $M_i$ associated to the puncture for $i=1 \dots k$. The operators $M_i$ are in the bifundamental representation of the $i$-th and $i+1$-th $SU(N)$ group for positive sign punctures and in the bifundamental of the $i+1$-th and $i$-th $SU(N)$ group for negative sign punctures. The color is defined by assigning charges to $M_i$ under the Cartan of $SU(k)\times SU(k)$. We parametrize the Cartan by $U(1)_{\beta_l}$ for one $SU(k)$ and $U(1)_{\gamma_i}$ for the other.
 For positive punctures the $M_i$ are charged plus one under one of the $U(1)_{\beta}$ and minus one under one of the $U(1)_\gamma$, and each $M_i$  is charged under different symmetries. The choice of the $U(1)_{\beta}$ and $U(1)_\gamma$ symmetries under which each of the $M_i$ operators are charged constitute the color of the puncture. We thus can think of the color, as discussed in previous sections here,  as defined by two permutations modulo cyclic transformations, that is the color index takes value in $\left(\sigma_\beta,\sigma_\gamma\right)\in S_k\times S_k/{\mathbb Z}_k$, where here $S_k$ is the symmetric group.
  For negative punctures the $U(1)_\beta$ charges are negative and $U(1)_\gamma$ are positive.
 For puncture of color $\left(1\,,\; 1\right)$ the charges of $M_i$ are plus one under $U(1)_{\beta_i}$ and minus one under $U(1)_{\gamma_i}$. Punctures of opposite orientation are mirror images under the reflection of the affine Dynkin diagram of $A$.
The tubes we have defined  have two maximal punctures of different color. 
We illustrate this in Figure \ref{tubexst}. 

Because of the superpotential terms, it is clear that the only fields contributing to linear anomalies are the flip fields. Thus defining the flux as in the previous section, we have here $n_i= 1/2$. The flux $Q_i$ under $U(1)_{\beta_i}$ is given by the sum of charges of the flip fields under the symmetry divided by $N\, k$.
Because flip fields are not charged under $U(1)$ symmetries we deduce that the linear anomaly in it is zero. Moreover, because of the supepotentials only the flip fields contribute to  $U(1)_{t\beta_j}^3$ anomalies. In particular we deduce that the linear anomaly in any $U(1)$ is the same as cubic up to a factor of $N^2$. This agrees with the six dimensional prediction. In general we state as argued in {\bf Conjectures} that the anomalies agree with the six dimensional prescription if we glue together $l$ tubes if,
$
\prod_{i=1}^l \sigma^{t_i}\equiv\sigma = 1\,.$ That is if the two colors of maximal punctures are identical. If $\sigma\neq 1$ the anomalies agree for symmetries fixed by $\sigma$.  We have verified this statement in numerous cases but did not obtain a rigorous proof.

We will discuss several examples in some detail next.

\subsubsection*{$k=2$}

\begin{figure}
\begin{center}
\includegraphics[scale=1.26]{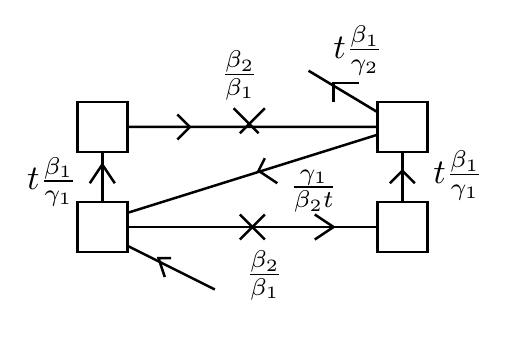}
\caption{The tube for $k=2$ with flux $(1,-1)$ for the $\beta$ symmetry. This can be obtained through closing punctures of free trinions and through the five dimensional computation. The two
punctures are of the same color and sign.}\label{tuhjluy}
\end{center}
\end{figure}

Let us consider the case of $k=2$. Here the only choice of $\sigma^t$ is the identity as $2$ splitting to two non vanishing numbers is $1+1$. This case is identical to the one we obtain by the closing of minimal punctures procedure.  The basic tube appears in Figure  \ref{tuhjluy}. 

Let us compute the flux of the model. The $\beta_1$ charges of the flip fields are $2N$ and those of the $\beta_2$ fields are $-2N$ dividing by $kN$ we obtain that the flux is $1$ in $\beta_1$ and $-1$ in $\beta_2$. This is is exactly the flux associated to this tube in \cite{Bah:2017gph}, and  it was checked 
that all anomalies agree with the six dimensional prescription. Moreover, it was verified in examples that the symmetry observed in the supersymmetric partition functions agrees with the expected symmetry implied by the value of the flux. 

\begin{figure}
\begin{center}
\includegraphics[scale=.9]{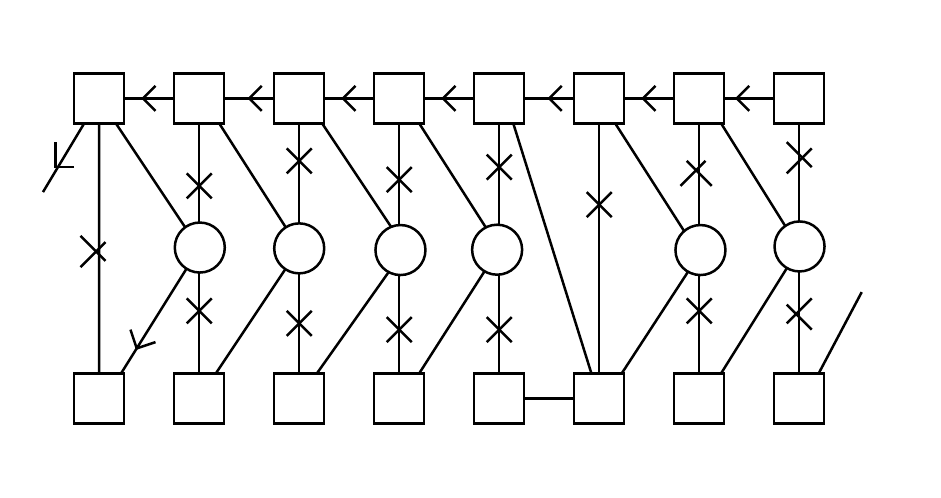}
\caption{Tube with flux $(0,0,\dots,-1,0,\dots,0,1,\dots)$ obtained from closing minimal punctures. The two maximal punctures are of the same color and sign.}\label{tyredf}
\end{center}
\end{figure}

Let us here quote a generalization of this tube following the procedure of closing punctures. The tube with two maximal punctures of the same color and same sign and with flux $1$ for one of the $\beta_i$ and $-1$ for another while zero for the rest is depicted in Figure \ref{tyredf}. With this tube any integer flux model can be constructed. Our construction will go beyond this by constructing models with fractional fluxes.

\subsubsection*{The tube with $\sigma^t=(2\, 3\, \dots k)$}

Let us discuss the example of tube with $\sigma^t =(2\, 3\, \dots k)$ for general $k$. This model  fixes one of the $\beta$ symmetries and without generality we can choose it to be $\beta_1$. The flux of this model computed from flip fields is $(1, -\frac1k, - \frac1k, \dots,-\frac1k,-\frac2k)$. As the permutation fixes $\beta_1$ and all of $\gamma_i$, the anomalies involving these symmetries, the R symmetry and the $U(1)$ agree with six dimensional computation. We can glue several such tubes together to form a torus in such a way that $\beta_1$ is always fixed. The flux is  fractional for general number of tubes, however for $l$ a multiple of  $k-1$ it is a multiple of $(k-1, -1,-1,\dots,-1)$, and in this case all the symmetries are preserved for the torus and anomalies agree with the computation in six dimensions. It is also easy to see that the quiver in this case is equivalent to a triangulation of the torus with $l$ triangles wrapping one cycle and $k$ another, with one side  of each triangle flipped. The flipped sides form $k$ lines wrapping the cycle with $l$ triangles. This is also the quiver that one would obtain if one glues together tubes that one naively corresponds to all same sign boundary conditions.

\subsubsection*{$k=3$ and $k=4$}

\begin{figure}
\begin{center}
\includegraphics[scale=0.87]{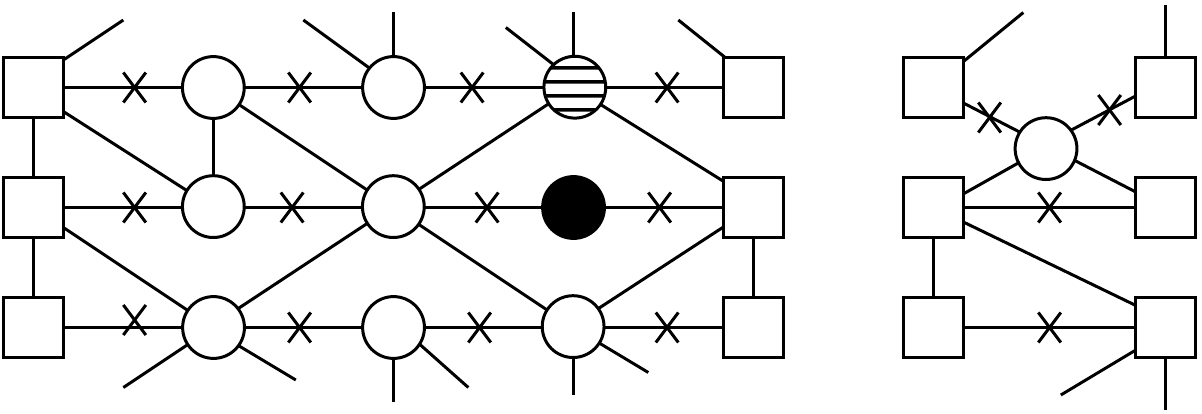}
\caption{On the left we have a combination of four tubes as in \eqref{kehf}. Note that one of the nodes, denoted in black, has $N_f=N$ and therefore its dynamics leads to it being Higgsed and the groups connected to it identified. Following this, the node denoted with dashed lines becomes $2N=N_f$. After performing Seiberg duality on this node, as well as a sequence of similar manipulations, we obtain the quiver on the right hand side. This is the same as the one obtained by closing punctures in \cite{Bah:2017gph} as in Figure \ref{tyredf}, if one flips the sign of one of the punctures as our tubes have punctures of different signs.
}\label{shgfywd}
\end{center}
\end{figure}

Let us here also discuss the two less obvious cases in some detail. Let us first take $k=3$. We have one type of tube as we can split $3=2+1$, which is the tube discussed in the previous subsection. We can define a similar tube, but with the permutation and flux in the $\gamma$ symmetries. We then have the freedom of gluing them together in a variety of ways. For example, we can take the following tubes,

\be
\sigma^{t_a}=(2 3)\,,  
\;\;\,
\sigma^{t_b}= (1 3)\;,
\;\;\, 
  \sigma^{t_c} = (2 3)\;,
  \;\;\,
  \sigma^{t_e} =(1 2)\,.
\ee We have that $\sigma^{t_a}\sigma^{t_b}\sigma^{t_c}\sigma^{t_e} =1$ and thus all anomalies are expected to agree with six dimensions. Moreover, the flux is,

\be\label{kehf}
(1,-\frac13,-\frac23)+(-\frac13,1,-\frac23)+(1,-\frac13,-\frac23)+(-\frac23,-\frac13,1)= (1,0,-1)\,.
\ee This is a flux one cab  obtain from closing punctures as in \cite{Bah:2017gph}, and as the two have same anomalies and expected symmetry they should be dual to each other. This should be possible to show using Seiberg duality \cite{Seiberg:1994pq}, see figure 
\ref{shgfywd}.

For $k=4$ we already have a richer variety of constructions. In addition to the $(123)$ tube we also have tubes associated with the $(12)(34)$ and $(13)(24)$ permutations.   The former has flux $(\frac34,\frac14,-\frac14,-\frac34)$ and the latter tube flux $(\frac12,-\frac12,\frac12,-\frac12)$. We can read off these fluxes easily from the flip fields. If we glue the first tube to itself we obtain the flux $(1,1,-1,-1)$, while doing the same for the second gives the flux $(1,-1,1,-1)$. We thus can construct these tubes from the ones we obtained by closing punctures and verify that the two constructions agree upon making use of dualities.

\subsection{Examples of $D$}

The discussion here will follow the general ideas of the previous section. In particular we start by defining the color of the punctures. The punctures have $SU(2N)^k \times SU(N)^4$ symmetry for $D_{k+3}$ case. We have $k+3$ operators associated to the puncture and we will denote them as $M^a_{1,2}$, $M^b_{1,2}$ and $M_{i}$ with $i\in\{1, . . . , k-1\}$. The punctures have a color which is defined by a label expected to take value in $W_{D_{k+3}}\times W_{D_{k+3}}$, which is the product of the two Weyl groups of $D_{k+3}$, possibly moded by some discrete symmetry. The tubes then can be viewed as associated to an element of the Weyl group of one of the two $D_{k+3}$ groups.

We parametrize one of the $D_{k+3}$ by $\beta_i$ and another by $\gamma_i$. We choose the fugacities so that the vector representation character is,

\be
{\bf 2k+6}_\gamma =\sum_{i=1}^{k+3} \gamma_i^{\mp2 N}\,,\;\qquad\, {\bf 2k+6}_\beta =
\sum_{j=1}^{k+3}
\beta_j^{\mp2 N}\, .
\ee

\begin{figure}
\begin{center}
\includegraphics[scale=0.87]{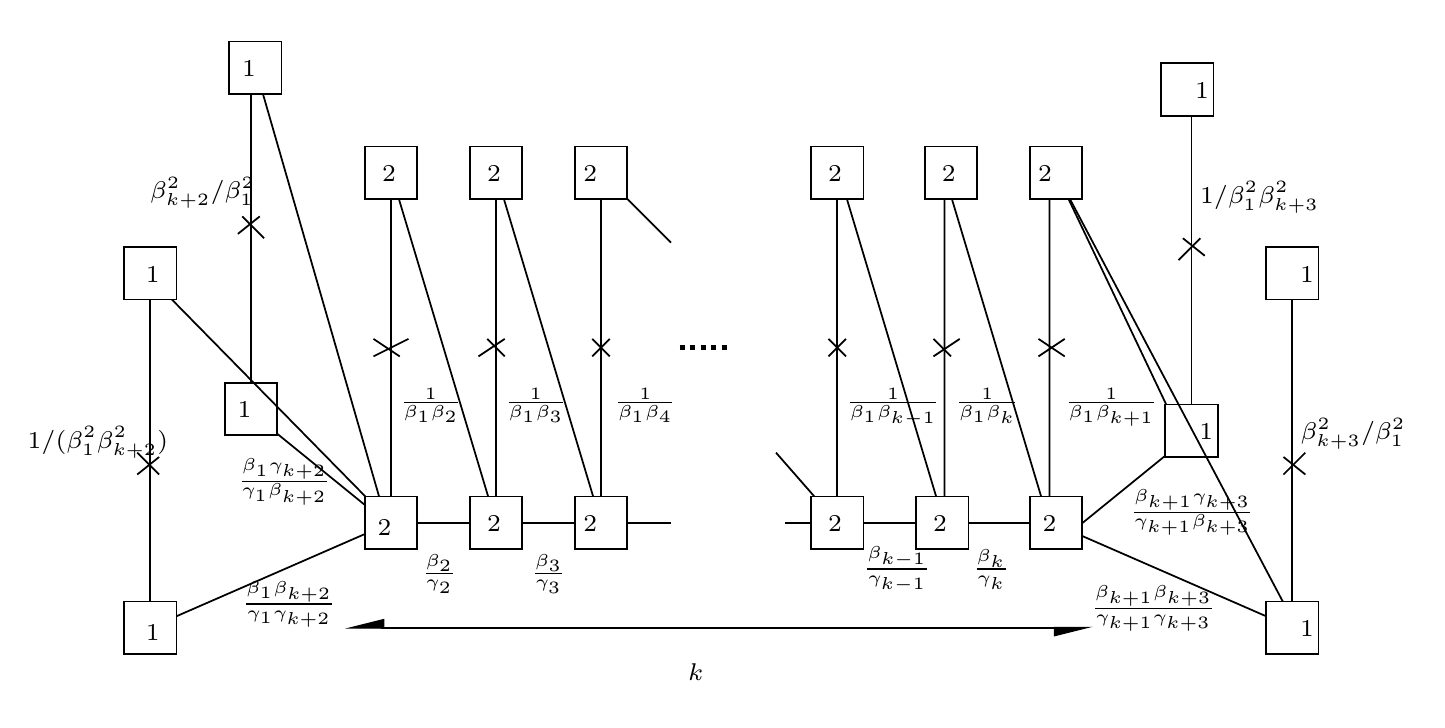}
\caption{The tube with all plus boundary conditions. We only show the charges for horizantal lines, which are the $M_i$, and for the flipped fields. The other charges are determined by the triangular superpotentials. Here flavor groups denoted by $2$ are $SU(2N)$ and by $1$ are $SU(N)$. Note that for $N=1$ the flip and flipped fields on the edges form mass terms and decouple.  The two punctures are of opposite sign and we glue them with $S$ gluing. We can see that the linear anomaly of this theory comes only from flip fields.
}\label{tubets}
\end{center}
\end{figure}

The basic tube of Figure \ref{tubets} acts on color by permuting clockwise $\beta_1 . . . \beta_{k+1}$ and by taking $\beta_{k+2}$ to $1/\beta_{k+2}$ and $\beta_{k+3}$ to $1/\beta_{k+3}$.  We can compute the charges of the flip fields to be,

\be
\beta_1:\quad  2N (k+4)\,,\;\;\, \beta_{k+3},\, \beta_{k+2}\;:\, 0\,, \;\;\, \beta_{l\neq 1,k+3,k+2} \,: 2N\,.
\ee All other symmetries have zero charge. In particular let us now glue $k+1$ such tubes together to torus. If $k$ is odd then all symmetries are preserved, and if $k$ is even $\beta_{k+3}$ and $\beta_{k+2}$ are broken because of the Weyl ${\mathbb Z}_2$ action of the tube. The charges then are, 

\be
 \beta_{l\neq k+3,k+2} \,: 2N(2k+4)\,,\;\;\qquad\, \beta_{k+3},\, \beta_{k+2}\;:\, 0\,, \;\;\, 
\ee We note that $h^\vee$ for $D_{k+3}$ is $2k+4$ and thus following our usual logic we identify the flux as being proportional to $2N$. Checking other anomalies we find that the flux  associated to the torus is one in $\frac1{k+1}\sum_{j=1}^{k+1}U(1)_{\beta_j}$. In particular this means that to compute the flux of a theory we compute the charge of the flip fields and divide by $2N (2k+4)$. 

We can choose different boundary conditions for the various fields. The tube will implement then action of various elements of the Weyl symmetry group. This will involve rotation of $\beta_i$ and flips. We will discuss this in detail in some cases.

\subsection*{Affine quiver}

We can take the tube above and glue the two punctures together. The theory one obtains is the ${\cal N}=2$ D shaped affine quiver with the adjoint (or more correctly bifundamentals of same group) fields flipped. All $\beta$ symmetries are broken save the diagonal combination of $\beta_i\neq k+3, k+2$. This $U(1)$ corresponds to the symmetry under which the adjoints in ${\cal N}=2$ are charged. All $\gamma$ symmetries survive and we expect the symmetry to enhance to $D_{k+3}$. Note that for $k=1$ and $N=2$ this is the symmetry of the Lagrangian. For other $k$ and $N$ the Lagrangian exhibits only the $U(1)$ symmetries. However, because of ${\cal N}=2$ dualities the index will be organized in representations of $D_{k+3}$. Moreover, the models have ${\cal N}=1$ conformal manifold on which the symmetry can enhance to $D_{k+3}$. Note that the dimension of the conformal manifold is $2k+6$
with all the symmetry preserved. This means that the index at order $q p$, which does not depend on any flavor, is $k+2$.
If the symmetry enhances to $D_{k+3}$ that means we will have the contribution of the currents for $D_{k+3}\times U(1)$. It is then conceivable that we have $k+3$ marginal operators which are singlets of the symmetry and another marginals in the adjoint of the non abelian group.

\subsubsection*{$D_4$}

\begin{figure}
\begin{center}
\includegraphics[scale=0.82]{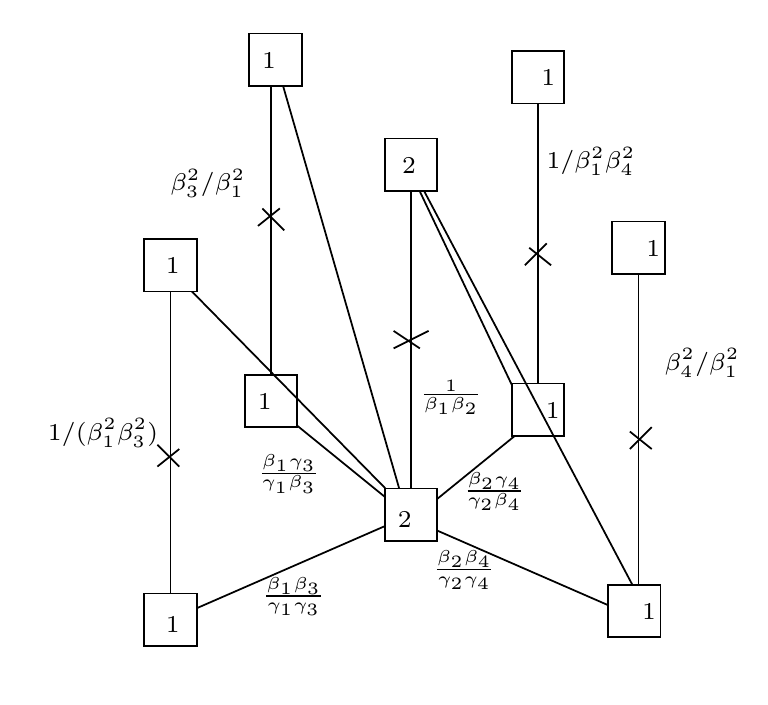}
\includegraphics[scale=0.82]{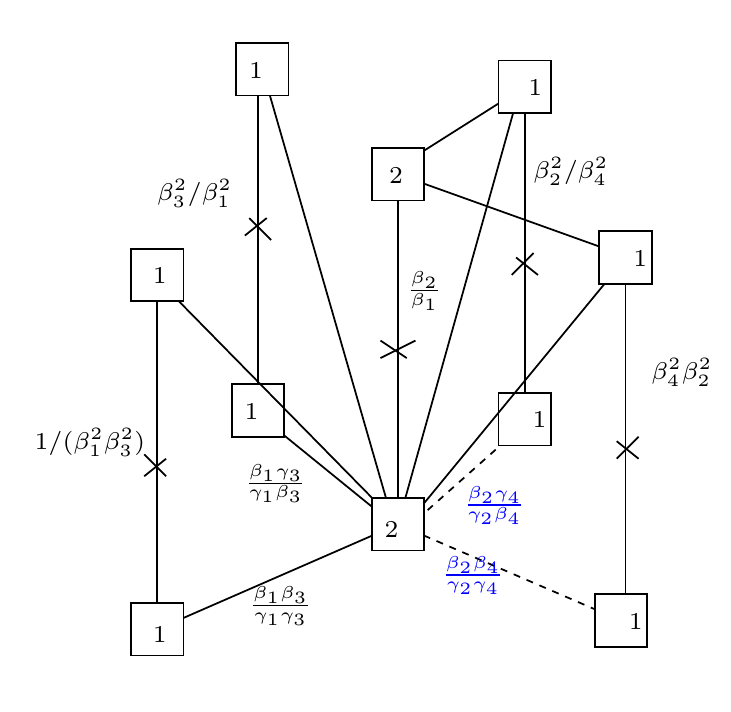}
\includegraphics[scale=0.82]{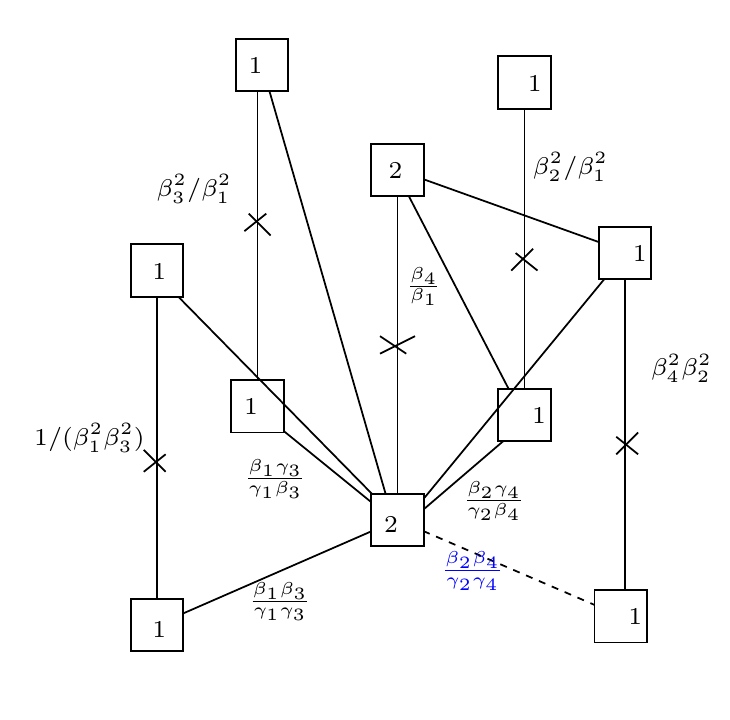}
\caption{Three different tubes for the $D_4$ case.  The two punctures are of opposite sign and different colors. The dotted line represents the $M_i$ which in the cases the line appears are composites.
The groups labeled by $1$ are $SU(N)$ and groups labeled by $2$ are $SU(2N)$.
}
\label
{ftyuw}
\end{center}
\end{figure}

Let us discuss the case of $D_4$.  In Figure \ref{ftyuw} we depict three different tubes for this case. The tube on the left corresponds to flux $(\frac56,\frac16,0,0)$ in $\beta$, the tube on the right to flux $(\frac12,-\frac12,0,0)$, and the tube on the bottom to flux $(\frac23,-\frac13,0,-\frac13)$. The tubes correspond to the following Weyl symmetry,

\be
&\frac56,\frac16,0,0& \;: \qquad\, \; \beta_1 \leftrightarrow \beta_2,\; \, \beta_3\to 1/\beta_3\,, \;\, \,
, \beta_4\to 1/\beta_4\,,\\
&\frac12,-\frac12,0,0& \;: \qquad\, \; \beta_1 \leftrightarrow 1/\beta_2,\; \, \beta_3\to 1/\beta_3\,, \;\, \,
, \beta_4\to 1/\beta_4\,,\nonumber\\
&\frac23,-\frac13,0,-\frac13 &\;: \qquad\, \; \beta_1 \leftrightarrow 1/\beta_4,\; \, \beta_3\to 1/\beta_3\,, \;\, \,
, \beta_2\to 1/\beta_2\,.\nonumber
\ee

For general values of $N$ we can easily compute the indices of some models corresponding to closed surfaces. For example let us glue two copies of the same tube together. All tubes of Figure \ref{ftyuw} will give equivalent theories. We can discuss the left tube which will give flux $(1,1,0,0)$ for $\beta$ and zero for $\gamma$. For general $N$ the gauge invariant operators are the baryons, flip fields, and operators corresponding to closed loops on the quiver. The baryons have large dimensions and flip fields are free fields for general $N$. The operators of smallest charge are then built from flipped fields winding the quiver and from operators corresponding to faces. 
 Additional operators contributing to the index are given by gaugino bilinears for each gauge group and by $\overline \psi_Q Q$ operators for each field. We have $26$ of the latter operators and have ten gaugino bilinears. We also have sixteen faces. The index is then, ignoring flip fields which are free,

\be
1+\frac5{\beta_1^2\beta_2^2} q^{\frac23}p^{\frac23}+\cdots\,.
\ee The superconformal R charge is the free one. The index at order $qp$ is vanishing. The order $qp$ in index computations using the superconformal R-symmetry counts the marginal operators minus the conserved currents for global symmetries \cite{Beem:2012yn}.
 The symmetry $\beta_1\beta_2$ is the symmetry which has the flux. The $D_4\times D_4$ symmetry is broken to $SU(2)^3\times U(1) \times SO(8)$. At this order of the index we see  the $U(1)$ symmetry.  At zero coupling we can count the dimension of the manifold of conformal couplings. The number of symmetries is $26$. The number of marginal operators is $26$. On a general point of the conformal manifold only eight symmetries are not broken. This indicates that the dimension of the conformal manifold is eight.  We expect then to have marginal operators in the adjoint of $SO(8)$ and $SU(2)^3\times U(1)$. These operators would give eight exactly marginal directions. Thus we conclude that it can be that on some point of the manifold the symmetry enhances. 
 
 We can also try to understand what are the states charged under the $\gamma$ symmetry. The generic states charged under these symmetries are baryonic operators built from bifundamental operators of two $SU(N)$ groups which are composites of two bifundamentals of $SU(2N) SU(N)$.
The contribution to the index of these is,

\be
q^{\frac{2N}3}p^{\frac{2N}3}\beta_1^N\beta_2^N\left[(\beta_3^N\beta_4^N+\frac1{\beta_3^N}\frac1{\beta_4^N}){\bf 8}_s+(\frac{\beta_3^N}{\beta_4^N}+\frac{\beta_4^N}{\beta_3^N}) {\bf 8}_c \,\, \right]\, \, .\;\;\;
\ee Here,

\be
&&{\bf 8}_s = (\gamma_1^N\gamma_3^N)^{\pm1}(\gamma_2^N\gamma_4^N)^{\pm1}+(\gamma_1^N/\gamma_3^N)^{\pm1}(\gamma_2^N/\gamma_4^N)^{\pm1} 
\, ,\\
&&{\bf 8}_c =(\gamma_1^N /\gamma_3^N)^{\pm1}(\gamma_2^N\gamma_4^N)^{\pm1}+(\gamma_1^N\gamma_3^N)^{\pm1}(\gamma_2^N/\gamma_4^N)^{\pm1} \,  \;\, \; .
\ee
We see that the operators form representations of $SO(8)\times SU(2)^3 \times U(1)$. 
Note that $\beta_3^N\beta_4^N+\frac1{\beta_3^N\beta_4^N}$ and $\beta_3^N/\beta_4^N+\beta_4^N/\beta_3^N$ are characters of the two spinor representations of  $SU(2)\times SU(2)\sim SO(4)$.
Note that for  $N=1$ there are additional operators at low charges and this is the special case of the E-string which is discussed in detail in \cite{Kim:2017toz}.

We can also combine different tubes together. Note that because of the non trivial Weyl action on the color, the order of gluing tubes actually can matter. For example combining the two tubes on the left and then two tubes on the right the theory has flux $(2,0,0,0)$. However combining the left tube to the right one and then taking two copies of this gives different flux, $(0,0,0,0)$. This theory is singular. In the first case the symmetry is actually enhancing to $U(1)\times SO(6)\times SO(8)$, which further enhances to $U(1)\times SO(14)$ in the case of $N=1$.  
 
\subsubsection*{$D_5$ minimal}

\begin{figure}
\begin{center}
\includegraphics[scale=0.51]{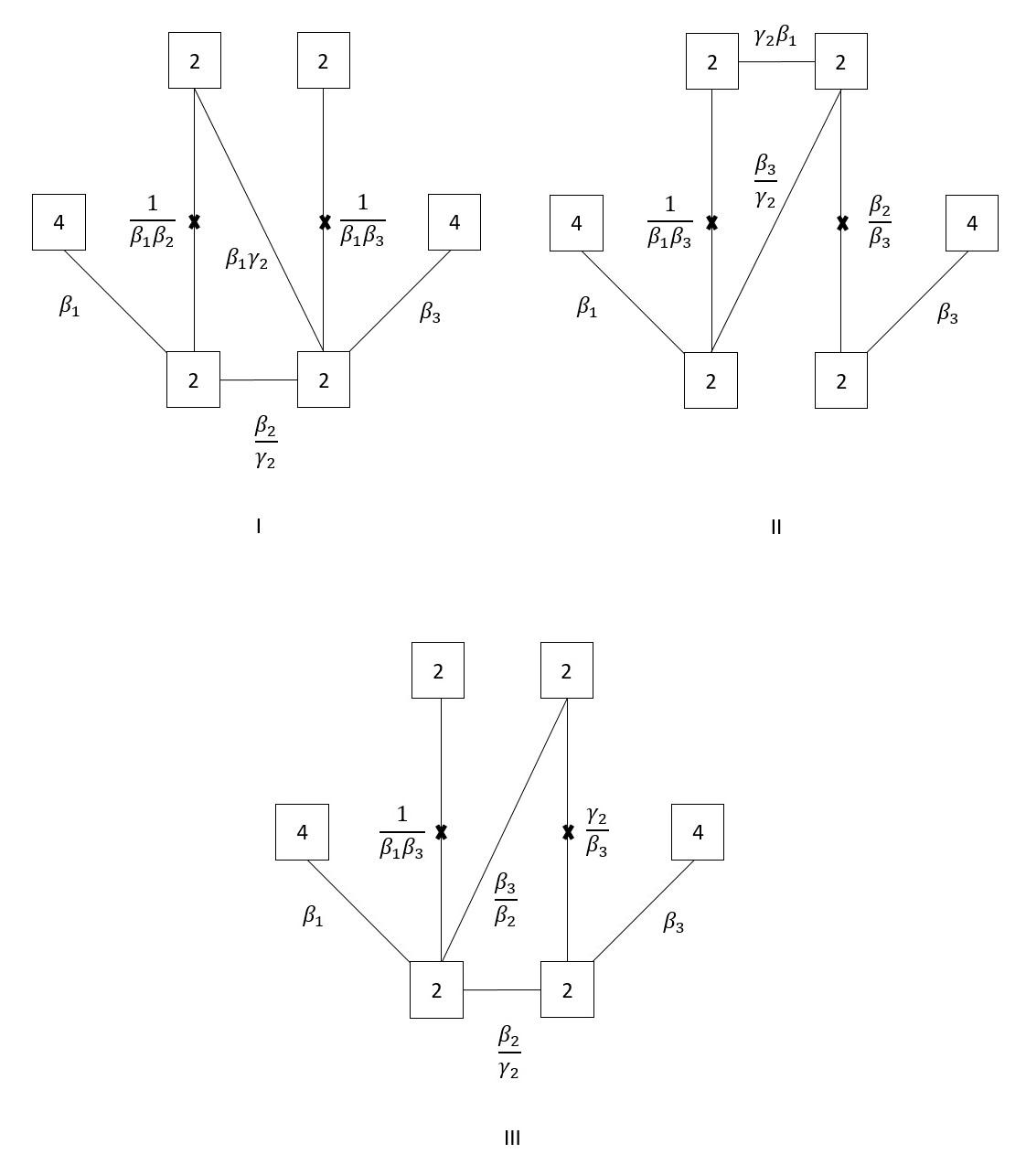}
\caption{A collection of tubes in the minimal $D_5$ case with no flux in the $SU(4)\times SU(4)$ groups rotating the flavors at the ends of the quiver.}
\label{D5Tubes}
\end{center}
\end{figure}

In this subsection we consider some examples for the case of minimal $D_5$. In Figure \ref{D5Tubes} we have drawn three tubes for this case, where we concentrate only on tubes with no flux in the $SU(4)\times SU(4)$ groups rotating the flavors at the ends of the quiver. Using our prescription, we associate with tube I the flux $(\frac{1}{2}, \frac{1}{4}, \frac{1}{4}, 0)$, with tube II the flux $(\frac{1}{4}, -\frac{1}{4}, \frac{1}{2}, 0)$ and with tube III the flux $(\frac{1}{4}, 0 , \frac{1}{2}, -\frac{1}{4})$. Here the fluxes are oriented as $(F_{\beta_1} , F_{\beta_2} , F_{\beta_3}, F_{\gamma_2})$, and for brevity we ignore the fluxes in $\beta_{4-5}, \gamma_1$ and $\gamma_{3-5}$ as these are zero for these tubes and for theories made of them. We note that to all of these tubes corresponds the same flux up to a Weyl transformation.

We next try to test these conjectures in various ways. As noted previously the anomalies for tubes generally do not match the $6d$ expectations, and these tubes are no different. However, it is possible that those for closed surfaces will work. To test these we next consider various closed surfaces that can be built from these tubes.

First we note that closing the tubes on to themselves leads to the same quiver for each tube. The quiver in question is an $\mathcal{N}=2$ $SU(2)\times SU(2)$ quiver gauge theory with a bifundamental hypermultiplet and two fundamental hypermultiplet for each of the $SU(2)$ gauge groups. Additionally there are chiral fields coming from the flipped bifundamental as well as the flipping fields. These theories correspond to the flux $(\frac{1}{3},\frac{1}{3},\frac{1}{3},0)$, up to a Weyl transformation. This comes about as when closing the tubes we are forced to identify the $3$ $U(1)$ groups with the flux, which forces it to distribute evenly between them leading to this structure.

The gluing breaks part of the global symmetry leaving us with a symmetry of rank $6$. This agrees with what the $6d$ expectation as for this value of flux to be consistent we must include center fluxes breaking the global symmetry to $U(1)\times SO(11)$. We can preform various consistency checks, particularly we can match anomalies which agree with the $6d$ expectations. We can also argue that the index should form characters of $U(1)\times SO(11)$ in the same manner as for the previous affine quivers.

We can also consider connecting each tube to itself to build theories associated with larger values of fluxes. However, in order to connect the tubes we need to cycle symmetries with flux in them, meaning that the fluxes of the resulting tube are not just twice that of the individual tubes. Particularly, when connecting three tubes we get to flux $(1,1,1,0)$, and those related by Weyl transformations for the other tubes. These can be closed to a torus without breaking symmetries with flux, and we can preform similar consistency checks on these theories as well, such as matching anomalies. More intricate checks are given by connecting two different tubes, and we next consider each in turn.   

First we consider gluing tubes $I$ and $II$. Due to the cycling of the global symmetry necessary when connecting the two tubes, we need to shift the fluxes for tube $II$. Summing the two fluxes, we associate with the resulting tube the flux $(\frac{1}{2}, \frac{1}{4}, \frac{1}{4}, 0) + (\frac{1}{2}, -\frac{1}{4}, \frac{1}{4}, 0) = (1, 0, \frac{1}{2}, 0)$.

\begin{figure}
\begin{center}
\includegraphics[scale=0.43]{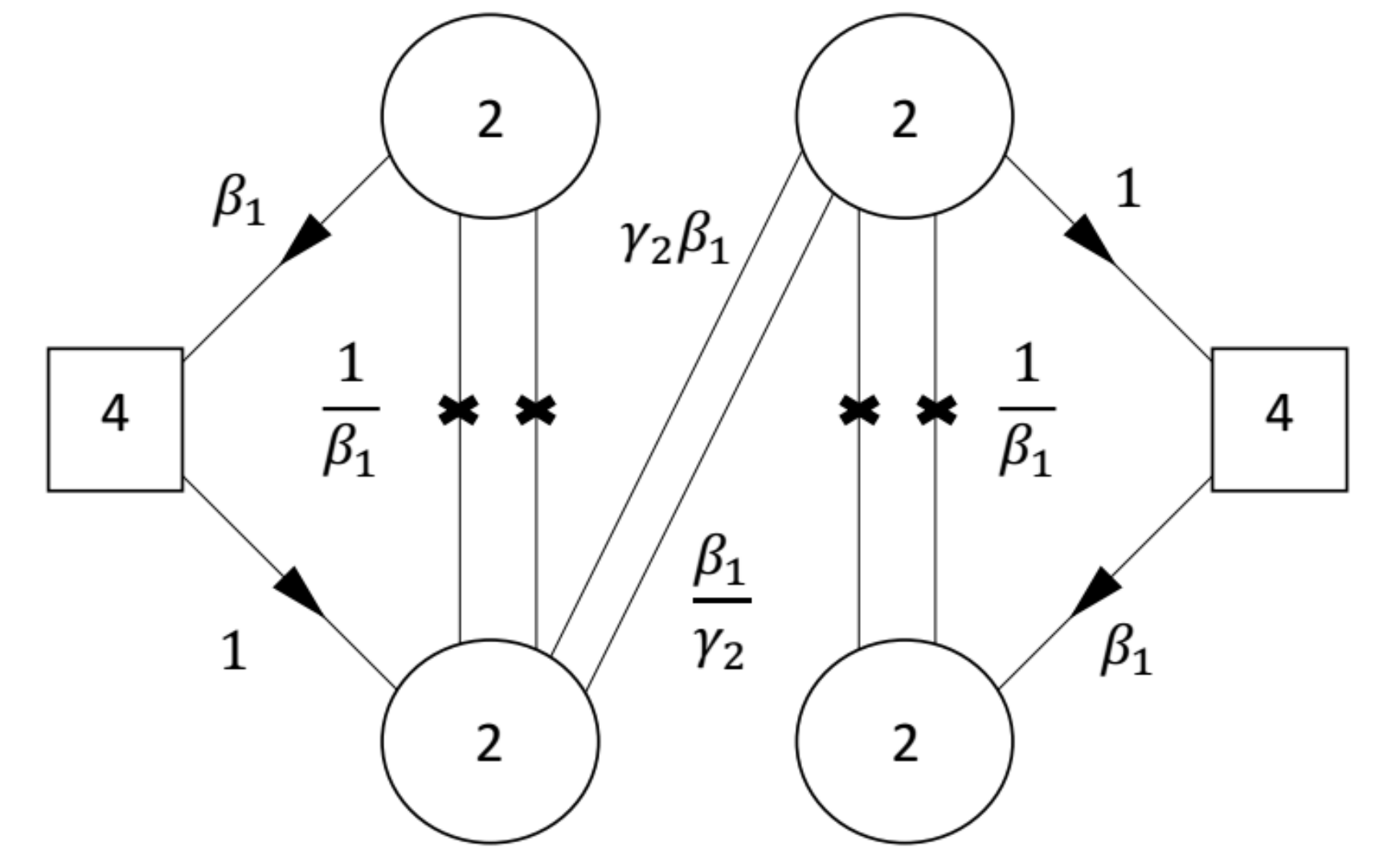}
\caption{The theory resulting from gluing the tubes $I$ and $II$ together.}
\label{QDIandII}
\end{center}
\end{figure}

We can next close the tube to a torus. When doing this we need to turn off the symmetries associated with $\beta_2$ and $\beta_3$. The resulting theory is shown in Figure \ref{QDIandII}. As we were required to turn $\beta_3$ off, we associate with this theory the flux $(1,0,0,0)$. We can test this in various ways. First we can compare anomalies where we find that they indeed match.

As a more intricate test we can consider the superconformal index. The general form of the flux is expected to preserve $U(1)\times SO(18)$ global symmetry, yet this specific value requires, for consistency, also a central flux element that breaks some of the global symmetry leading to the breakdown of $\beta_2$ and $\beta_3$. The resulting symmetry, while dependent on the choice of central element, is known to be at most $U(1)\times SO(15)$. We can try to test this by evaluating the superconformal index and see if the appearing operators can be merged so as to form characters of this symmetry.  

First we should consider the superconformal R-symmetry. Using a-maximazation we find it to be: $U(1)^{sc}_R = U(1)^{6d}_R - \sqrt{\frac{11}{45}} U(1)_{\beta_1}$. With this R-symmetry, we find no operators violating the unitarity bound and so no contradiction with this theory flowing to an interacting SCFT. We can then evaluate the index, where, for the purpose of the evaluation, we shall use the R-symmetry $U(1)^{6d}_R - \frac{1}{2} U(1)_{\beta_1}$, which is quite close to the superconformal one, as $\frac{1}{2} - \sqrt{\frac{11}{45}}\approx 0.0056$. We find:

\bea
I & = & 1 + (p q)^{\frac{1}{2}} \left( \frac{2}{\beta^2_1} + \beta^2_1 (4 + \chi[\bold{15}]_{SO(15)}) \right) + (p q)^{\frac{1}{2}} (p+q) \beta^2_1 (3 + \chi[\bold{15}]_{SO(15)}) \\ \nonumber & + & p q \left(\frac{3}{\beta^4_1} + 5 + \chi[\bold{15}]_{SO(15)} + \beta^4_1 (\chi[\bold{119}]_{SO(15)}+4\chi[\bold{15}]_{SO(15)}+9) \right) + ...
\eea 

Here we have already written the index in characters of the expected $U(1)\times SO(15)$ global symmetry, where: $\chi[\bold{15}]_{SO(15)} = 1 + \gamma^2_2 + \frac{1}{\gamma^2_2} + \chi[\bold{6},\bold{1}] + \chi[\bold{1},\bold{6}]$. This shows that the index can indeed be written in characters of $U(1)\times SO(15)$, at least to the evaluated order. 

We next consider gluing tubes $II$ and $III$. Due to the cycling of the global symmetry necessary when connecting the two tubes, we now need to shift the fluxes for tube $III$. Again summing the two fluxes, we associate with the resulting tube the flux $(\frac{1}{4}, -\frac{1}{4}, \frac{1}{2}, 0) + (\frac{1}{2}, 0 , \frac{1}{4}, - \frac{1}{4}) = (\frac{3}{4}, - \frac{1}{4}, \frac{3}{4}, - \frac{1}{4})$. 

\begin{figure}
\begin{center}
\includegraphics[scale=0.43]{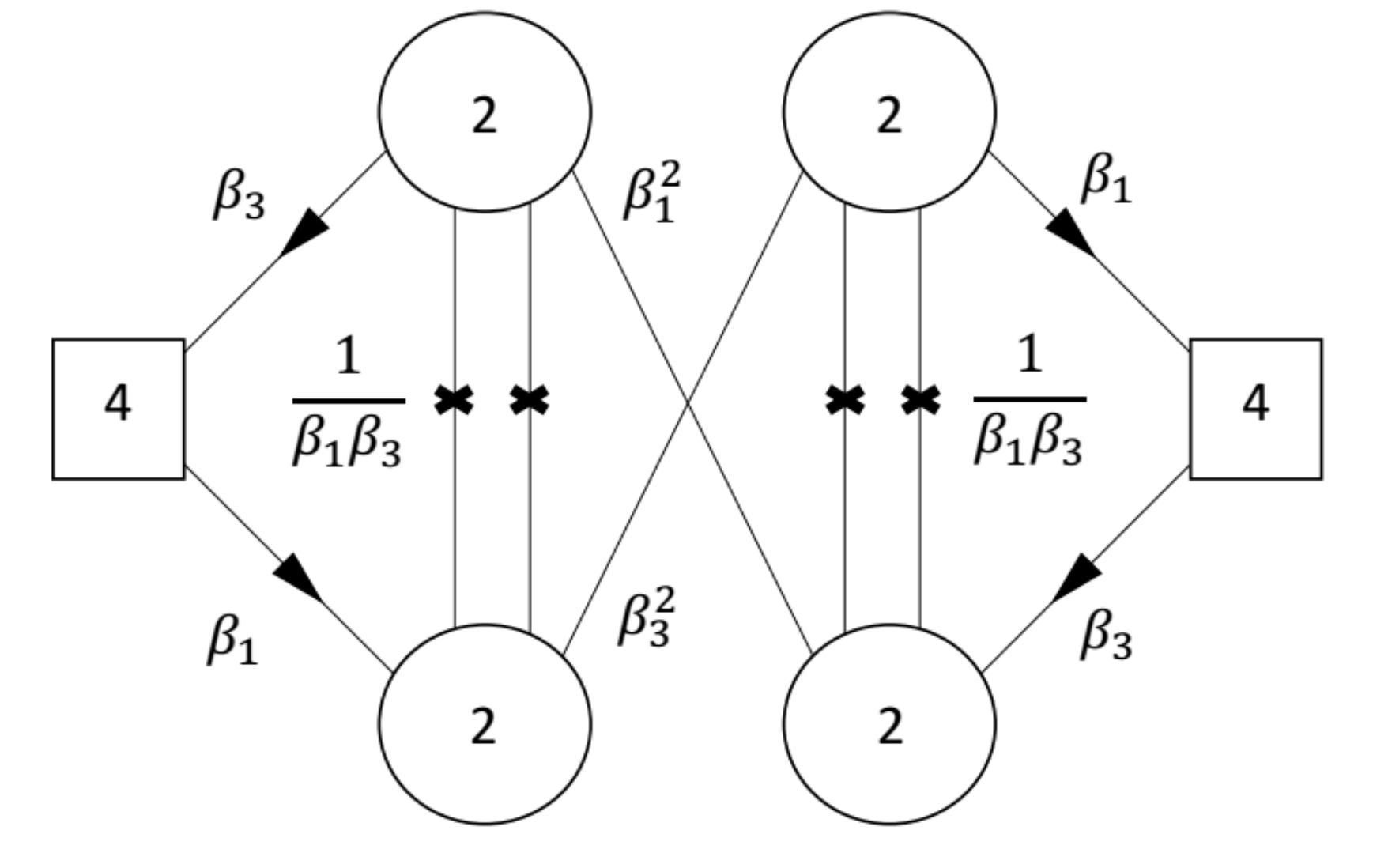}
\caption{The theory resulting from gluing the tubes $II$ and $III$ together.}
\label{QDIIandIII}
\end{center}
\end{figure}

We can next close the tube to a torus. When doing this we are forced to identify $\beta_3 = \frac{1}{\gamma_2}$ and $\beta_1 = \frac{1}{\beta_2}$. The resulting theory is shown in Figure \ref{QDIIandIII}. Due to the required identification, we associate with this theory the flux $(\frac{1}{2},-\frac{1}{2},\frac{1}{2},-\frac{1}{2})$. We next test this in various ways. The basic test is to compare anomalies against those expected from $6d$, where we indeed find that they match.

We can again consider evaluating the superconfomal index looking for character structure of the expected global symmetry. Here the structure of the flux is that of $U(1)\times SU(4) \times SO(12)$ preserving flux. However due to the fractional flux, part of the symmetry is broken so that at most $U(1)\times SU(2) \times SO(12)$ can be preserved. This is manifested in the construction by the identification that we were forced to perform upon closing the tube.  

We start by studying the superconformal R-symmetry using a-maximazation. We find it to be: $U(1)^{sc}_R = U(1)^{6d}_R - \sqrt{\frac{11}{126}} U(1)_{\beta_1} - \sqrt{\frac{11}{126}} U(1)_{\beta_3}$. With this R-symmetry, we find no operators violating the unitarity bound and so no contraction with this theory flowing to an interacting SCFT. we can then evaluate the index, where, for the purpose of the evaluation, we shall use the R-symmetry $U(1)^{6d}_R - \frac{1}{3} U(1)_{\beta_1} - \frac{1}{3} U(1)_{\beta_3}$, which is quite close to the superconformal one, as $\frac{1}{3} - \sqrt{\frac{11}{126}}\approx 0.04$. We find:

\bea
I & = & 1 + (p q)^{\frac{1}{3}} \beta^2_1 \beta^2_3( 3 + \chi[\bold{3}]_{SU(2)} ) \\ \nonumber & + & (p q)^{\frac{2}{3}} \left(\frac{2}{\beta^2_1 \beta^2_3} + \beta^4_1 \beta^4_3 ( \chi[\bold{5}]_{SU(2)} + 3\chi[\bold{3}]_{SU(2)} + 7 ) + \beta_1 \beta_3 \chi[\bold{2}]_{SU(2)} \chi[\bold{12}]_{SO(12)} \right) + ...
\eea 

Here we have already written the index in characters of the expected $U(1)\times SU(2) \times SO(12)$ global symmetry, where: $\chi[\bold{2}]_{SU(2)} = \frac{\beta_1}{\beta_3} + \frac{\beta_3}{\beta_1}$ and $\chi[\bold{12}]_{SO(12)} = \chi[\bold{6},\bold{1}] + \chi[\bold{1},\bold{6}]$. This shows that the index can indeed be written in characters of $U(1)\times SU(2) \times SO(12)$, at least to the evaluated order.

Finally we consider gluing tubes $I$ and $III$. Due to the cycling of the global symmetry necessary when connecting the two tubes, we again need to shift the fluxes for tube $III$. Summing the two fluxes, we associate with the resulting tube the flux $(\frac{1}{2}, \frac{1}{4}, \frac{1}{4}, 0) + (\frac{1}{2}, 0 , \frac{1}{4}, - \frac{1}{4}) = (1, \frac{1}{4}, \frac{1}{2}, - \frac{1}{4})$. 

\begin{figure}
\begin{center}
\includegraphics[scale=0.43]{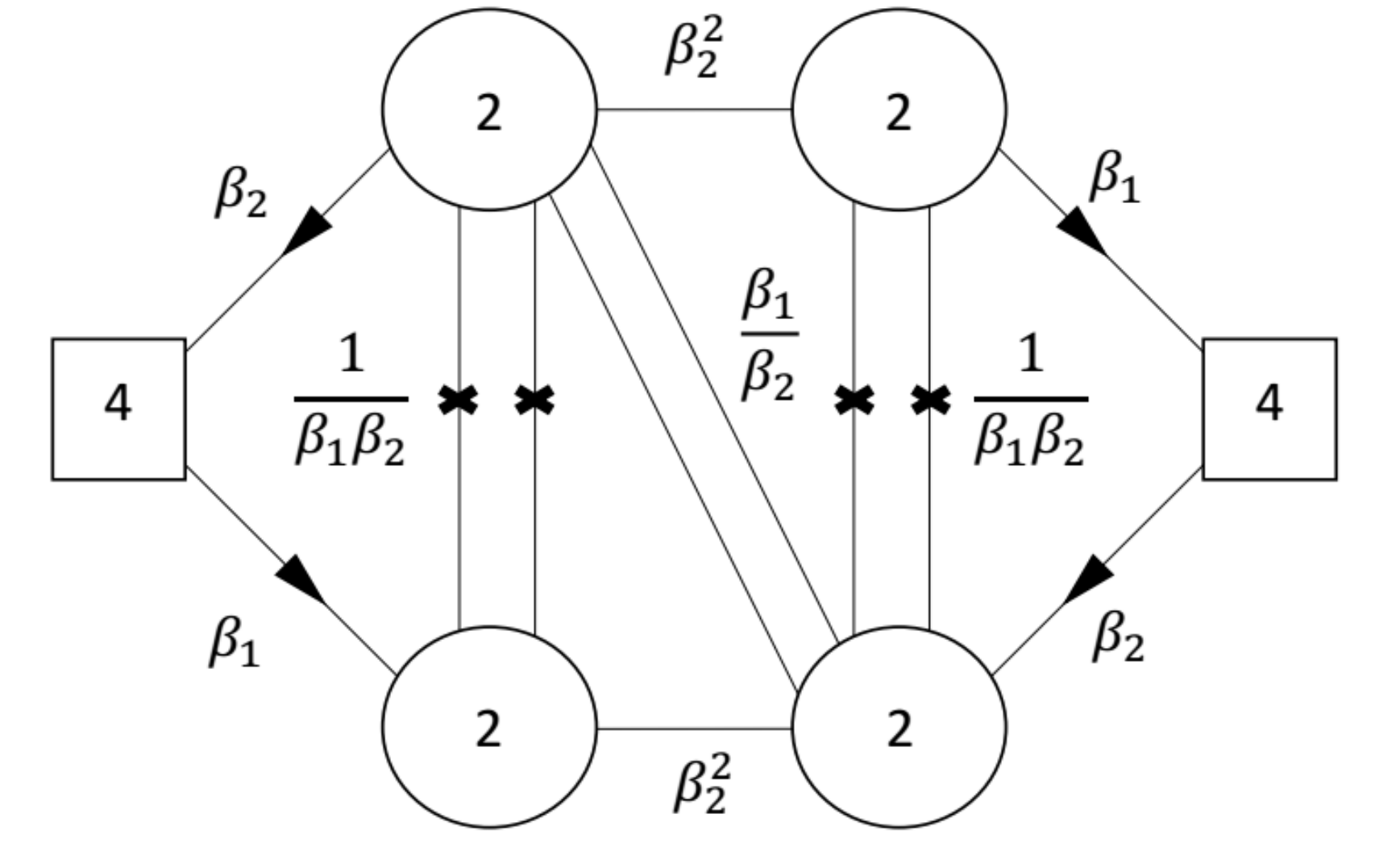}
\caption{The theory resulting from gluing the tubes $I$ and $III$ together.}
\label{QDIandIII}
\end{center}
\end{figure}

We can next close the tube to a torus. When doing this we are forced to identify $\beta_2 = \beta_3 = \frac{1}{\gamma_2}$. The resulting theory is shown in Figure \ref{QDIandIII}. Due to the required identification, we associate with this theory the flux $(1,\frac{1}{3},\frac{1}{3},\frac{1}{3})$. We can test this by comparing anomalies where we indeed find they match the $6d$ expectations.

\subsubsection*{Duality with $USp(2k)/SU(k+1)$ quivers}

The construction of minimal type $D$ conformal matter can also be approached from a different perspective, and comparing the two then leads to interesting physical phenomena. Particularly, the minimal type $D$ conformal matter has, besides the $SU(2)$ quiver description, two additional $5d$ gauge theory descriptions, as a $USp(2k)$ and an $SU(k+1)$ gauge theories with fundamental hypermultiplets. These can also be used to construct $4d$ theories in a similar manner to that which is done here, but by using a $5d$ domain wall extrapolating between the $USp(2k)$ and the $SU(k+1)$ descriptions. This construction was covered extensively in \cite{Kim:2018bpg}.  

A rather interesting aspect in this comparison is that we can construct the same compactification using different tubes. This should then give two dual descriptions of the same theory, that has at its heart the duality between the different $5d$ gauge theory descriptions of the minimal type $D$ conformal matter. The simplest case here is to use the tube with all plus boundary condition by gluing a multiple of $k+1$ of them to form a torus. When $k$ is odd then this compactification can be easily built from the tubes introduced in \cite{Kim:2018bpg}. This leads to a duality between a plane quiver theory of $SU(2)$ gauge groups and a circular quiver of alternating $USp(2k)$ and $SU(k+1)$ groups. This case was discussed in Appendix of \cite{Kim:2018bpg}.

From the constructions presented both here and in \cite{Kim:2018bpg} we can build a large number of different examples as in both cases we have ample tools to engineer torus compactifications with different values of flux. For instance we considered an example for $D_5$ involving the two tubes called $II$ and $III$. From these we can engineer a theory with flux $(1,-1,1,-1, 0,0,0,0,0,0)$, similarly to how we constructed the theory in Figure \ref{QDIIandIII}. 

We can also, using the tubes associated with the $USp/SU$ construction, build a torus compactification with the same flux. In fact, we can construct the torus compactification associated with flux $(\frac{1}{2},\frac{1}{2},\frac{1}{2},\frac{1}{2},0,0,0,0,0,0)$, which naively should be dual to the theory in Figure \ref{QDIIandIII}, as the fluxes are the same up to a Weyl transformation. However, as previously stated, this flux requires also a central element in the global symmetry to be consistently quantized, and the resulting theories differ in these central elements. Particularly,  for correct quantization a ${\mathbb Z}_2$ central flux is required. For the theory in Figure \ref{QDIIandIII}, this central flux is embedded in the center of the $SU(4)$ global symmetry. However, for the analogous theory in the $USp/SU$ construction, this central flux is embedded in the center of the $SO(12)$ global symmetry. 

\begin{figure}
\begin{center}
\includegraphics[scale=0.65]{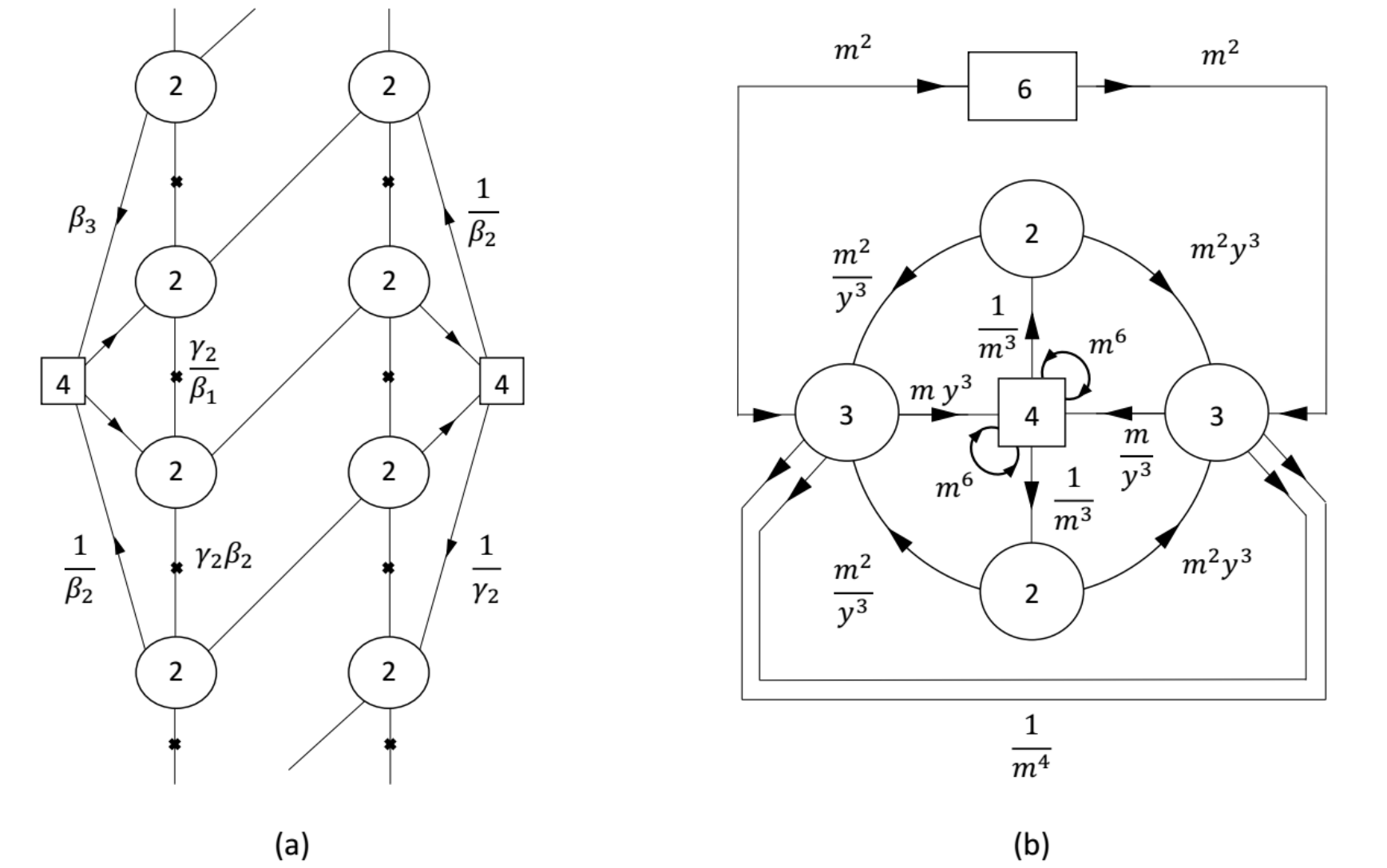}
\caption{Two theories expected to be dual since they both describe the same compactification of the minimal $(D_5,D_5)$ conformal matter. (a) The theory constructed from the tubes $II$ and $III$ that we introduced previously. We have written some of the charges of the fields via fugacities, the rest can be inferred from anomalies and the superpotentials. (b) The theory constructed from $USp(2k)/SU(k+1)$ domain wall discussed in \cite{Kim:2018bpg}. Here we used the simpler version given after Intriligator-Pouliot duality. We refer the reader to \cite{Kim:2018bpg} for the details. The gauge groups are $SU(3)$ and $SU(2)$ as it is the same as $USp(2)$. The circular double arrows connected to the central $SU(4)$ global symmetry group stand for chiral fields in the antisymmetric representation of said $SU(4)$ flavor symmetry group. These flip the gauge invariant states made from the $SU(2) \times SU(4)$ bifundamentals.}
\label{DualThr}
\end{center}
\end{figure}

As a result to get a duality we need to form a torus compactification with integer flux, the simplest case being $(1,-1,1,-1, 0,0,0,0,0,0)$. In Figure \ref{DualThr} we have drawn the two dual theories associated with this flux. Both are expected to have a $U(1)\times SU(4)\times SO(12)$ global symmetry, which is the symmetry preserved by the flux. For the theory in figure \ref{DualThr} (b), $U(1)_m \times SU(4)$ should map to the first part, while $U(1)_y \times SU(6)$ should enhance to $SO(12)$ as $\bold{12}_{SO(12)} = y^3 \bold{6}_{SU(6)} + \frac{1}{y^3} \bar{\bold{6}}_{SU(6)}$. For the theory in figure \ref{DualThr} (a), the combination $U(1)_{\beta_1} + U(1)_{\beta_3} - U(1)_{\beta_2} - U(1)_{\gamma_2}$ should map to the $U(1)$, the other $3$ combinations should build the $SU(4)$ as $\bold{4}_{SU(4)} = \sqrt{\frac{\beta_1 \gamma_2}{\beta_2 \beta_3}} (\beta_1 \beta_2 + \frac{1}{\beta_1 \beta_2 }) + \sqrt{\frac{\beta_2 \beta_3}{\beta_1 \gamma_2}} (\beta_3 \gamma_2 + \frac{1}{\beta_3 \gamma_2})$, and the two $SU(4)$ groups should build $SO(12)$ as $\bold{12}_{SO(12)} = \bold{6}_{SU(4)_1} + \bold{6}_{SU(4)_2}$. The exact relation between the $U(1)$ groups on both sides is expected to be: $U(1)_m = \frac{3}{2}(U(1)_{\beta_1} + U(1)_{\beta_3} - U(1)_{\beta_2} - U(1)_{\gamma_2})$, or in fugacities: $m^6 = \frac{\beta_1 \beta_3}{\beta_2 \gamma_2}$.  

Note that the theory one constructs using the $USp/SU$ domain wall has $SU(3)$
nodes with ten flavors. This is IR free theory and the way to understand the model is by first not gauging the $SU(3)$ groups but only the $USp(2)$ groups, see \cite{Kim:2018bpg} for a discussion in closely related case. 
Let us first discuss the theory with only the $SU(2)$ groups gauged and without the fields charged under $SU(6)$ and $SU(3)$.  This theory has the symmetries $U(1)_m$ and $U(1)_y$ with only the former mixing with the R-symmetry. The model is asymptotically free and after a maximization we obtain that the superconformal R-symmetry is  $R_0-0.00345q_m$, where $R_0$ assigns R-charge $3/5$ to fields charged under the $SU(2)$ with the rest fixed by the superpotential,  with no unitarity bound violating operators. It is then plausible that the theory flows to an interacting conformal fixed point. We now add the six fundamental fields for both $SU(3)$ flavor groups. Then we obtain that $ 
Tr (R SU(3)^2) = 6(\frac23-1)\frac12+(-3+2\sqrt{\frac{19}{51}})$
with the second term coming from the fixed point. We note that this term $+3$, which is the contribution to 
$Tr (R SU(3)^2)$ from the $SU(3)$ gauge field, is positive meaning that the $SU(3)$ group is asymptotically 
free at the fixed point with the addition of the six fundamental fields. We remind the reader that the beta function is proportional to $-Tr(U(1)_R SU(N)^2)$ with $R$ being the superconformal symmetry of the fixed point. We 
then flow to a fixed point with all operators above the unitarity bound, and the superpotential involving fields charged under $SU(6)$ and $SU(3)$ is marginal.  This implies that the theory makes sense as a sequence of flows starting from weakly couple UV theory.

We can test the duality in various ways. First we can compare anomalies, where we find the anomalies indeed match between the two theories, with the expected identification, and also match the $6d$ prediction. We can also compute and compare the superconformal index. We indeed find that it matches between the two theories, at least to the order we evaluated it. We also observe that it forms characters of the expected $U(1)\times SU(4)\times SO(12)$ global symmetry. Specifically, we find for the index:

\bea
I & = & 1 + 2 \frac{\beta_1 \beta_3}{\beta_2 \gamma_2} \chi[\bold{6}]_{SU(4)} (p q)^{\frac{1}{3}} \\ \nonumber & + & (p q)^{\frac{2}{3}} \left( \sqrt{\frac{\beta_1 \beta_3}{\beta_2 \gamma_2}}\chi[\bold{4}]_{SU(4)} \chi[\bold{12}]_{SO(12)} + \frac{\beta^2_1 \beta^2_3}{\beta^2_2 \gamma^2_2}(3\chi[\bold{20'}]_{SU(4)} + \chi[\bold{15}]_{SU(4)} + 3) \right) + ...
\eea 

Here we have used the notation of Figure \ref{DualThr} (a), the transformation to the notation of the other theory can be done using the relations given above. We have also used the R-symmetry $U(1)^{6d}_R - \frac{1}{3}(U(1)_{\beta_1} + U(1)_{\beta_3} - U(1)_{\beta_2} - U(1)_{\gamma_2})$, which is close to the superconformal R-symmetry which is $U(1)^{6d}_R - \sqrt{\frac{11}{126}}(U(1)_{\beta_1} + U(1)_{\beta_3} - U(1)_{\beta_2} - U(1)_{\gamma_2})$. There are no operators violating the unitary bound with respect to the superconformal R-symmetry.

Finally we note that the first two terms in the index are exactly as expected from the compactification of $6d$ theories based on the reasoning of \cite{sbfh} (see also Appendix E in \cite{Kim:2017toz}). The third term is just the self-product of the first term. 

\subsection{Examples of $E$}

Let us now give some illustrative computations for the compactifications of $E$ conformal matter. As the gauge groups in the relevant quiver diagrams become of large dimensions even in the minimal case, there are very few computations one can perform explicitly. We will thus restrict to checking anomalies and verifying indices in limiting cases of the minimal conformal matter.

\subsection*{$E_6$}

\begin{figure}[htbp]
\center
\;\;\; \includegraphics[width=0.31\textwidth]{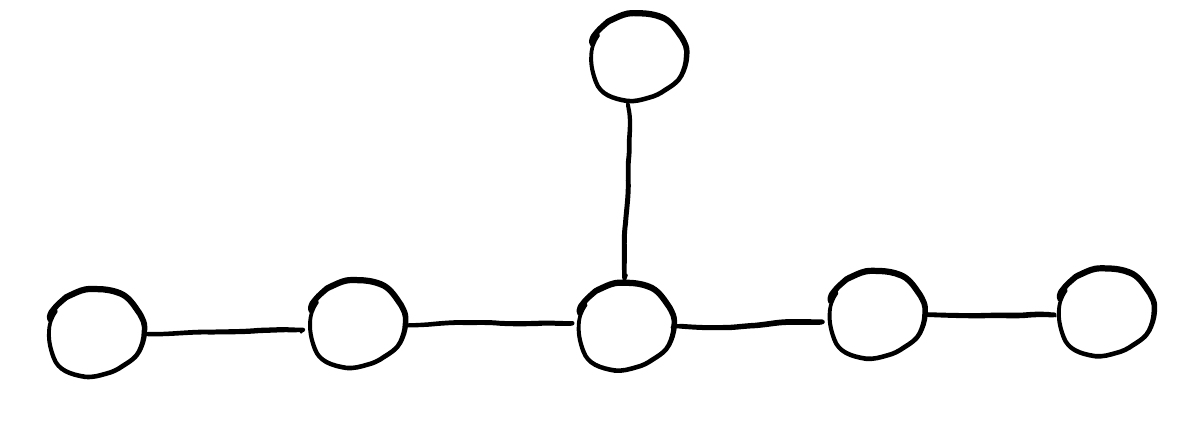}\,\;  \includegraphics[width=0.31\textwidth]{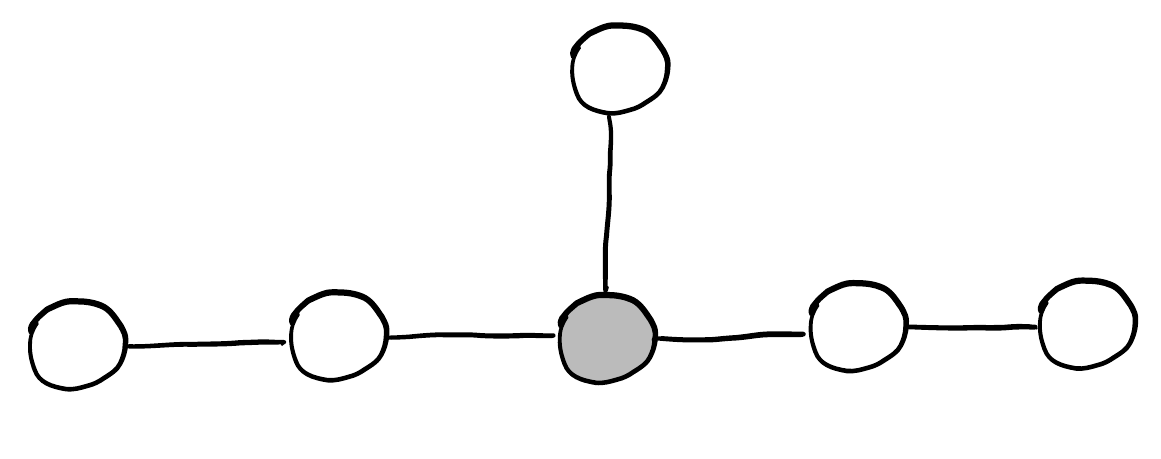}
\caption{The Dynkin diagram of $E_6\times E_6$  with the shaded node corresponding to the node with the flux.}
\label{eadui}
\end{figure}

A typical tube is depicted in Figure \ref{fig:E6-tubess}.  
As discussed in previous sections, the flux associated to the tubes is such that gluing the tube to itself six times one obtains integer flux preserving the full symmetry of the theory unbroken by the flux, which in this case is $E_6\times SU(3)\times SU(3)\times SU(2)\times U(1)$. That is the flux of the combined model is in the $U(1)$ corresponding to the central node of one of the $E_6$ groups. See Figure \ref{eadui}.

\begin{figure}[htbp]
\center
\includegraphics[width=0.36\textwidth]{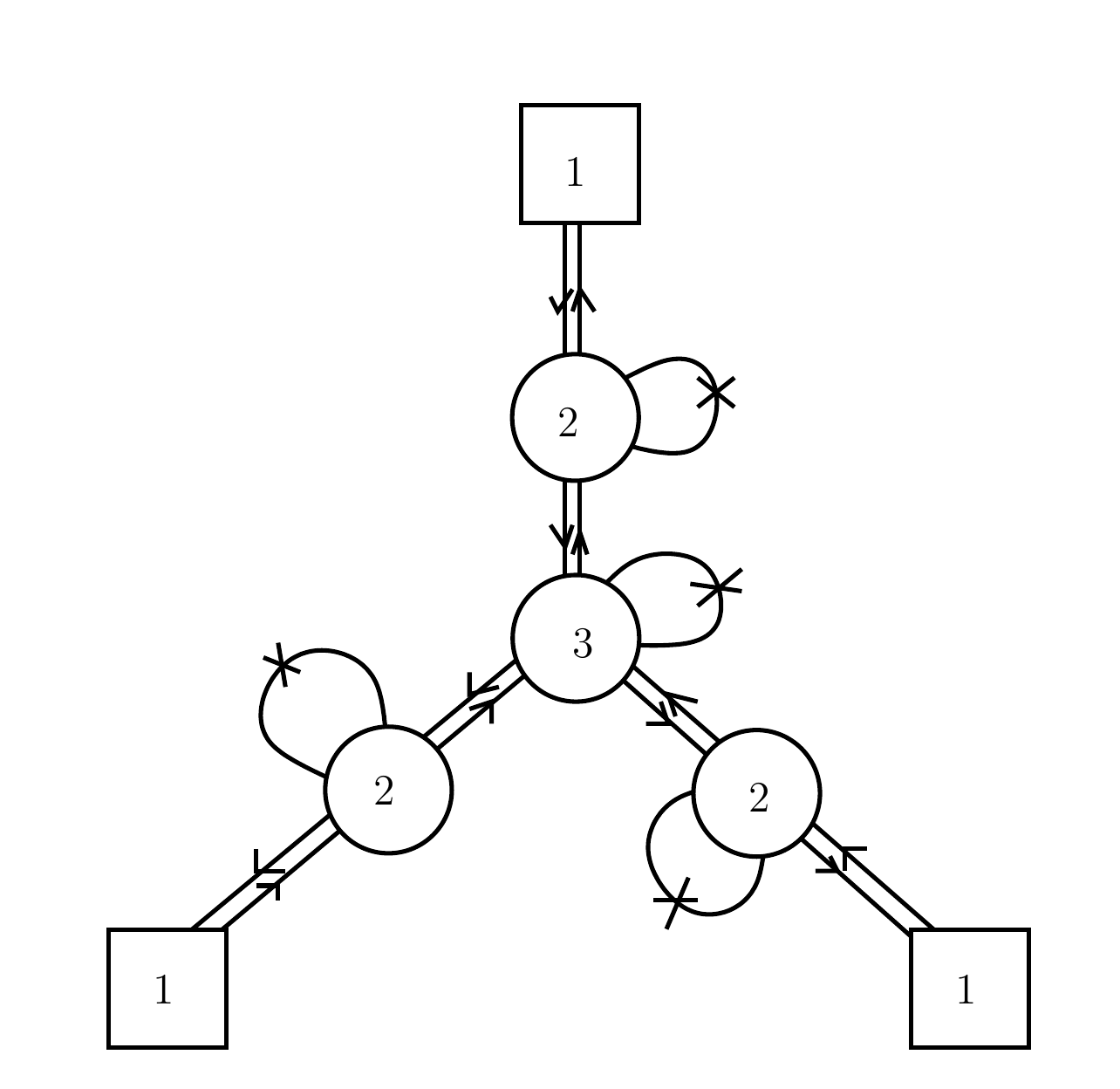} 
\caption{Torus with $1/6$ units of flux in the $U(1)$ corresponding to the central node. The lines from vertex to itself are adjoint plus a singlet.}
\label{mieb}
\end{figure}

Gluing the tube to itself we obtain the affine quiver of Figure \ref{mieb}. The Figure is for the minimal case. For non minimal the groups become $SU(lN)$ with the $l$ label appearing in the figure and all are gauge nodes. We will discuss only the minimal case in what follows.
\
Note that the flipping of the baryonic operator is irrelevant for the $SU(3)$ gauge group. We can perform naive a maximization ignoring this and find that this agrees with the six dimensional computation. 
 Computing the index we can see that protected states organize in $U(1)\times E_6$ representations.  The rest of the symmetry is broken by the fractional flux, where out of one of the two $E_6$ symmetries only the $U(1)$ with the flux is not broken.
Parametrizing the $U(1)$ in the end of the legs of the quiver as $b_i$ and the $U(1)$ under which the bifundamentals in the middle are charged by $ a_i$, we also have an additional $U(1)$ we denote by $t$ under which the bifundamental fields are charged half and the adjoint fields charged one. The simplest operators charged under $a$ and $b$ symmetries appear at $ (     q   p)^{\frac43}$ in the index computation and are in the following representations,

\be
({\bf 3}\,  , \, \overline{\bf 3}\, , \, {\bf 1})+({\bf 3}\,  , \, {\bf 1}\,, \, \overline{\bf 3})+(\overline {\bf 3}\,  , \, {\bf 3}\, , \, {\bf 1})+(\overline {\bf 3}\,  , \, {\bf 1}\,, \, {\bf 3})+({\bf 1}\,  ,  \, {\bf 3}\,  , \, \overline{\bf 3})+({\bf 1} \, ,  \, \overline{\bf 3}\,  , \, {\bf 3})\, ,
\ee where ${\bf 3}_l=a_l^2+\frac1{a_l}(b_l^{-1}+b_l)$.  The above representations  naturally form ${\bf 27}{\oplus} \overline {\bf 27}$ of $E_6$, they have $U(1)_t$ charge two. We can compute the index in a limit. Note that without the singlet fields this is an ${\cal N}=2$ model and it has an \cite{Gadde:2011uv} HL limit.  In terms of six dimensional R charge this corresponds to keeping $ q\, p\,\, t$ fixed while sending $q ,\, p \,, 1/t$ to zero.
Keeping the singlet fields in the bifundamentals does not spoil the limit but the flip fields give singular contributions. As the flip fields are free we can compute the index without these. We obtain,

\be&&
1+2t+(3+\overline {\bf 27} +  {  \bf   27}) t^2 +(-1+2\,{\bf 27}   +2\,   \overline {  \bf 27  }+{   \bf  78 } ) t^3  +\\&& (  -  7 +  (Sym^2(3+\overline {  \bf 27}+{\bf 27})) -3\,\;{  \bf 78}-4\;\overline {  \bf 27 }-4\, {  \bf 27} ) t^4+ .. .\,.\nonumber
\ee
Note that all the operators form $E_6\times U(1)$ representations. We also mention that this model does not actually possess an $E_6$ symmetric point on its conformal manifold.
This can be shown as the theory is conformal and all the exactly marginal deformations are ${\cal N}=2$. This is analogous to a similar statement in \cite{Kim:2017toz}. We do not have any contradictions for theories with higher amounts of flux having this symmetry.

\

\begin{figure}[htbp]
\center
\includegraphics[width=0.5\textwidth]{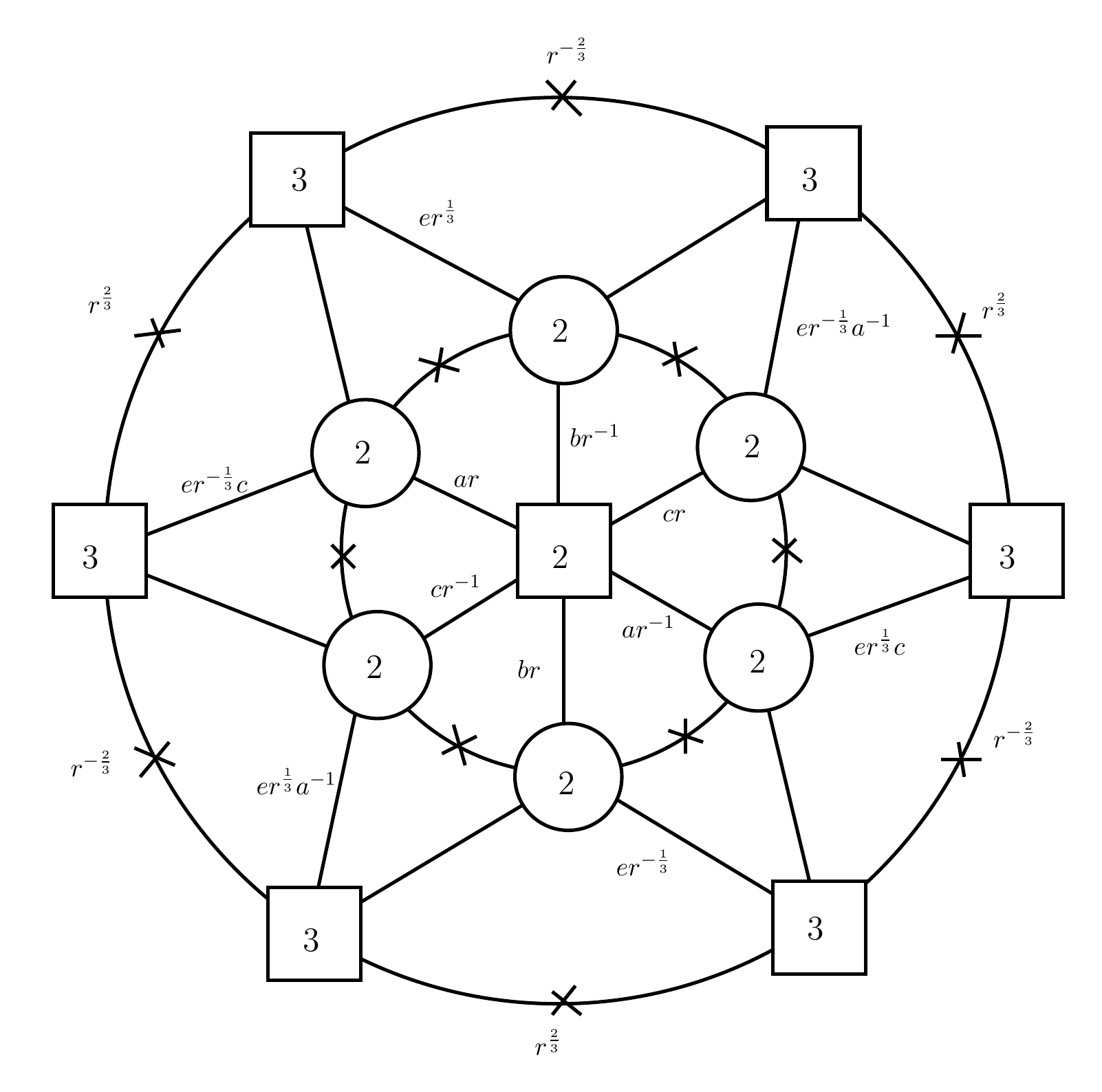} 
\caption{The leg of torus with one unit of flux. The links of the quiver without the flip fields are charges with charge half under $U(1)_t$, the fields with the flip are charged minus one. We denoted the other charges on the quiver with the charges of links with no labels derivable from superpotentials. The $SU(2)$ explicitly visible in the quiver is $SU(2)_f$.}
\label{suebalyut}
\end{figure}

We can combine the tubes to form integer value of flux. Gluing six two punctured spheres we obtain torus with flux one. We expect the full symmetry to be visible there. The theory can be composed of three copies of the one in Figure \ref{suebalyut} by gluing them along the perimeter. The anomaly conditions will identify
\be\label{condchyuit} 
\prod_{i=1}^3a_i=\prod_{j=1}^3 c_j=\prod_{l=1}^3 b_l=1\,,
\ee with $(a_n,b_n,c_n)$ being the symmetries of the three copies, and $a_n c_n b_n=1$.
The symmetry is then given by the three $SU(2)$s of the three copies, three $U(1)_{e_n}$, one $U(1)_r$, one $U(1)_t$, three copies of $U(1)_aU(1)_bU(1)_c$ subject to five constraints. The rank of the symmetry is twelve as expected.
The index in the limit above is given as follows,
\be
&&1+({\bf 2},{\bf 3},{\bf 3}; {\bf 1}) t +(-({\bf 1}+{\bf 3},{\bf \overline 3},{\bf \overline 3};{\bf1})-12+({\bf 1}+{\bf 3},{\bf 6},{\bf \overline 3};{\bf1})+({\bf 1}+{\bf 3},{\bf \overline 3},{\bf 6};{\bf1}))t^2+\\
&&+2({\bf 1},{\bf 3},{\bf 1};{\bf 27})t^2+2({\bf 1},{\bf 1},{\bf 3};{\bf \overline{27}}) t^2+\dots\,. \nonumber
\ee Here we have the representations of $(SU(2)_r, SU(3)_\alpha, SU(3)_\gamma; E_6)$. The characters are,

\be 
&&{\bf 2}_{SU(2)_r}=\frac1{r^2}+r^2    \, ,\,\,\qquad  {\bf 27}_{E_6} = ({\bf \overline 3}_1, {\bf 3}_2,{\bf 1}_3)+({\bf 1}_1,{\bf    \overline 3}_2  \,  ,{\bf 3}_3)+({\bf 3}_1 \, ,  \, {\bf 1} _2 ,  \, {\bf \overline 3}_3)  \, , \\
&& {\bf 3}_i = {e'_i}^{-2}+e'_i(\frac1{f_i}+f_i)  \,  ,    \qquad    e_i =
e_i'(a_i^2 b_i)^{\frac13}  \,  , \qquad  {\bf 3}_\alpha =\alpha_1+\alpha_2+\frac1{\alpha_2\alpha_1}   
\,  , {\bf 3}_\gamma =\gamma_2+\gamma_1+\frac1{\gamma_2\gamma_1}   \, ,\nonumber\\
&& a_1= (\gamma_2\gamma_1\alpha_1\alpha_2)^{-\frac12}  ,  \,  \qquad b_1= (\alpha_2\gamma_2)^{\frac12}   \,  ,  \qquad  b_2 = (\gamma_1/\alpha_1\alpha_2)^{\frac12}   \,  ,                                           \qquad  a_2=(\alpha_1\gamma_2)^{\frac12}      \,  \nonumber  .         
\ee One can actually understand some of the terms in the index from six dimensions. Note that under $U(1)_t$ which has the flux we have, \be
{\bf 78}={\bf 3}_\alpha {\bf 3}_{\gamma} (t^{-2} +t {\bf 2}_r)+{\bf \overline  3_\alpha}{\bf \overline 3_\gamma} (t^2 + t^{-1} {\bf 2}_r)+{\bf 2}_r(t^{-3}+t^3)+ {\bf 3}_r+{\bf 8}_\gamma+{\bf 8}_\alpha+1\,.
\ee We expect this term to contribute at order $q p$ with the six dimensional R-symmetry with the multiplicities determined by flux and charges under $t$ (see \cite{Kim:2017toz} appendix E for this statement which summarizes the results of \cite{sbfh}). In our limit we see that  the term ${\bf 3}_\alpha{\bf 3}_\gamma  \, {\bf 2}_r$ should survive and contribute to the index at order $t$, which we observe. The states with ${\bf \overline 3}_\gamma{\bf \overline 3}_\alpha t^2$ should contribute with multiplicity $2$. Note that these are divergent in the limit we take and they are captured exactly by the flip fields in the smaller circles of the three legs of the quiver. The states with $t^3 {\bf 2}_r$ contribute with multiplicity three and also are divergent and they are captured by flip fields coming from the large circle in the center of the quiver. The remaining states in the adjoint of $E_6$ vanish in the limit.  The states we see in representations of the $E_6$ which is invariant under the flux is also easy to understand. The six dimensional theory has operators in ${\bf 27}_a\otimes {\bf \overline 27}_b\; \oplus \;    {\bf \overline 27}_a   \,\otimes {\bf 27}_b$ at R-charge four.  The operators which survive the limit have charge two under $U(1)_t$ and thus appear with factor of two precisely as the representations appearing in the index. 

Let us compute for completeness the flux of a tube following from our assignment of symmetries. Note that we have two factors of $E_6$ which we will denote as $\beta$ and $\gamma$. The flux is only in one of them. Reading off the charges of the flip fields appearing in the quiver of Figure \ref{suebalyut} we obtain that the only non vanishing charge is in $t$ and is equal to $144$. We need to add also the contribution of the flip fields flipping the bifundamenta baryons between pairs of $SU(1)$ groups. Although in this case the baryons are just the fields and the flipping removes them, the prescription of counting the charges is to take all the flip fields.
We also note that the  symmetries satisfy \eqref{condchyuit}. This theory is built from six tubes so we have to divide by six to obtain flux of a single tube and then further divide by $24$ which sets the coefficient $n_t=\frac32$ in \eqref{sfdrtyet} .  To obtain the flux for a single tube we need to read off the charges of a wedge in the Figure.

\

\begin{figure}
\center
\includegraphics[width=0.58\textwidth]{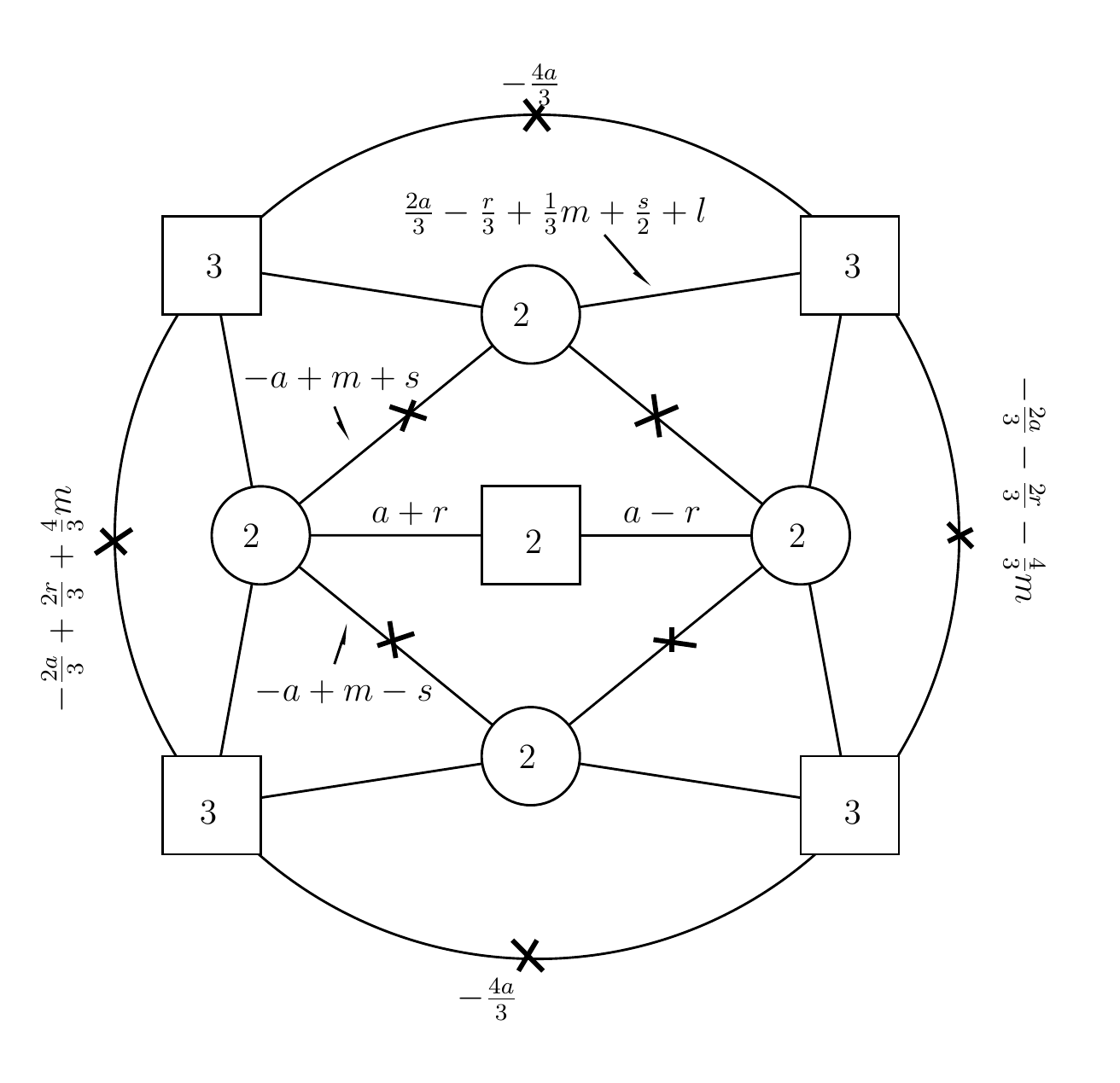} 
\caption{Leg of torus theory with one unit of flux breaking symmetry to $SU(6) U(1) E_6$. Three copies of this model are glued to by gauging the diagonal $SU(3)$ symmetries. The charges under different symmetries are deduced from the superpotentials associated to the faces. The flux is for the $U(1)_a$ symmetry. The $SU(2)$ flavor symmetry with $U(1)_l$ symmetry enhances to $SU(3)$ for each leg and the three $SU(3)$s from the three legs enhance to $E_6$.}
\label{coebeuyr}
\end{figure}

\begin{figure}[htbp]
\center
\includegraphics[width=0.35\textwidth]{figure/eadyu.pdf}\,\;  \includegraphics[width=0.35\textwidth]{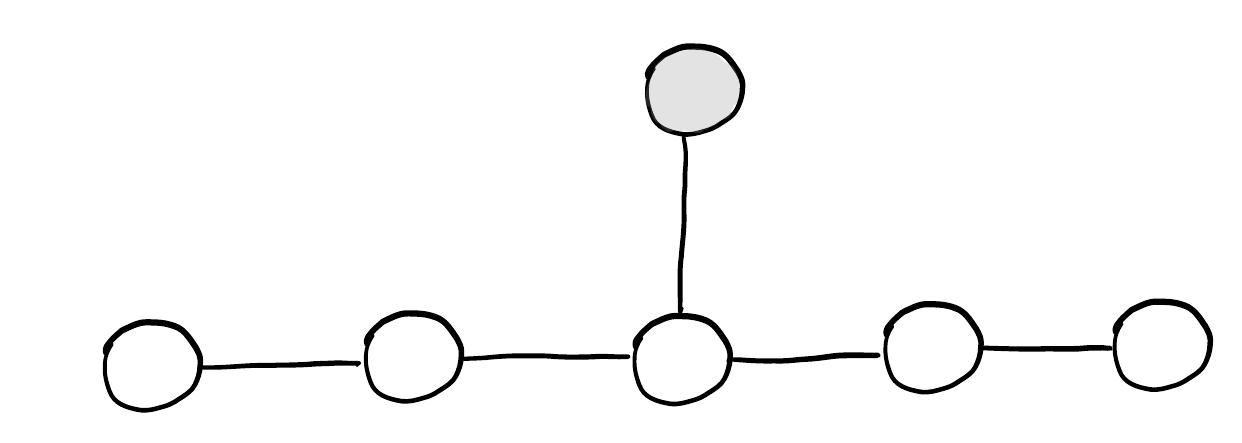}
\caption{The Dynkin diagram of $E_6\times E_6$  with the shaded node corresponding to node with flux.}
\label{easduetyt}
\end{figure}

One can consider other choices of the flux. For example taking the flux to be such that one of the $E_6$ factors is broken to $U(1)\times SU(6)$ we claim the torus with unit flux is in Fig. \ref{coebeuyr}. 
 The node of the Dynkin diagram with the flux is depicted in Figure \ref{easduetyt}.

We can decompose the torus to two equal tubes such that each has fractional flux. We can also further decompose the tubes to two different ones having different flux.   One can check that the anomalies of the model agree with six dimensions and that the index forms the representations of the symmetry, at least in similar limits as the one we discussed here.

\

\subsection*{ $E_7$ and $E_8$}

The basic tube here has the form of the affine Dynkin diagram of the $E_7$ and $E_8$ group. The gauge structure here is more involved than in other cases so explicit checks of the claims are harder to perform. Thus in this section we will restrict to discussing the affine quivers which are obtained by combining a single tube to form a torus. The flux in both cases is to the $U(1)$ corresponding to the central node of one of the groups. 
 The value of the flux is $\frac1{12}$ for the $E_7$ case and $\frac1{30}$ for the $E_8$ case.

\begin{figure}[htbp]
\begin{center}
\includegraphics[scale=0.68]{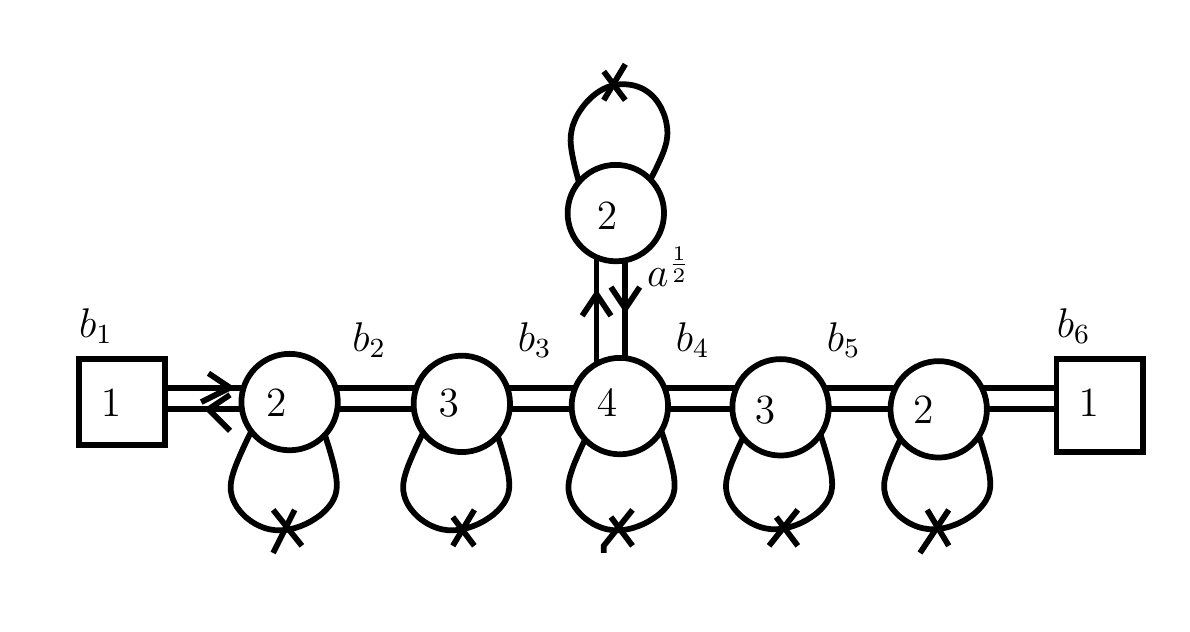}
\caption{The quiver for the compactification of minimal $E_7$ conformal matter on torus with flux $1/12$ to the $U(1)$ corresponding to the central node of one of the two $E_7$ symmetry factors.
The fugacities denote the $U(1)$ symmetries under which the bifundamental fields are charged. As usual all adjoints are charged under an additional $U(1)$ with charge minus one and the bifundamentals are charged $1/2$.
}
\label{wedgeryut}
\end{center}
\end{figure}

A theory, preserving all the symmetry, can be constructed when we combine multiples of the basic theory to get non fractional flux. The symmetry in the $E_7$ case is $SU(4)\times SU(3)\times SU(2)\times U(1)\times E_7$ and is $SU(5)\times SU(2)\times SU(3)\times  E_8\times U(1)$ for the case of $E_8$ (minimal) conformal matter.

We can compute the index in the limit we have studied. Let us quote the result for $E_7$. Without the flip fields, but with the additional singlets with same charges as the adjoints, this is given by,

\be
1+t+t^2+({\bf 56}+1)t^3+({\bf 133}+{\bf 56}-5)t^4+ ...\,.
\ee We see that the index forms representations of $E_7$. We also see that the first $E_7$ representation is the ${\bf 56}$ which enters at order $t^3$. This is in accordance with our discussion in section 2, where we noted that these classes of theories have operators in the bifundamental representation, which for the case of $E_7$ means one in the $({\bf 56},{\bf 56})$. Furthermore, in the minimal case considered here it is expected to contribute with R-charge $6$ under the $U(1)$ R-symmetry inherited from $6d$. When converted to the index limit used here, this indeed gives an operator contributing at order $t^3$.        

The way the $E_7$ representations arise is as follows. We decompose $E_7$ to $SO(12)\times SU(2)$. The $SU(2)$ Cartan is the $U(1)_{\it a}$ symmerty appearing  in Figure \ref{wedgeryut}. The other $U(1)$ symmetries map to the Cartan of $SO(12)$. We denote $v_i$ to be Cartan of $SO(12)$ so that the vector is $$\sum_{c=1}^6 v_c^{\pm1}.$$ Then the map of charges is,

\be
&& b_2 =(v_4 v_1^2 v_6)^{\frac16}\, , \;\;\, b_5=(\frac{v_2}{v_5 v_3^2})^{\frac16}\,, \;\;\, \qquad 
 b_4 =(\frac{v_2 v_3^2}{v_5})^{\frac16}\,,\\
 &&                                        b_3=(\frac{v_4 v_6}{v_1^2})^{\frac16} \, , \;\;\,\qquad \, b_6= \sqrt{v_2 v_5}\, ,\;\;\,\qquad b_1=\sqrt{\frac{v_4}{v_6}} \, . \;\, \;\qquad \;\,\; \nonumber
\ee In general we would then make this assignment to the $\gamma$ copy of $E_7$ to the $M_i$ operators of the tube and then derive the charges with respect to $\beta$ copies according to the boundary conditions.

\begin{figure}[htbp]
\begin{center}
\includegraphics[scale=0.68]{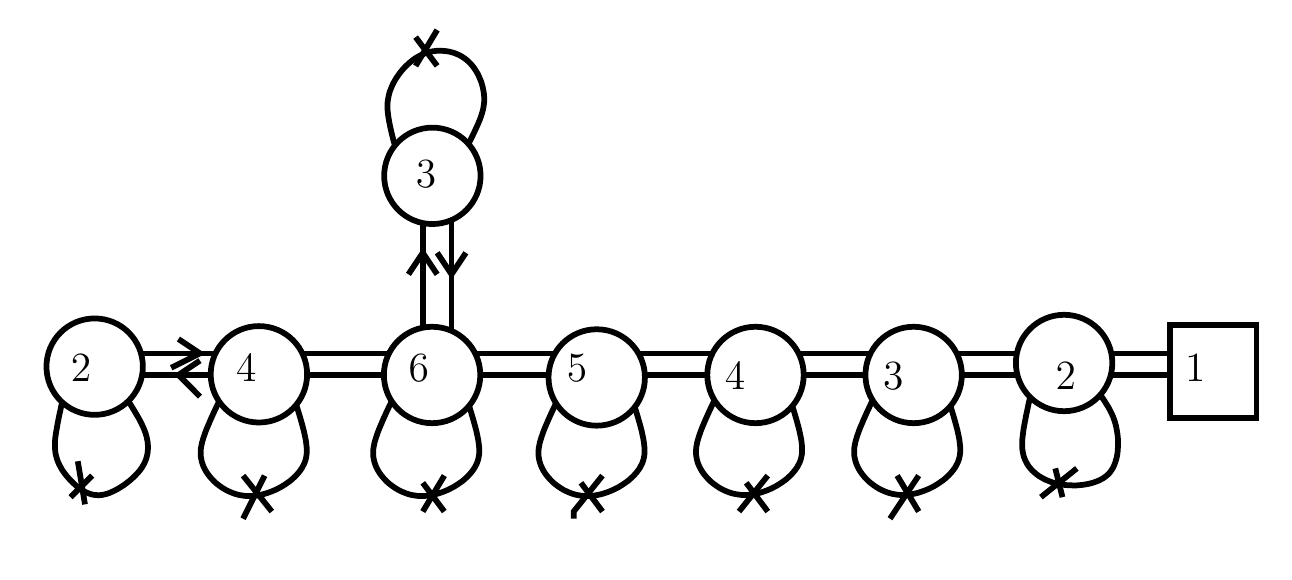}
\caption{The affine quiver diagram corresponding to compactification on torus of $E_8$ minimal conformal matter with flux $1/30$ for the $U(1)$ corresponding to central node of one of the two $E_8$ factors.}
\end{center}
\end{figure}

%%%%%%%%%%%%%%%%%%%%%%%%%%%%%%%%

\section*{Acknowledgments}
We like to thank SCGP summer workshop 2017 for hospitality during part of this work.
  The research of HK and CV is supported in part by NSF grant PHY-1067976. HK is supported in part by the National Research Foundation of Korea (NRF) Grant 2018R1D1A1B07042934. GZ is supported in part by  World Premier International Research Center Initiative (WPI), MEXT, Japan.  The research of SSR was  supported by Israel Science Foundation under grant no. 1696/15 and by I-CORE  Program of the Planning and Budgeting Committee.
  \vfill\eject

%%%%%%%%%%%%%%%%%%%%%%%%%%%%%%%%

%%%%%%%%%%%%%%%%%%%%%%%%%%%%%%%%
\bibliographystyle{./aug/ytphys}
\bibliography{./aug/refs}

\providecommand{\href}[2]{#2}\begingroup\raggedright\begin{thebibliography}{10}

\bibitem{HMRV}
J.~J. Heckman, D.~R. Morrison, T.~Rudelius, and C.~Vafa, ``{Atomic
  Classification of 6D SCFTs},''
  \href{http://dx.doi.org/10.1002/prop.201500024}{{\em Fortsch. Phys.}
  {\bfseries 63} (2015) 468--530},
\href{http://arxiv.org/abs/1502.05405}{{\ttfamily arXiv:1502.05405 [hep-th]}}.
%%CITATION = ARXIV:1502.05405;%%.

\bibitem{ZHTV}
M.~Del~Zotto, J.~J. Heckman, A.~Tomasiello, and C.~Vafa, ``{6d Conformal
  Matter},'' \href{http://dx.doi.org/10.1007/JHEP02(2015)054}{{\em JHEP}
  {\bfseries 02} (2015) 054},
\href{http://arxiv.org/abs/1407.6359}{{\ttfamily arXiv:1407.6359 [hep-th]}}.
%%CITATION = ARXIV:1407.6359;%%.

\bibitem{Bhardwaj:2015xxa}
L.~Bhardwaj, ``{Classification of 6d $ \mathcal{N}=\left(1,0\right) $ gauge
  theories},'' \href{http://dx.doi.org/10.1007/JHEP11(2015)002}{{\em JHEP}
  {\bfseries 11} (2015) 002},
\href{http://arxiv.org/abs/1502.06594}{{\ttfamily arXiv:1502.06594 [hep-th]}}.
%%CITATION = ARXIV:1502.06594;%%.

\bibitem{Gaiotto:2009we}
D.~Gaiotto, ``{${\cal N}=2$ dualities},''
\href{http://arxiv.org/abs/arXiv:0904.2715}{{\ttfamily arXiv:arXiv:0904.2715
  [hep-th]}}.
%%CITATION = 0904.2715;%%.

\bibitem{Benini:2009mz}
F.~Benini, Y.~Tachikawa, and B.~Wecht, ``{Sicilian gauge theories and N=1
  dualities},'' \href{http://dx.doi.org/10.1007/JHEP01(2010)088}{{\em JHEP}
  {\bfseries 01} (2010) 088},
\href{http://arxiv.org/abs/0909.1327}{{\ttfamily arXiv:0909.1327 [hep-th]}}.
%%CITATION = ARXIV:0909.1327;%%.

\bibitem{Bah:2012dg}
I.~Bah, C.~Beem, N.~Bobev, and B.~Wecht, ``{Four-Dimensional SCFTs from
  M5-Branes},'' \href{http://dx.doi.org/10.1007/JHEP06(2012)005}{{\em JHEP}
  {\bfseries 06} (2012) 005},
\href{http://arxiv.org/abs/1203.0303}{{\ttfamily arXiv:1203.0303 [hep-th]}}.
%%CITATION = ARXIV:1203.0303;%%.

\bibitem{Gaiotto:2015usa}
D.~Gaiotto and S.~S. Razamat, ``{$ \mathcal{N}=1 $ Theories of Class $
  {\mathcal{S}}_k $},'' \href{http://dx.doi.org/10.1007/JHEP07(2015)073}{{\em
  JHEP} {\bfseries 07} (2015) 073},
\href{http://arxiv.org/abs/1503.05159}{{\ttfamily arXiv:1503.05159 [hep-th]}}.
%%CITATION = ARXIV:1503.05159;%%.

\bibitem{Razamat:2016dpl}
S.~S. Razamat, C.~Vafa, and G.~Zafrir, ``{4d ${\cal N}=1$ from 6d (1,0)},''
\href{http://arxiv.org/abs/1610.09178}{{\ttfamily arXiv:1610.09178 [hep-th]}}.
%%CITATION = ARXIV:1610.09178;%%.

\bibitem{Bah:2017gph}
I.~Bah, A.~Hanany, K.~Maruyoshi, S.~S. Razamat, Y.~Tachikawa, and G.~Zafrir,
  ``{4d $ \mathcal{N}=1 $ from 6d $\mathcal{N}=\left(1,0\right) $ on a torus
  with fluxes},'' \href{http://dx.doi.org/10.1007/JHEP06(2017)022}{{\em JHEP}
  {\bfseries 06} (2017) 022},
\href{http://arxiv.org/abs/1702.04740}{{\ttfamily arXiv:1702.04740 [hep-th]}}.
%%CITATION = ARXIV:1702.04740;%%.

\bibitem{Kim:2017toz}
H.-C. Kim, S.~S. Razamat, C.~Vafa, and G.~Zafrir, ``{E-String Theory on Riemann
  Surfaces},''
\href{http://arxiv.org/abs/1709.02496}{{\ttfamily arXiv:1709.02496 [hep-th]}}.
%%CITATION = ARXIV:1709.02496;%%.

\bibitem{Kim:2018bpg}
H.-C. Kim, S.~S. Razamat, C.~Vafa, and G.~Zafrir, ``{$D$-type Conformal Matter
  and $SU/USp$ Quivers},''
\href{http://arxiv.org/abs/1802.00620}{{\ttfamily arXiv:1802.00620 [hep-th]}}.
%%CITATION = ARXIV:1802.00620;%%.

\bibitem{OSTYKafffguy}
K.~Ohmori, H.~Shimizu, Y.~Tachikawa, and K.~Yonekura, ``{6D $\mathcal{N}=(1,0)$
  Theories on $T^2$ and Class S Theories: Part I},''
  \href{http://dx.doi.org/10.1007/JHEP07(2015)014}{{\em JHEP} {\bfseries 07}
  (2015) 014},
\href{http://arxiv.org/abs/1503.06217}{{\ttfamily arXiv:1503.06217 [hep-th]}}.
%%CITATION = ARXIV:1503.06217;%%.

\bibitem{OSTYKdf}
K.~Ohmori, H.~Shimizu, Y.~Tachikawa, and K.~Yonekura, ``{6d
  $\mathcal{N}=\left(1,\;0\right) $ theories on S$^{1}$/T$^{2}$ and class S
  theories: part II},'' \href{http://dx.doi.org/10.1007/JHEP12(2015)131}{{\em
  JHEP} {\bfseries 12} (2015) 131},
\href{http://arxiv.org/abs/1508.00915}{{\ttfamily arXiv:1508.00915 [hep-th]}}.
%%CITATION = ARXIV:1508.00915;%%.

\bibitem{DelZotto:2015rca}
M.~Del~Zotto, C.~Vafa, and D.~Xie, ``{Geometric engineering, mirror symmetry
  and $6{\mathrm{d}}_{\left(1,0\right)}\to
  4{\mathrm{d}}_{\left(\mathcal{N}=2\right)} $},''
  \href{http://dx.doi.org/10.1007/JHEP11(2015)123}{{\em JHEP} {\bfseries 11}
  (2015) 123},
\href{http://arxiv.org/abs/1504.08348}{{\ttfamily arXiv:1504.08348 [hep-th]}}.
%%CITATION = ARXIV:1504.08348;%%.

\bibitem{Nardoni:2016ffl}
E.~Nardoni, ``{4d SCFTs from negative-degree line bundles},''
\href{http://arxiv.org/abs/1611.01229}{{\ttfamily arXiv:1611.01229 [hep-th]}}.
%%CITATION = ARXIV:1611.01229;%%.

\bibitem{Fazzi:2016eec}
M.~Fazzi and S.~Giacomelli, ``{$\mathcal{N} = 1$ superconformal theories with
  $D_N$ blocks},''
\href{http://arxiv.org/abs/1609.08156}{{\ttfamily arXiv:1609.08156 [hep-th]}}.
%%CITATION = ARXIV:1609.08156;%%.

\bibitem{HZ2018}
A.~Hanany and G.~Zafrir, ``{Discrete Gauging in Six Dimensions},''
  \href{http://arxiv.org/abs/hep-th/1804.08857}{{\ttfamily
  arXiv:hep-th/1804.08857}}.

\bibitem{HM2018}
A.~Hanany and N.~Mekareeya, ``{The Small $E_8$ Instanton and the Kraft Procesi
  Transition},'' \href{http://arxiv.org/abs/hep-th/1801.01129}{{\ttfamily
  arXiv:hep-th/1801.01129}}.

\bibitem{OSTY}
K.~Ohmori, H.~Shimizu, Y.~Tachikawa, and K.~Yonekura, ``{Anomaly Polynomial of
  General 6D SCFTs},'' \href{http://dx.doi.org/10.1093/ptep/ptu140}{{\em PTEP}
  {\bfseries 2014} no.~10, (2014) 103B07},
\href{http://arxiv.org/abs/1408.5572}{{\ttfamily arXiv:1408.5572 [hep-th]}}.
%%CITATION = ARXIV:1408.5572;%%.

\bibitem{Chan:2000qc}
C.~S. Chan, O.~J. Ganor, and M.~Krogh, ``{Chiral compactifications of 6-D
  conformal theories},''
  \href{http://dx.doi.org/10.1016/S0550-3213(00)00706-9}{{\em Nucl. Phys.}
  {\bfseries B597} (2001) 228--244},
\href{http://arxiv.org/abs/hep-th/0002097}{{\ttfamily arXiv:hep-th/0002097
  [hep-th]}}.
%%CITATION = HEP-TH/0002097;%%.

\bibitem{Gaiotto:2015una}
D.~Gaiotto and H.-C. Kim, ``{Duality walls and defects in 5d $ \mathcal{N}=1 $
  theories},'' \href{http://dx.doi.org/10.1007/JHEP01(2017)019}{{\em JHEP}
  {\bfseries 01} (2017) 019},
\href{http://arxiv.org/abs/1506.03871}{{\ttfamily arXiv:1506.03871 [hep-th]}}.
%%CITATION = ARXIV:1506.03871;%%.

\bibitem{Horava:1996ma}
P.~Horava and E.~Witten, ``{Eleven-dimensional supergravity on a manifold with
  boundary},'' \href{http://dx.doi.org/10.1016/0550-3213(96)00308-2}{{\em Nucl.
  Phys.} {\bfseries B475} (1996) 94--114},
\href{http://arxiv.org/abs/hep-th/9603142}{{\ttfamily arXiv:hep-th/9603142
  [hep-th]}}.
%%CITATION = HEP-TH/9603142;%%.

\bibitem{Horava:1995qa}
P.~Horava and E.~Witten, ``{Heterotic and type I string dynamics from
  eleven-dimensions},''
  \href{http://dx.doi.org/10.1016/0550-3213(95)00621-4}{{\em Nucl. Phys.}
  {\bfseries B460} (1996) 506--524},
  \href{http://arxiv.org/abs/hep-th/9510209}{{\ttfamily arXiv:hep-th/9510209
  [hep-th]}}.
[,397(1995)].
%%CITATION = HEP-TH/9510209;%%.

\bibitem{Tachikawa:2015mha}
Y.~Tachikawa, ``{Instanton Operators and Symmetry Enhancement in 5D
  Supersymmetric Gauge Theories},''
\href{http://arxiv.org/abs/1501.01031}{{\ttfamily arXiv:1501.01031 [hep-th]}}.
%%CITATION = ARXIV:1501.01031;%%.

\bibitem{Yonekura:2015ksa}
K.~Yonekura, ``{Instanton operators and symmetry enhancement in 5d
  supersymmetric quiver gauge theories},''
  \href{http://dx.doi.org/10.1007/JHEP07(2015)167}{{\em JHEP} {\bfseries 07}
  (2015) 167},
\href{http://arxiv.org/abs/1505.04743}{{\ttfamily arXiv:1505.04743 [hep-th]}}.
%%CITATION = ARXIV:1505.04743;%%.

\bibitem{Romelsberger:2005eg}
C.~Romelsberger, ``{Counting chiral primaries in N = 1, d=4 superconformal
  field theories},''
  \href{http://dx.doi.org/10.1016/j.nuclphysb.2006.03.037}{{\em Nucl. Phys.}
  {\bfseries B747} (2006) 329--353},
\href{http://arxiv.org/abs/hep-th/0510060}{{\ttfamily arXiv:hep-th/0510060
  [hep-th]}}.
%%CITATION = HEP-TH/0510060;%%.

\bibitem{Kinney:2005ej}
J.~Kinney, J.~M. Maldacena, S.~Minwalla, and S.~Raju, ``{An Index for 4
  dimensional super conformal theories},''
  \href{http://dx.doi.org/10.1007/s00220-007-0258-7}{{\em Commun. Math. Phys.}
  {\bfseries 275} (2007) 209--254},
\href{http://arxiv.org/abs/hep-th/0510251}{{\ttfamily arXiv:hep-th/0510251
  [hep-th]}}.
%%CITATION = HEP-TH/0510251;%%.

\bibitem{Beem:2012yn}
C.~Beem and A.~Gadde, ``{The ${\cal N}=1$ superconformal index for class $S$
  fixed points},'' \href{http://dx.doi.org/10.1007/JHEP04(2014)036}{{\em JHEP}
  {\bfseries 04} (2014) 036},
\href{http://arxiv.org/abs/1212.1467}{{\ttfamily arXiv:1212.1467 [hep-th]}}.
%%CITATION = ARXIV:1212.1467;%%.

\bibitem{Dolan:2008qi}
F.~A. Dolan and H.~Osborn, ``{Applications of the Superconformal Index for
  Protected Operators and q-Hypergeometric Identities to N=1 Dual Theories},''
  \href{http://dx.doi.org/10.1016/j.nuclphysb.2009.01.028}{{\em Nucl. Phys.}
  {\bfseries B818} (2009) 137--178},
\href{http://arxiv.org/abs/0801.4947}{{\ttfamily arXiv:0801.4947 [hep-th]}}.
%%CITATION = ARXIV:0801.4947;%%.

\bibitem{Hanany:2015pfa}
A.~Hanany and K.~Maruyoshi, ``{Chiral theories of class $ \mathcal{S} $},''
  \href{http://dx.doi.org/10.1007/JHEP12(2015)080}{{\em JHEP} {\bfseries 12}
  (2015) 080},
\href{http://arxiv.org/abs/1505.05053}{{\ttfamily arXiv:1505.05053 [hep-th]}}.
%%CITATION = ARXIV:1505.05053;%%.

\bibitem{Franco:2015jna}
S.~Franco, H.~Hayashi, and A.~Uranga, ``{Charting Class $\mathcal S_k$
  Territory},'' \href{http://dx.doi.org/10.1103/PhysRevD.92.045004}{{\em Phys.
  Rev.} {\bfseries D92} no.~4, (2015) 045004},
\href{http://arxiv.org/abs/1504.05988}{{\ttfamily arXiv:1504.05988 [hep-th]}}.
%%CITATION = ARXIV:1504.05988;%%.

\bibitem{Seiberg:1994pq}
N.~Seiberg, ``{Electric - magnetic duality in supersymmetric nonAbelian gauge
  theories},'' \href{http://dx.doi.org/10.1016/0550-3213(94)00023-8}{{\em Nucl.
  Phys.} {\bfseries B435} (1995) },
\href{http://arxiv.org/abs/hep-th/9411149}{{\ttfamily arXiv:hep-th/9411149
  [hep-th]}}.
%%CITATION = HEP-TH/9411149;%%.

\bibitem{sbfh}
C.~Beem, S.~S. Razamat, and G.~Zafrir, ``{{\it to appear}},''.

\bibitem{Gadde:2011uv}
A.~Gadde, L.~Rastelli, S.~S. Razamat, and W.~Yan, ``{Gauge Theories and
  Macdonald Polynomials},''
  \href{http://dx.doi.org/10.1007/s00220-012-1607-8}{{\em Commun. Math. Phys.}
  {\bfseries 319} (2013) 147--193},
\href{http://arxiv.org/abs/1110.3740}{{\ttfamily arXiv:1110.3740 [hep-th]}}.
%%CITATION = ARXIV:1110.3740;%%.

\end{thebibliography}\endgroup
%%%%%%%%%%%%%%%%%%%%%%%%%%%%%%%%

\end{document}